\documentclass[useAMS,usenatbib]{mn2e}

\usepackage[ansinew]{inputenc}
\usepackage{graphicx}
\usepackage{pdflscape}
\usepackage{longtable} 
\def\aj{AJ}

\def\apj{ApJ}
\def\apjl{ApJ}
\def\apjs{ApJS}

\def\aap{A\&A}

\def\mnras{MNRAS}
\def\pasp{PASP}

\def\physrep{Physics Reports} 

\title[A Spectral Atlas of HII Galaxies]{A Spectral Atlas of HII Galaxies in the Near-Infrared}
\author[Martins et al.]{Lucimara P. Martins$^{1}$\thanks{E-mail:
lucimara.martins@cruzeirodosul.edu.br}, Alberto Rodr\'{\i}guez-Ardila$^2$, Suzi Diniz$^{1,4}$,\newauthor  Ruth Gruenwald$^{3}$, and Ronaldo de Souza$^{3}$\\
$^{1}$NAT - Universidade Cruzeiro do Sul, Rua Galvao Bueno, 868, S\~ao Paulo, SP, Brazil\\
$^{2}$Laborat\'orio Nacional de Astrof\'isica/MCT, Rua dos Estados Unidos 154, CEP~37501-064. Itajub\'a, MG, Brazil\\
$^{3}$Instituto Astron\^omico e Geof\'isico - USP, Rua do Mat\~ao, 1226, S\~ao Paulo, SP\\
$^{4}$Universidade Federal do Rio Grande do Sul - IF, Departamento de Astronomia, CP 15051, 91501-970, Porto Alegre, RS, Brasil.}

\begin{document}


\pagerange{\pageref{firstpage}--\pageref{lastpage}} \pubyear{2009}

\maketitle

\label{firstpage}

\begin{abstract}

Recent models show that TP-AGB stars should dominate the NIR spectra of
populations 0.3 to 2 Gyr old, leaving unique signatures that can be used
to detect young/intermediate stellar population in galaxies.
However, no homogeneous database of star-forming galaxies is available
in the NIR to fully explore and apply these results.
With this in mind, we study the NIR spectra of a sample of 23 H\,{\sc ii}
and starburst galaxies, aimed at characterizing the most prominent spectral
features (emission and
absorption) and continuum shape in the 0.8-2.4~$\mu$m region of these
objects. Five normal galaxies are also observed as a control sample. Spectral indices
are derived for the relevant absorption lines/bands and a comparison with optical
indices of the same sample of galaxies available in the literature is made.
We found no correlation between the optical and the NIR indexes. This is
probably due to the differences in aperture between these two sets of data. That result is
further supported by the absence or weakness of emission lines in the NIR
for a subsample galaxies, while in the optical the emission lines are strong and clear,
which means that the ionisation source in many of these galaxies is not
nuclear, but circumnuclear or located in hot spots outside the nucleus.
We detected important signatures predicted for a stellar population
dominated by the TP-AGBs, like CN 1.1 $\mu$m and CO 2.3$\mu$m. In
at least one galaxy (NGC~4102) the CN band at 1.4 $\mu$m was detected for
the first time. We also detect TiO and ZrO bands in the region 0.8-1~$\mu$m that have
never been reported before in extragalactic sources. The shape of the continuum
emission is found to be strongly correlated to the presence/lack of emission lines. An
observational template for the star-forming galaxies is derived to be used as a benchmark of
stellar population(s) in starburst galaxies against which to compare near-IR spectroscopy of
different types of galaxies, especially those with AGN activity and/or those at
high-redshift.

\end{abstract}


\section{Introduction}

The study of star-formation systems is of extreme importance 
to our understanding of galaxy 
formation and evolution. Age-dating the stellar populations of
galaxies provides us with a cosmic timescale that is independent
of cosmological models as well as with a mean of reconstructing their
star-formation history. It also allows to investigate on issues such as
the propagation of the star formation, how it is triggered and the relationship
between starbursts and nuclear activity. 

In this context, H\,{\sc ii}/star-forming galaxies are excellent and the closest 
laboratories available for studying the above phenomena in
extragalactic sources. Because of the proximity of many of them,
high-angular resolution can be obtained at almost all wavelengths.
Moreover, they can be seen as ``scaled-down'' versions of
the starbursts seen at high redshifts
\citep{bernard-salas+09}.

Therefore, the study of local 
H\,{\sc ii}/star-forming galaxies is a need, as these objects were much 
more numerous in the past \citep{blain+02, elbaz+03}, and
are believed to give rise to the bulk of the cosmic infrared
background radiation \citep{bernard-salas+09}.

Starburst features in the optical region are nowadays considerably
well known and studied, and have been a fundamental tool to identify 
star-formation in galaxies
\citep[eg.][]{kennicutt+88, kennicutt92, worthey+97, balogh+97, gu+06}. 
However, in many cases, the use
of this knowledge is not always possible due either to severe dust obscuration,
or the strength of an active galactic nucleus (AGN) in the optical region. In this case, the near-infrared
region (NIR, hereafter) might be of unprecedented value. At NIR
wavelengths stellar photospheres usually remain the dominant 
sources of light, and galaxy spectra are shaped by red supergiants (RSG) 
shortly after starbursts, and then by giants of the first and of the 
asymptotic giant branches (AGB). Indeed, evolved intermediate-mass (2 $-$ 10 M$_\odot$) 
stars have been shown to contribute significantly to integrated NIR
fluxes, even when they represent a negligible contribution to the stellar mass 
\citep{persson+83,frogel+90, melbourne+12,lyubenova+12}. 

The use of the NIR to study stellar populations is not new. 
\citet{rieke+80} studied the stellar population of NGC\,253 and
M\,82 using spectroscopy at this wavelength range. They 
detected a strong 2.2~$\micron$ CO band, suggesting the presence of a dominant 
population of red giants and supergiants in the
nuclear region of these sources. Many other
authors have been using the NIR wavelength region to study
stellar populations in galaxies, mostly based on the 2.2~$\micron$ CO band
 \citep[eg][]{origlia+93,oliva+95,engelbracht+98,lancon+01,riffel+08,riffel+09, melbourne+10}. 
Photometric methods in this region have also been used 
 \citep[eg.][]{moorwood+82,hunt+03,micheva+12}.

It was \citet{heisler+99} who presented
the first systematic study of H\,{\sc ii}/starburts galaxies in the
NIR, covering the interval 1-2.4$\micron$. Their sample, composed of
13 of such galaxies, was characterised by a peak of the continuum 
emission at 60$\micron$. They attribute the relatively warm colours of
these objects to the fact that
they are undergoing their first major episode of
massive star formation. Based on the colours found from
a sample of AGNs, they proposed an evolutionary scenario
in which the starburst represents the early stages that eventually evolves
into Seyfert when the starburst has faded and collapsed to form and/or fuel 
an AGN. Since then, many works were devoted to the study of samples composed
of H\,{\sc ii}/star-forming  and normal galaxies as well as AGNs
\citep[][to name a few]{goldader+97, coziol+01, 
reunanen+02,reunanen+03, alonso-herrero+00, 
cesetti+09, bendo+04, riffel+08,riffel+09,martins+10,kotilainen+12}.
Up to our knowledge, no homogeneous sample of local H\,{\sc ii}/star-forming 
galaxy spectra ($z<$0.02) has been published in the literature in the
NIR, deterring efforts to gather a clearer view of these sources.
The discussion becomes highly relevant in the light of
recent results obtained by \citet{melbourne+12}, which shows that
TP-AGB sequences can account for as much as 17\% of the 1.6$\mu$m
flux for local galaxies and for a much larger fraction at higher redshifts.
Indeed, these authors show that at high-$z$, the summation of all IR luminous
sources can be expected to be as high as 70\%, most of
which comes from TP-AGB and red helium burning stars (RHeBs).
Other approaches \citep{zibetti+12} claim that the two main signatures
of the presence of TP-AGB stars predicted by the \citet{maraston05}'s models,
CN at 1.41$\mu$m and C$_2$ at 1.77$\mu$m, are not detected in a 
sample of 16 post-starburst galaxies.

With all this in mind, we obtained NIR spectra of a sample
of galaxies known to have young stellar populations from optical
observations,
suitable for studying their most important spectral features and
that can be used as proxies for stellar population studies.
We also include ``normal" galaxies in our
sample as a control sample. This study aims
to complement efforts made by other authors in the optical \citet{ho+95}
and in the mid-infrared \citep{bernard-salas+09} to characterised
these sources.
In \S 2 we present the details of our observations and reduction
process; in \S 3 we show the measurement of NIR indexes; in \S 4
we present the measurement of the emission lines; in \S 5 we discuss and correlate
the results from the previous sections, in \S 6 we analyse the continuum shape
and in \S 7 we present our conclusions.
 
\section{Observations and Reduction}

The galaxies studied here are a subset of those presented in the magnitude-limited
optical spectroscopic survey of nearby bright galaxies of \citet[hereafter HO95]{ho+95}. They observed 
over 486 galaxies having $B_T\leq$12.5 mag and $\delta>0\deg$. 
Sources falling into category (2) defined by \citet[hereafter HO97]{ho+97} as those composed 
of ``nuclei dominated by emission lines from regions of active star formation (H\,{\sc ii} or starburst
nuclei)" were prone to observation. In addition, five galaxies classified as ``normal"
in the optical, dominated by old-stellar populations and with no detected emission lines, 
were included as a control sample. 
It is important to mention that \citet{kotilainen+12} observed a sample
of "normal" spiral galaxies at NIR wavelengths. They were not used here
as a comparison because of the the difference in slit sizes 
and because they only have H and K bands. 
Although our initial list of objects was composed of nearly 
100 galaxies, only a small fraction of them could be observed during our available observing time. 
In this sense, our sample is not complete as the targets were selected according to their 
availability  in the sky. However, they are representative of their class as all of them were 
extracted from a magnitude-limited catalogue, and have the same distribution of morphology, 
redshifts and H$\alpha$ flux as the original sample.
 
All spectra were obtained at the NASA 3m Infrared
Telescope Facility (IRTF) in two observing runs. The first one
during the period of October\,31 $-$ November\,2, 2007, and the second one
on the nights of April 23-25, 2008. Table~\ref{obs:tab} shows
the log of observations for all galaxies. The
SpeX spectrograph \citep{rayner+03} was used in the short
cross-dispersed mode (SXD, 0.8-2.4 $\mu$m).  The detector
consists of a 1024x1024 ALADDIN~3 InSb array with a spatial scale
of 0.15"/ pixel. A 0.8"x 15" slit oriented in the north-south direction
was used, providing a spectral resolution, on average, of
320 km\,s$^{-1}$. This value was determined both from the arc lamp and 
the sky line spectra and was found to vary very little with wavelength along 
the observed spectra. The seeing varied from night to night but on average, 
most objects were observed
under 1$\arcsec$ seeing conditions.

Observations were done nodding in an object-sky-sky-object  pattern.  
The sky position was usually several arcminutes from the galaxy nucleus
free of extended emission or background stars.
After each galaxy a telluric star, close in airmass to the former, was
observed to remove telluric features and to perform the flux calibration.
The spectral reduction, extraction and wavelength calibration
procedures were performed using SPEXTOOL, the in-house
software developed and provided by the SpeX team for
the IRTF community \citep{cushing+04}. Telluric
features removal and flux calibration were done using XTELLCOR,
another software available by the SpeX team. The different orders were 
merged into a single 1D spectrum from 0.8$\mu$m to 2.4$\mu$m using the 
XMERGEORDERS routine. After this procedure, the IDL routine "Xlightloss", also 
written by the SpeX team, was employed. It corrects an input spectrum 
for slit losses relative to the standard star used for flux calibration. 
This program is useful if either the object or the standard were not 
observed with the slit at the parallactic angle. Differential refraction is,
indeed, the main source of uncertainty in flux calibration and it was
minimised following the above procedure. Finally, the spectra were 
corrected for Galactic extinction using the \citet{cardelli+89} law
and the extinction maps of \citet{schlafly+11}.
 
Table~\ref{obs:tab} shows the final sample, composed of 28 galaxies, 
23 of them classified as H\,{\sc ii}/star-forming galaxies. The remaining
objects are classified as non-starforming 
galaxies for comparison. In addition to the nuclear spectrum, a different 
number of off-nuclear apertures were extracted for each galaxy, depending on 
the size of the extended emission across the slit. Informations for each 
object are also presented in Table~\ref{obs:tab}.

\begin{table*}
\centering
\caption{ Log of observations}

\begin{tabular}{@{}ccccccccccc@{}}
\hline
Galaxy &     z  & Morphology &  Exp. Time (s) & Seeing(\arcsec)&  Apertures & Class & Opt& W{\tiny (G band)}$^a$ & log F(H$_\alpha$)$^a$& Reference\\
       &        &            &                &                &   (b)      & (c)   & (d)&  (e)      & (f)   &\\
\hline

NGC 0221  &-0.0007 & cE2        &    10x150 &  1.4 & nuc + 3 & N & yes & 4.54 & -     & 1 \\
NGC 0278  & 0.0021 & SAB(rs)b   &    12x150 &  0.8 & nuc + 3 & H & yes & 0.50 & -13.45 & 1 \\
NGC 0514  & 0.0082 & SAB(rs)c   &    12x180 &  0.8 & nuc + 2 & H & yes & 3.86 & -14.64 & 1 \\
NGC 0674  & 0.0104 & SAB(r)c    &    12x180 &  1.7 & nuc + 2 & H & yes & 4.38 & -14.31 & 1 \\
NGC 0783  & 0.0173 & SBc        &    10x180 &  0.8 & nuc + 2 & H & yes & 1.98 & -13.54 & 1 \\
NGC 0864  & 0.0052 & SAB(rs)c   &    10x180 &  0.9 & nuc + 2 & H & yes & 1.03 & -12.88 & 1 \\
NGC 1174  & 0.0091 & SB(r)bc    &    10x180 &  0.9 & nuc + 2 & H & yes & 3.04 & -13.07 & 1 \\
NGC 1232  & 0.0053 & SAB(rs)c   &    8x180  &  1.3 & nuc     & N & no  &   -  & -    & 2 \\
NGC 1482  & 0.0064 & SA0+ pec   &    4x180  &  1.3 & nuc +2  & H & no  &   -  & -    & 3 \\
NGC 2339  & 0.0074 & SAB(rs)bc  &    6x180  &  1.1 & nuc + 2 & H & yes & 0.00 & -13.07 & 1 \\
NGC 2342  & 0.0176 & S pec      &    7x180  &  1.0 & nuc     & H & yes & 0.67 &  -12.83& 1 \\
NGC 2903  & 0.0019 & SAB(rs)bc  &    12x180 &  0.6 & nuc + 7 & H & yes & 0.57 &  -12.63& 1\\ 
NGC 2950  & 0.0045 & (R)SB(r)0  &    10x180 &  1.1 & nuc + 4 & N & yes & 4.98 &  -    & 1\\
NGC 2964  & 0.0044 & SAB(r)bc   &    12x180 &  1.0 & nuc + 2 & H & yes & 1.03 &  -12.75& 1 \\
NGC 3184  & 0.0020 & SAB(rs)cd  &    12x180 &  0.6 & nuc + 2 & H & yes & 1.37 &  -13.12& 1 \\
NGC 4102  & 0.0028 & SAB(s)b    &    10x180 &  0.8 & nuc + 2 & H & yes & 1.94 &  -12.52& 1\\
NGC 4179  & 0.0042 & S0         &    8x180  &  1.3 & nuc + 2 & N & yes & 4.89 &  -14.30& 1\\
NGC 4303  & 0.0052 & SAB(rs)bc  &    10x180 &  0.9 & nuc + 6 & H & yes & 3.60 &  -12.84& 1\\
NGC 4461  & 0.0064 & SB(s)0+    &    16x180 &  1.0 & nuc + 2 & N & yes & 4.67 &  -     & 1\\
NGC 4845  & 0.0041 & SA(s)ab    &    10x180 &  0.8 & nuc + 6 & H & yes & 4.18 & -13.61 & 1\\
NGC 5457  & 0.0008 & SAB(rs)cd  &    10x180  &  2.0 & nuc + 2 & H & yes & 1.03 & -13.33 & 1\\
NGC 5905  & 0.0113 & SB(r)b  &    10x180 &   0.8 & nuc + 2 & H & yes & 2.15 & -13.13 & 1\\
NGC 6181  & 0.0079 & SA(rs)c  &    9x180  &   0.7 & nuc + 2 & H & yes & 4.43 & -13.57 & 1\\
NGC 6946  & 0.0195 & SAB(rs)cd  &    16x180 &   1.1 & nuc     & H & yes & 0.32 & -13.01 & 1\\
NGC 7080  & 0.0161 & SB(r)b  &    10x150 &   0.8 & nuc + 2 & H & yes & -    & -13.49 & 1\\
NGC 7448  & 0.0073 & SA(rs)bc  &    12x180 &   1.1 & nuc + 2 & H & yes & 0.42 & -14.01& 1\\
NGC 7798  & 0.0080 & SBc        &    8x180   &  1.0 & nuc + 2 & H & yes & 0.04 & -13.16& 1\\
NGC 7817  & 0.0077 & SAbc  &    12x180 &   1.1 & nuc + 2 & H & yes & 3.57 & -13.25& 1\\

\hline
\end{tabular}
\raggedright
References:
(1) \citet{ho+95}
(2) \citet{kennicutt+88}
(3) \citet{coziol+98}
\\
(a)Adopted from \citet{ho+97}.
(b) Number of apertures extracted from each galaxy. (c) H means HII galaxy and 
N means normal galaxy. (d) Does it have observations in the optical? 
(e) Measured G band. (f) Measured H$\alpha$ flux.
\label{obs:tab}
\end{table*}

\section{Near-Infrared Spectral Indexes} \label{indices}

The final reduced NIR spectra for all the galaxies are presented in the appendix, 
together with the optical spectra observed by HO95 for comparison.
Note that for some extractions, the region between $\sim$0.8--1$\mu$m is
not shown. This is because of the very small S/N blueward of 1$\mu$m.
Sometimes, if the S/N was small but it was still possible to detect the shape
of the continuum emission, the region blueward of 1$\mu$m was painted
in light colour to ease the visualisation of the spectra. In these figures
the apertures are marked on the right side of each spectra. Apertures denoted by "N" 
are extracted to the north of the nucleus and the ones marked with "S" are to the
south of the nucleus. The numbers following the letters "N" or "S" increase with the
distance to the centre.

A simple visual inspection of the NIR galaxy spectra 
reveals a large diversity in terms of both continuum shapes and strength of the emission
and absorption lines/bands. Conspicuous emission features that appear most frequently in the
objects are [S\,{\sc iii}]\,0.953$\micron$, 
He\,{\sc i}\,1.083$\micron$, Pa$\gamma$, [Fe\,{\sc ii}]\,1.257$\micron$, 
Pa$\beta$, H$_2$ 2.121$\mu$m and Br$\gamma$. Pa$\alpha$ was not measured
for many galaxies because it falls partially or totally in the gap between
the $H$- and $K$-bands. Note also that some H\,{\sc ii}/starburst spectra 
lack of emission lines. This diversity strongly contrasts to what is seen in the
optical region: all H\,{\sc ii}/star-forming galaxies in the HO95's catalogue 
show a similar continuum shape and strong to moderate emission lines
(see discussion in section 5). 

It is easy to see that all galaxy spectra are dominated by stellar absorption 
lines/bands, from the very blue to the red end of the observed spectral 
region. CaT at 0.85$\micron$ and CN at 1.1$\micron$, very prominent in most sources, 
as well as Si\,{\sc i}, Mg\,{\sc i}, Na\,{\sc i}, Ca\,{\sc i} and CO, are the most 
common features observed. All of them are marked in the plots. Note that these 
absorption lines are also detected in normal galaxies but their strength
seems to differ between the later and the former. We also found evidence of
absorption features/bands not previously reported in extragalactic sources.
An analysis of all the spectral characteristics is made in Sections~\ref{indices}
and~\ref{continuum}.

The observed continuum emission is also found to vary considerably from source to source. 
It can be seen, for instance, that there are objects dominated by a very steep 
continuum (NGC\,7448, for example), with the flux continuously increasing with 
decreasing wavelength. Other galaxies display a continuum emission that rises 
steeply with increasing wavelength up to $\sim1.1\micron$, then the continuum 
gets flatter in the $H$-band, and finally, it decreases steeply with wavelength 
in the $K$-band. NGC\,6946 in an example of that category. This issue will be
explored in more detail in Sect.~\ref{continuum}.

Some of the most important signatures predicted for a stellar population dominated 
by TP-AGBs \citep{maraston05} are the CN\,1.1 $\micron$ and the CO 2.3$\micron$. These features 
can be clearly seen in most of our spectra by a simple visual inspection (for example, for NGC~2339,
NGC~2903, NGC~4102 or NGC~6946). 
CN~1.4 $\micron$ and C$_2$~1.768 $\micron$, also expected to be strong in TP-AGBs, although more difficult
to observe. The former falls in a region of very poor atmospheric transmission, 
between the $J-$ and the $H-$bands and rarely any information falling in this region can
be recovered after division by the telluric star. In only one galaxy of our sample (NGC\,4102),
we have secure evidence of its detection. This can be seen in Figure~14, at the appendix,
where all the bands are marked.
 This is the first detection of CN\,1.4$\mu$m in an extragalactic object.

Observations of C$_2$~1.768 $\micron$ are possible provided the redshift 
is small enough. This is because of a gap between the $H-$ and $K-$bands in SpeX, which
in laboratory wavelength starts at 1.80$\mu$m and extends up to 1.88$\mu$m. 
C$_2$ have been detected only in NGC\,5102 \citep{miner+11}. In our sample,
evidence of C$_2$~1.768 $\micron$ is found in NGC\,1232, NGC\,1482, NGC\,2339, NGC\,2950,
NGC\,4179, NGC\,4303, NGC\,4845 and NGC\,6946 (see Figures in the Appendix). Note, however, 
that the region around that feature is noisy and with the present data it is not possible 
to fully warrant its presence. 

Another spectral region of interest is the one located between 0.8 and 1$\micron$. 
It contains at least four absorption bands in addition to the well-studied CaT system  
at 0.850, 0.854 and 0.866\,$\mu$m \citep{garcia-rissman+05, vega+09}. 
A visual inspection of the spectra reveals the presence of broad absorption features at 
0.82, 0.88, 0.92, and $\sim1\micron$ in some objects. These absorptions are rather complex,
varying in shape and strength from source to source. 
 No information have been published in the literature on identified bands
on that spectral region in galaxies.   
 
\begin{figure*} 
\includegraphics[width=160mm]{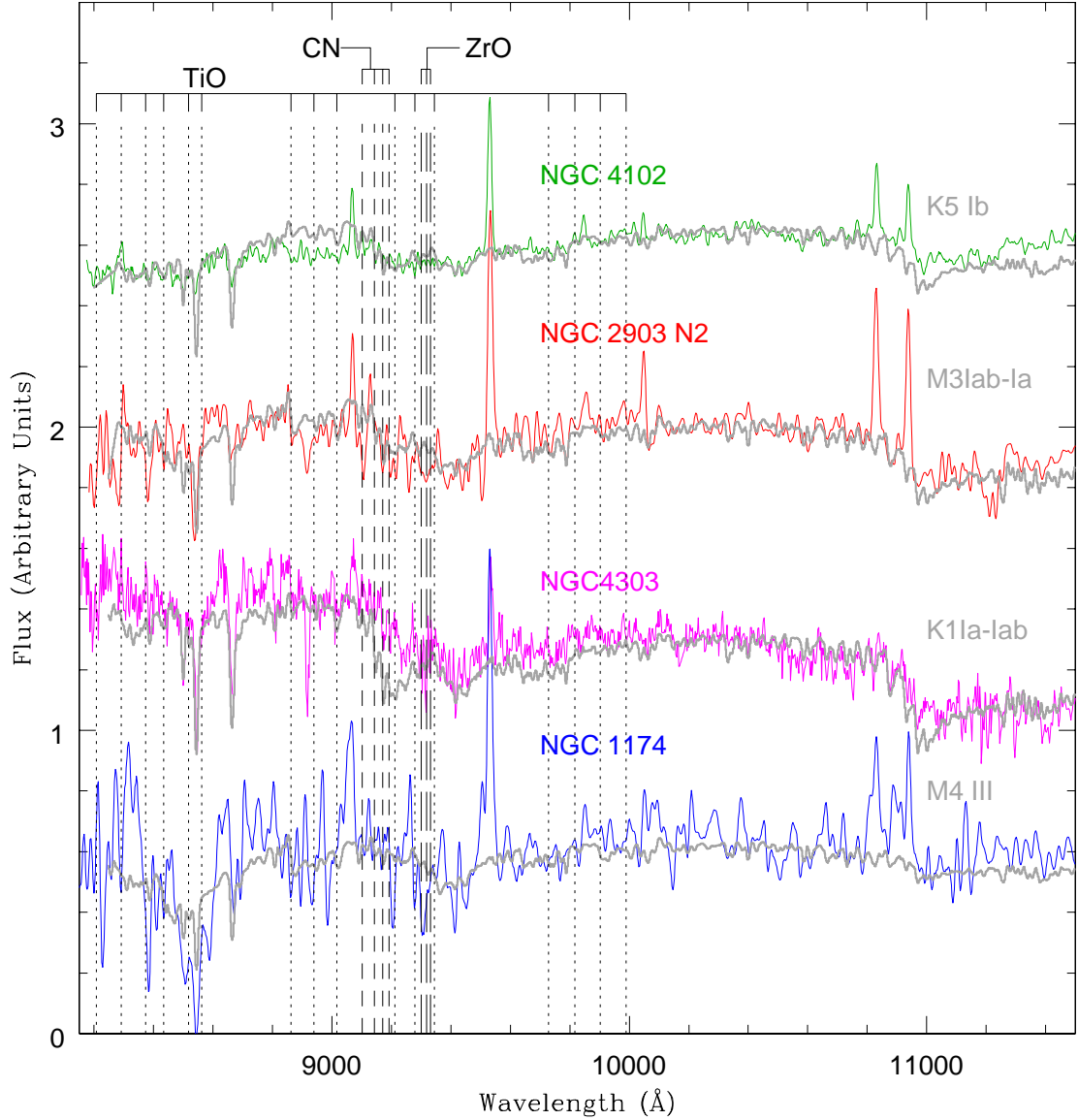}
\caption{Zoom of the observed spectra in the region 0.8--1.1\,$\micron$ for 
four sources of the sample. Overlaid to each galaxy continuum is the stellar 
template (in thick grey) from the library of R09 that best matches the observed
continuum. The vertical lines mark the position of the band heads of TiO (dotted); 
CN (dashed) and ZrO (long-dashed). }
\label{fig:tiobands}
\end{figure*}

Based on the spectral library of cool stars published recently by \citet[][hereafter R09]{rayner+09},
we identified the absorptions at 0.82, 0.88 and 0.92$\mu$m as due to TiO. This molecule 
displays numerous absorption systems in the spectra of late-type stars over the 
0.8 $-$1.5\,$\mu$m wavelength range (see, for instance, Figures 7-9, 12, 28-30 of R09). 
Several TiO band heads starting at 0.82$\mu$m are listed in their Table\,10,
which coincides in position to the ones observed in our sample. It also easy to see that these
features dramatically change with the spectral type of the star. In addition
to TiO, other absorption features attributed to CN and ZrO are also identified
by R09 in the spectra of M giant stars and 
carbon stars. A cross-correlation of some of R09's stars with our data allow us to
confirm that all these features (TiO, CN and ZrO) are indeed present in the
galaxy spectra. Figure~\ref{fig:tiobands} shows a zoom around the 0.8$-$1~$\mu$m region for the 
objects where the above features are clearly detected. For comparison, we also 
include the stellar spectrum that closely matches the observed galaxy continuum
in the 0.8-1.12\,$\mu$m region. Note that the agreement is excellent, with most stellar
absorptions reproduced by our data. 

 The identification of TiO and ZrO has never been reported before 
in extragalactic sources. CN\,1.1$\mu$m has been reported \citep{riffel+08} but
not CN~0.92$\mu$m. At this point, it is important to note that the stellar spectra
reproduced in Figure~\ref{fig:tiobands} is only for illustrative purposes and by
no means represents the galaxy continua. A stellar population synthesis 
is necessary in order to carefully model the observed continuum. This is out of 
the scope of this paper, but will be presented in a following paper (Martins et al. 2013, in 
preparation). However, it calls the attention that the spectral type of the
stars plotted in Figure~\ref{fig:tiobands} correspond to giant and super giants
of K and M classes. The latter ones, during the AGB evolution, give origin
to the CN features and are crucial to model the stellar population in the NIR.

In order to quantify the above characteristics we followed \citet{riffel+09} and compute 
the absorption features' equivalent widths, W$\lambda$, measured according to the definitions
of \citet{riffel+08}. This process involves the definition of a central bandpass
covering the spectral feature of interest, and two other
bandpasses at the red and blue sides which are used to trace
a local continuum level through a linear fit to the mean values
in both bands. The corresponding value of each equivalent width with the relative
errors for all galaxies of the sample (nuclear and off-nuclear apertures)  
are presented in Table~\ref{indices}. The dashes in the table are due to the fact 
that for a few objects we were not able to measure one or more values because of 
the absence of the absorption/emission lines or a low signal-to-noise ratio.
Only one of these galaxies was found in previous studies of the 
same wavelength region, NGC~4179 in \citet{kotilainen+12}, which is one of the normal galaxies 
of our sample. Our index measurements are in general agreement with theirs (our bandpasses
are a little bigger, so our indexes are a little larger).

We found no correlation between these indexes and the W$_{(G band)}$ measured from the optical. 
This is an indication that we might be looking at different stellar populations in the optical and in 
the infrared. We also tried other optical indicators (namely, L$_{H\alpha}$, W$_{H\beta}$, NII/$H\alpha$)
but again found no correlation. 
We also have tried diagrams involving the NIR absorption lines (or bands) but only
weak correlations were found. The same result was found by \citet{riffel+09} for Seyfert 
galaxies.

\clearpage
\onecolumn
\begin{center}
\small
\setlength{\tabcolsep}{3pt}
\begin{longtable}{ccccccccc}
\caption[]{Equivalent widths in the NIR measured in \AA - J and H band} \label{indices} \\

\hline \multicolumn{1}{|c}{\textbf{}} & \multicolumn{1}{c}{\textbf{}} 
     & \multicolumn{1}{c}{\textbf{CN}} & \multicolumn{1}{c}{\textbf{AlI}} & \multicolumn{1}{c}{\textbf{NaI}} & \multicolumn{1}{c}{\textbf{SiI}} 
     & \multicolumn{1}{c}{\textbf{MgI}} & \multicolumn{1}{c}{\textbf{SiI}} 
     & \multicolumn{1}{c|}{\textbf{CO}} 
\\ \hline

 \multicolumn{1}{|c}{\textbf{Galaxy}} & \multicolumn{1}{c}{\textbf{Ap}} 
     & \multicolumn{1}{c}{\textbf{1.100}}& \multicolumn{1}{c}{\textbf{1.1250}} & \multicolumn{1}{c}{\textbf{1.1395}} & \multicolumn{1}{c}{\textbf{1.2112}} 
     & \multicolumn{1}{c}{\textbf{1.5771}}& \multicolumn{1}{c}{\textbf{1.5894}} 
     & \multicolumn{1}{c|}{\textbf{1.6175}} 
\\ \hline \hline
\endfirsthead

\multicolumn{9}{c}%
{{\bfseries \tablename\ \thetable{} -- continued from previous page}} \\
\\
\hline \multicolumn{1}{|c}{\textbf{}} & \multicolumn{1}{c}{\textbf{}} 
      & \multicolumn{1}{c}{\textbf{CN}}& \multicolumn{1}{c}{\textbf{AlI}} & \multicolumn{1}{c}{\textbf{NaI}} & \multicolumn{1}{c}{\textbf{SiI}} 
     & \multicolumn{1}{c}{\textbf{MgI}} & \multicolumn{1}{c}{\textbf{SiI}} 
     & \multicolumn{1}{c|}{\textbf{CO}} 
\\ \hline

\multicolumn{1}{|c}{\textbf{Galaxy}} & \multicolumn{1}{c}{\textbf{Ap}} 
      & \multicolumn{1}{c}{\textbf{1.100}}& \multicolumn{1}{c}{\textbf{1.1250}} & \multicolumn{1}{c}{\textbf{1.1395}} & \multicolumn{1}{c}{\textbf{1.2112}} 
     & \multicolumn{1}{c}{\textbf{1.5771}}& \multicolumn{1}{c}{\textbf{1.5894}} 
     & \multicolumn{1}{c|}{\textbf{1.6175}} 
\\ \hline \hline
\endhead

\hline \hline

\endfoot

\hline \hline
\endlastfoot                                                                                                    							
NGC 0221 & nuc &   14.61$\pm$0.36 &           ------    &  1.14$\pm$0.04 &    2.68$\pm$0.08 &    4.16$\pm$0.14 &  2.80$\pm$0.03 &  4.23$\pm$0.45  \\
         & S1  &   16.21$\pm$0.44 &           ------    &         ------ &    2.94$\pm$0.14 &    4.14$\pm$0.18 &  2.46$\pm$0.09 &  4.74$\pm$0.45  \\
         & S2  &   11.65$\pm$0.75 &           ------    &  2.18$\pm$0.10 &    2.81$\pm$0.11 &    4.10$\pm$0.23 &  2.08$\pm$0.09 &  5.29$\pm$0.52  \\
         & N1  &   12.96$\pm$0.59 &           ------    &         ------ &    3.55$\pm$0.06 &    4.61$\pm$0.13 &  2.45$\pm$0.04 &  5.35$\pm$0.55  \\
\hline        
NGC 0278 & nuc &   12.41$\pm$0.43 &           ------    &         ------ &    1.58$\pm$0.07 &    5.30$\pm$0.28 &  2.87$\pm$0.13 &  5.36$\pm$0.14  \\
         & S1  &        ------    &    2.48$\pm$0.51    &         ------ &           ------ &    5.32$\pm$0.36 &  4.45$\pm$0.06 &  8.08$\pm$0.28  \\
         & S2  &        ------    &           ------    &  1.46$\pm$0.68 &    6.02$\pm$0.32 &    4.84$\pm$0.56 &  2.49$\pm$0.07 & 16.06$\pm$0.25  \\
         & N1  &        ------    &           ------    &  6.08$\pm$0.68 &    1.72$\pm$0.13 &    8.10$\pm$0.71 &  4.91$\pm$0.06 &  6.15$\pm$0.23  \\
\hline        
NGC 0514 & nuc &    3.79$\pm$0.56 &    3.86$\pm$0.37    &  3.34$\pm$0.28 &    3.10$\pm$0.14 &    1.04$\pm$0.34 &  2.93$\pm$0.16 &  5.56$\pm$0.30  \\
         & S1  &        ------    &    7.68$\pm$0.44    &         ------ &   22.29$\pm$0.35 &           ------ & 12.15$\pm$0.34 &	  ------ 1   \\
         & N1  &        ------    &           ------    &  9.52$\pm$2.04 &   12.31$\pm$0.23 &           ------ &	 ------ & 13.62$\pm$2.14  \\
\hline        
NGC 0674 & nuc &   23.72$\pm$2.59 &    1.14$\pm$0.18    &  2.03$\pm$0.24 &    4.81$\pm$0.03 &    6.96$\pm$0.45 &  5.39$\pm$0.65 &  2.18$\pm$0.44  \\
         & N1  &        ------    &   10.50$\pm$1.26    &         ------ &   11.48$\pm$0.30 &    9.54$\pm$0.82 &  9.20$\pm$1.10 &  5.08$\pm$0.67  \\
         & S1  &        ------    &           ------    &         ------ &   14.62$\pm$0.22 &    1.61$\pm$0.87 &  6.13$\pm$0.57 &	  ------  \\
\hline        
NGC 0783 & nuc &   31.28$\pm$0.47 &           ------    &  2.37$\pm$0.03 &           ------ &    7.28$\pm$0.34 &  3.53$\pm$0.05 &  4.78$\pm$0.56  \\
         & S1  &   10.09$\pm$3.72 &    1.23$\pm$0.15    &         ------ &           ------ &   13.33$\pm$0.15 &  2.22$\pm$0.10 &  4.97$\pm$1.45  \\
         & N1  &        ------    &           ------    &  2.93$\pm$0.84 &    1.86$\pm$0.22 &    9.62$\pm$0.90 &	 ------ & 16.41$\pm$0.39  \\
\hline        
NGC 0864 & nuc &   13.58$\pm$1.51 &           ------    &  5.67$\pm$0.34 &           ------ &    4.97$\pm$0.24 &  1.80$\pm$0.02 &  6.65$\pm$0.05  \\
         & S1  &        ------    &           ------    & 17.02$\pm$0.93 &           ------ &    5.06$\pm$0.56 &	 ------ &	  ------     \\
         & N1  &        ------    &           ------    &         ------ &           ------ &   13.24$\pm$0.69 & 14.76$\pm$0.92 &	  ------ 1  \\
\hline        
NGC 1174 & nuc &        ------    &           ------    &  2.06$\pm$0.14 &    5.16$\pm$0.12 &    4.83$\pm$0.53 &  1.89$\pm$0.06 &  7.40$\pm$0.58  \\
         & S1  &        ------    &           ------    &  5.95$\pm$0.93 &   23.25$\pm$0.19 &    6.86$\pm$0.56 &  9.11$\pm$0.22 &  8.69$\pm$0.47  \\
         & N1  &        ------    &           ------    &         ------ &   29.57$\pm$0.30 &   17.85$\pm$1.00 &  3.92$\pm$1.46 &	  ------  \\
\hline        
NGC 1232 & nuc &   62.14$\pm$3.22 &    2.91$\pm$0.79    &  6.31$\pm$0.48 &    4.19$\pm$0.09 &           ------ &  1.65$\pm$0.02 &  8.09$\pm$0.56  \\
\hline        
NGC 1482 & nuc &   18.58$\pm$1.50 &           ------    &  5.73$\pm$0.30 &    1.15$\pm$0.06 &    3.87$\pm$0.71 &  2.86$\pm$0.06 &  5.49$\pm$0.21  \\
         & S1  &   11.78$\pm$2.87 &           ------    & 15.23$\pm$1.22 &           ------ &    7.13$\pm$1.33 &  4.69$\pm$0.84 & 13.42$\pm$0.18  \\
         & N1  &        ------    &           ------    & 16.04$\pm$1.50 &    3.66$\pm$0.51 &           ------ &	 ------ & 16.46$\pm$1.10  \\
\hline        
NGC 2339 & nuc &   18.44$\pm$0.48 &    1.44$\pm$0.11    &  1.29$\pm$0.13 &    3.60$\pm$0.13 &    2.79$\pm$0.07 &  3.14$\pm$0.01 &  4.89$\pm$0.34  \\
         & S1  &        ------    &    1.38$\pm$0.24    &  1.36$\pm$0.35 &   12.37$\pm$0.11 &    3.37$\pm$0.26 &  3.30$\pm$0.06 &  6.69$\pm$0.52  \\
         & N1  &        ------    &           ------    &         ------ &           ------ &    2.70$\pm$0.61 &  4.43$\pm$0.05 &  6.23$\pm$0.19     \\
\hline        
NGC 2342 & nuc &   39.16$\pm$1.03 &           ------    &  2.76$\pm$0.10 &           ------ &    6.36$\pm$0.11 &  3.44$\pm$0.01 &  5.69$\pm$0.43  \\
\hline        
NGC 2903 & nuc &   17.76$\pm$0.25 &    2.49$\pm$0.28    &         ------ &    2.74$\pm$0.04 &    3.35$\pm$0.10 &  3.09$\pm$0.02 &  5.94$\pm$0.50  \\
         & S1  &   13.80$\pm$1.00 &           ------    &         ------ &    3.48$\pm$0.02 &    3.13$\pm$0.15 &  2.84$\pm$0.14 &  5.23$\pm$0.70  \\
         & S2  &   17.28$\pm$0.42 &    3.07$\pm$0.59    &  1.17$\pm$0.12 &    1.72$\pm$0.11 &    3.25$\pm$0.16 &  2.31$\pm$0.02 &  6.19$\pm$0.63  \\
         & S3  &   16.15$\pm$0.91 &    1.33$\pm$0.52    &         ------ &           ------ &           ------ &  2.64$\pm$0.03 &  4.27$\pm$0.65  \\
         & N1  &   12.20$\pm$0.27 &    3.08$\pm$0.48    &  1.34$\pm$0.07 &    1.63$\pm$0.04 &    4.67$\pm$0.02 &  2.98$\pm$0.01 &  6.10$\pm$0.53  \\
         & N2  &   18.92$\pm$0.25 &    1.92$\pm$0.29    &  3.32$\pm$0.05 &    2.45$\pm$0.06 &    3.49$\pm$0.06 &  3.30$\pm$0.02 &  8.62$\pm$0.61  \\
         & N3  &   29.85$\pm$0.38 &    4.20$\pm$1.09    &         ------ &    1.28$\pm$0.14 &    2.60$\pm$0.10 &  1.77$\pm$0.08 &  6.56$\pm$0.81  \\
         & N4  &   19.66$\pm$1.71 &    4.36$\pm$0.62    &         ------ &    5.33$\pm$0.10 &    5.16$\pm$0.20 &  2.72$\pm$0.09 & 10.31$\pm$1.30  \\
\hline        
NGC 2950 & nuc &   18.63$\pm$0.69 &           ------    &  2.81$\pm$0.10 &    3.01$\pm$0.05 &    4.82$\pm$0.18 &  2.90$\pm$0.02 &  6.81$\pm$0.39  \\
         & S1  &   17.38$\pm$0.42 &           ------    &  3.36$\pm$0.22 &    4.00$\pm$0.08 &    2.88$\pm$0.26 &  2.54$\pm$0.07 &  8.28$\pm$0.32  \\
         & S2  &    8.40$\pm$1.71 &           ------    &  2.82$\pm$0.29 &    2.08$\pm$0.04 &    3.72$\pm$0.20 &  1.89$\pm$0.02 &  6.65$\pm$0.47  \\
         & N1  &   22.27$\pm$1.18 &           ------    &  2.67$\pm$0.21 &    2.71$\pm$0.05 &    4.68$\pm$0.31 &  2.64$\pm$0.03 &  6.43$\pm$0.18  \\
         & N2  &   10.09$\pm$1.66 &           ------    &  4.63$\pm$0.64 &    6.00$\pm$0.09 &    4.13$\pm$0.20 &  2.62$\pm$0.02 &  8.34$\pm$0.19  \\
\hline        
NGC 2964 & nuc &   29.67$\pm$2.23 &    4.35$\pm$1.01    &         ------ &    5.96$\pm$0.05 &    7.68$\pm$0.05 &  2.95$\pm$0.02 &  3.47$\pm$0.38     \\
         & S1  &        ------    &   16.10$\pm$1.93    &         ------ &    3.90$\pm$0.29 &           ------ &  7.50$\pm$1.11 &  5.94$\pm$1.51     \\
         & N1  &        ------    &          ------     &         ------ &   22.06$\pm$0.22 &    3.06$\pm$0.24 &  8.28$\pm$0.12 &  9.67$\pm$1.78     \\
\hline        
NGC 3184 & nuc &        ------    &    7.25$\pm$0.54    &         ------ &    4.59$\pm$0.26 &           ------ &	 ------ &  7.56$\pm$0.93  \\
         & S1  &        ------    &    9.67$\pm$0.97    &         ------ &    8.81$\pm$0.66 &           ------ &  3.82$\pm$0.39 & 12.00$\pm$1.45  \\
         & N1  &        ------    &   24.55$\pm$5.12    &         ------ &    7.54$\pm$0.46 &           ------ &	 ------ &  6.26$\pm$0.71     \\
\hline        
NGC 4102 & nuc &   23.20$\pm$0.90 &           ------    &         ------ &    2.58$\pm$0.14 &    1.05$\pm$0.14 &  2.32$\pm$0.03 &  5.98$\pm$0.54  \\
         & S1  &   21.65$\pm$0.47 &           ------    &  1.05$\pm$0.44 &    2.31$\pm$0.03 &    5.27$\pm$0.08 &  2.50$\pm$0.05 &  4.97$\pm$0.33  \\
         & N1  &   15.76$\pm$0.96 &           ------    &  2.13$\pm$0.30 &    2.51$\pm$0.08 &    6.92$\pm$0.09 &  1.61$\pm$0.07 &  7.46$\pm$0.20  \\
\hline        
NGC 4179 & nuc &   28.84$\pm$0.94 &           ------    &  2.08$\pm$0.15 &    3.29$\pm$0.14 &    3.97$\pm$0.32 &  2.63$\pm$0.03 &  5.29$\pm$0.25  \\
         & S1  &    8.20$\pm$1.00 &           ------    &  2.48$\pm$0.30 &           ------ &    6.03$\pm$0.21 &  1.84$\pm$0.10 &  7.74$\pm$0.70  \\
         & N1  &   42.99$\pm$1.03 &    4.58$\pm$1.48    &         ------ &    2.33$\pm$0.14 &    1.64$\pm$0.45 &  1.61$\pm$0.04 &  6.17$\pm$0.03  \\
\hline        
NGC 4303 & nuc &   11.78$\pm$0.95 &           ------    &  2.74$\pm$0.08 &    1.68$\pm$0.06 &    4.40$\pm$0.29 &  3.02$\pm$0.03 &  6.36$\pm$0.19  \\
         & S1  &   10.74$\pm$0.54 &           ------    &  3.16$\pm$0.22 &    2.24$\pm$0.09 &    1.68$\pm$0.15 &  3.34$\pm$0.04 &  6.02$\pm$0.55  \\
         & S2  &    9.71$\pm$0.30 &           ------    &  3.64$\pm$0.40 &    2.97$\pm$0.21 &           ------ &  1.95$\pm$0.03 &  6.38$\pm$0.51  \\
         & S3  &        ------    &           ------    &  3.93$\pm$0.61 &    2.06$\pm$0.09 &           ------ &  2.74$\pm$0.44 &  9.54$\pm$0.77  \\
         & N1  &   18.88$\pm$0.51 &           ------    &  3.51$\pm$0.25 &           ------ &    3.00$\pm$0.22 &  3.82$\pm$0.07 &  5.37$\pm$0.29  \\
         & N2  &   19.67$\pm$1.85 &           ------    &  3.53$\pm$0.35 &    4.93$\pm$0.13 &           ------ &  3.49$\pm$0.04 &  9.87$\pm$0.35  \\
         & N3  &        ------    &           ------    & 13.05$\pm$1.38 &    7.08$\pm$0.40 &           ------ &	 ------ & 11.47$\pm$0.89  \\
\hline        
NGC 4461 & nuc &    7.85$\pm$1.78 &           ------    &  2.41$\pm$0.14 &           ------ &    6.79$\pm$0.58 &  6.43$\pm$0.05 &  4.07$\pm$0.02  \\
         & N1  &        ------    &           ------    &  9.28$\pm$0.53 &           ------ &    3.62$\pm$0.28 &  5.76$\pm$0.03 &  4.11$\pm$0.11  \\
         & S1  &        ------    &           ------    &  1.93$\pm$0.72 &    7.93$\pm$0.67 &    5.91$\pm$0.46 &  6.34$\pm$0.50 &  2.08$\pm$0.55  \\
\hline        
NGC 4845 & nuc &   16.40$\pm$1.10 &           ------    &  2.87$\pm$0.21 &    4.80$\pm$0.08 &    6.16$\pm$0.26 &  3.38$\pm$0.02 &  5.05$\pm$0.16  \\
         & S1  &   12.84$\pm$0.97 &           ------    &  5.08$\pm$0.39 &    4.73$\pm$0.17 &    4.83$\pm$0.28 &  3.02$\pm$0.29 &  4.55$\pm$0.37  \\
         & S2  &        ------    &           ------    &  7.00$\pm$0.85 &    3.08$\pm$0.43 &    3.79$\pm$0.27 &  2.24$\pm$0.24 &  7.08$\pm$0.24  \\
         & S3  &        ------    &           ------    &  5.73$\pm$0.82 &           ------ &           ------ &	 ------ & 12.88$\pm$0.45     \\
         & N1  &   17.65$\pm$1.53 &    1.33$\pm$0.39    &         ------ &    1.53$\pm$0.03 &    6.62$\pm$0.25 &  2.66$\pm$0.02 &  5.83$\pm$0.29  \\
         & N2  &        ------    &           ------    &  1.43$\pm$0.21 &    8.77$\pm$0.07 &    6.07$\pm$0.39 &  1.86$\pm$0.18 &  8.95$\pm$0.40  \\
         & N3  &        ------    &           ------    & 13.80$\pm$1.26 &    9.72$\pm$0.47 &    6.07$\pm$0.96 &	 ------ & 12.31$\pm$1.10  \\
\hline        
NGC 5457 & nuc &        ------    &   12.52$\pm$1.13    &         ------ &    8.29$\pm$0.14 &    2.33$\pm$0.06 &  2.37$\pm$0.06 &  4.15$\pm$0.02  \\
         & S1  &        ------    &   18.98$\pm$2.52    &         ------ &    1.97$\pm$0.91 &    6.13$\pm$0.84 &	 ------ & 11.96$\pm$0.02  \\
         & N1  &        ------    &   28.41$\pm$1.48    & 14.97$\pm$0.14 &           ------ &    9.37$\pm$0.40 &	 ------ &  5.52$\pm$0.76  \\
\hline        
NGC 5905 & nuc &        ------    &           ------    &  1.65$\pm$0.03 &           ------ &    6.50$\pm$0.16 &  2.68$\pm$0.02 &  5.65$\pm$0.22  \\
         & S1  &        ------    &    1.39$\pm$0.19    &         ------ &    1.67$\pm$0.09 &    6.15$\pm$0.31 &  3.64$\pm$0.03 &  8.03$\pm$0.43  \\
         & N1  &        ------    &    1.33$\pm$0.25    &  1.48$\pm$0.08 &    2.84$\pm$0.27 &    7.47$\pm$0.30 &  4.90$\pm$0.03 &  3.00$\pm$0.32  \\
\hline         
NGC 6181 & nuc &   30.23$\pm$1.24 &    1.73$\pm$0.16    &  2.03$\pm$0.19 &    1.24$\pm$0.14 &    5.69$\pm$0.18 &  3.21$\pm$0.02 &  4.55$\pm$0.32  \\
         & S1  &   23.57$\pm$0.91 &           ------    &  1.37$\pm$0.17 &    7.67$\pm$0.32 &    2.70$\pm$0.05 &  4.76$\pm$0.06 &  5.99$\pm$0.61  \\
         & N1  &   18.84$\pm$3.00 &    2.60$\pm$0.43    &  5.33$\pm$0.75 &           ------ &           ------ &  5.02$\pm$0.04 &  4.72$\pm$0.48  \\
\hline        
NGC 6946 & nuc &   19.35$\pm$0.41 &    1.09$\pm$0.11    &         ------ &    1.02$\pm$0.04 &    5.15$\pm$0.26 &  3.10$\pm$0.03 &  4.99$\pm$0.55  \\
\hline        
NGC 7080 & nuc &   13.23$\pm$0.51 &           ------    &         ------ &           ------ &    3.56$\pm$0.20 &  2.68$\pm$0.02 &  6.76$\pm$0.24  \\
         & S1  &   16.33$\pm$2.34 &           ------    &         ------ &           ------ &    6.19$\pm$0.48 &  2.50$\pm$0.07 &  9.13$\pm$0.68  \\
         & N1  &   10.38$\pm$2.66 &           ------    &  3.78$\pm$0.23 &           ------ &    7.16$\pm$0.09 &  4.35$\pm$0.02 &  6.38$\pm$0.25  \\
\hline        
NGC 7448 & nuc &   50.53$\pm$1.89 &           ------    &  4.94$\pm$0.47 &    4.21$\pm$0.21 &    2.83$\pm$0.30 &  4.57$\pm$0.03 &  3.21$\pm$0.22  \\
         & S1  &        ------    &           ------    &  5.11$\pm$0.65 &    5.62$\pm$0.17 &           ------ &  3.44$\pm$0.40 & 12.82$\pm$0.52  \\
         & N1  &        ------    &           ------    & 11.27$\pm$0.89 &           ------ &           ------ &  3.21$\pm$0.93 &  5.18$\pm$0.37  \\
\hline
NGC 7798 & nuc &   17.71$\pm$0.37 &           ------    &         ------ &    2.20$\pm$0.01 &    3.97$\pm$0.05 &  4.49$\pm$0.02 &  5.50$\pm$0.33  \\  
         & S1  &        ------    &           ------    &  3.33$\pm$0.76 &           ------ &    3.72$\pm$0.11 &  1.30$\pm$0.16 &  5.13$\pm$0.19  \\
         & N1  &        ------    &           ------    &       ------   &   12.61$\pm$0.30 &           ------ &  1.93$\pm$0.13 &  7.78$\pm$0.68  \\
\hline        
NGC 7817 & nuc &   21.32$\pm$0.67 &           ------    &  4.65$\pm$0.49 &           ------ &    5.61$\pm$0.31 &  1.76$\pm$0.03 &  5.69$\pm$0.32  \\ 
         & S1  &        ------    &           ------    & 15.09$\pm$1.62 &           ------ &   14.02$\pm$1.21 &  3.31$\pm$0.23 &  4.87$\pm$0.11  \\
         & N1  &        ------    &           ------    & 16.03$\pm$1.74 &           ------ &   12.91$\pm$1.66 &	 ------ &	  ------  \\
        
\hline

\end{longtable}
\end{center}
\twocolumn

\clearpage
\onecolumn
\begin{center}
\small
\setlength{\tabcolsep}{3pt}
\begin{longtable}{ccccccc}
\caption[]{Equivalent widths in the NIR measured in \AA - K band} \label{indices} \\

\hline \multicolumn{1}{|c}{\textbf{}} & \multicolumn{1}{c}{\textbf{}} 
     & \multicolumn{1}{c}{\textbf{NaI}} & \multicolumn{1}{c}{\textbf{CaI}}
     & \multicolumn{1}{c}{\textbf{CO$_1$}} & \multicolumn{1}{c}{\textbf{CO$_2$}} & \multicolumn{1}{c|}{\textbf{CO$_3$}}
\\ \hline

 \multicolumn{1}{|c}{\textbf{Galaxy}} & \multicolumn{1}{c}{\textbf{Ap}} 
     & \multicolumn{1}{c}{\textbf{2.2063}} & \multicolumn{1}{c}{\textbf{2.2655}}
     & \multicolumn{1}{c}{\textbf{2.2980}} & \multicolumn{1}{c}{\textbf{2.3255}} & \multicolumn{1}{c|}{\textbf{2.3545}}
\\ \hline \hline
\endfirsthead

\multicolumn{7}{c}%
{{\bfseries \tablename\ \thetable{} -- continued from previous page}} \\
\\
\hline \multicolumn{1}{|c}{\textbf{}} & \multicolumn{1}{c}{\textbf{}} 
     & \multicolumn{1}{c}{\textbf{NaI}} & \multicolumn{1}{c}{\textbf{CaI}}
     & \multicolumn{1}{c}{\textbf{CO$_1$}} & \multicolumn{1}{c}{\textbf{CO$_2$}} & \multicolumn{1}{c|}{\textbf{CO$_3$}}
\\ \hline

\multicolumn{1}{|c}{\textbf{Galaxy}} & \multicolumn{1}{c}{\textbf{Ap}} 
     & \multicolumn{1}{c}{\textbf{2.2063}} & \multicolumn{1}{c}{\textbf{2.2655}}
     & \multicolumn{1}{c}{\textbf{2.2980}} & \multicolumn{1}{c}{\textbf{2.3255}} & \multicolumn{1}{c|}{\textbf{2.3545}}
\\ \hline \hline
\endhead

\hline \hline

\endfoot

\hline \hline
\endlastfoot

NGC 0221 & nuc &       4.55$\pm$0.14 &    7.43$\pm$0.01 &   14.70$\pm$0.20 &  9.58$\pm$0.28 &	17.23$\pm$0.22 \\
         & S1  &       5.43$\pm$0.28 &    6.18$\pm$0.19 &   14.56$\pm$0.32 &  9.49$\pm$0.44 &	16.60$\pm$0.26 \\
         & S2  &       6.63$\pm$0.41 &   11.28$\pm$2.18 &   13.49$\pm$1.07 & 12.72$\pm$0.33 &	19.38$\pm$0.18 \\
         & N1  &       6.15$\pm$0.26 &    5.60$\pm$0.28 &   13.94$\pm$0.21 &  8.95$\pm$0.31 &	17.98$\pm$0.19 \\
\hline        
NGC 0278 & nuc &       6.02$\pm$0.02 &  	 ------ &   19.52$\pm$0.53 &  3.04$\pm$0.37 &	13.81$\pm$0.66 \\
         & S1  &   	      ------ &    7.56$\pm$1.61 &   18.92$\pm$1.78 &  6.88$\pm$1.48 &	24.16$\pm$1.16 \\
         & S2  &      16.37$\pm$0.15 &    3.95$\pm$4.38 &   32.12$\pm$5.13 &	     ------ &	43.84$\pm$1.60 \\
         & N1  &   	      ------ &  	 ------ &   41.93$\pm$0.47 & 17.98$\pm$0.46 &	 8.45$\pm$0.88 \\
\hline        
NGC 0514 & nuc &       8.34$\pm$0.38 &    9.11$\pm$0.48 &   15.95$\pm$1.42 &	     ------ &	29.09$\pm$1.65 \\
         & S1  &   	      ------ &   26.53$\pm$1.70 &   18.82$\pm$7.64 &	     ------ &		------    \\
         & N1  &      13.14$\pm$0.11 &   18.23$\pm$0.47 &	    ------ &	     ------ &	20.79$\pm$8.60 \\
\hline        
NGC 0674 & nuc &       4.80$\pm$0.29 &    7.45$\pm$0.50 &   11.44$\pm$0.57 & 14.13$\pm$0.53 &	24.69$\pm$0.30 \\
         & N1  &   	      ------ &   16.93$\pm$2.28 &    6.88$\pm$1.29 & 25.74$\pm$2.12 &	12.86$\pm$4.84 \\
         & S1  &   	      ------ &   17.98$\pm$0.71 &	    ------ & 29.96$\pm$0.61 &	63.33$\pm$0.14 \\
\hline        
NGC 0783 & nuc &       6.83$\pm$0.12 &    2.84$\pm$0.49 &   18.13$\pm$0.26 & 10.43$\pm$0.34 &	25.25$\pm$0.26 \\
         & S1  &       4.47$\pm$0.92 &    1.05$\pm$0.66 &	    ------ & 36.43$\pm$0.14 &	51.63$\pm$0.04 \\
         & N1  &      11.88$\pm$0.75 &  	 ------ &    5.44$\pm$2.45 & 16.51$\pm$3.17 &	15.92$\pm$6.41 \\
\hline        
NGC 0864 & nuc &       8.05$\pm$0.19 &    4.33$\pm$0.57 &   19.86$\pm$0.38 &  9.51$\pm$0.40 &	 7.78$\pm$0.78 \\
         & S1  &   	      ------ &   12.51$\pm$0.84 &   22.58$\pm$3.07 & 27.86$\pm$0.45 &		------    \\
         & N1  &       2.58$\pm$1.92 &  	 ------ &	     ------&	  ------    &	       ------	 \\
\hline        
NGC 1174 & nuc &       3.97$\pm$0.12 &  	 ------ &   17.54$\pm$0.10 &  9.89$\pm$0.14 &	16.47$\pm$0.37 \\
         & S1  &       3.02$\pm$0.98 &  	 ------ &   16.91$\pm$2.15 &	     ------ &	 9.60$\pm$4.15 \\
         & N1  &       5.55$\pm$0.42 &  	 ------ &   47.83$\pm$1.82 & 33.42$\pm$1.28 &	49.51$\pm$0.27 \\
\hline        
NGC 1232 & nuc &       9.38$\pm$0.45 &  	 ------ &	     ------&  6.49$\pm$0.30 &	 7.18$\pm$0.44 \\
\hline        
NGC 1482 & nuc &       6.93$\pm$0.17 &  	 ------ &   19.65$\pm$0.10 & 13.70$\pm$0.08 &	20.20$\pm$0.60 \\
         & S1  &       2.38$\pm$1.34 &  	 ------ &   14.59$\pm$0.74 & 10.95$\pm$1.19 &	10.58$\pm$2.32 \\
         & N1  &       9.18$\pm$0.67 &  	 ------ &   13.07$\pm$1.06 &  6.13$\pm$1.49 &	18.47$\pm$2.51 \\
\hline        
NGC 2339 & nuc &       5.68$\pm$0.14 &    1.50$\pm$0.66 &   20.15$\pm$0.08 & 14.49$\pm$0.05 &	20.10$\pm$0.31 \\
         & S1  &   	      ------ &  	 ------ &   24.12$\pm$0.48 & 14.66$\pm$0.78 &	14.15$\pm$1.30 \\
         & N1  &   	      ------ &   12.06$\pm$1.30 &   13.90$\pm$1.17 & 16.50$\pm$0.43 &		------    \\
\hline        
NGC 2342 & nuc &       3.99$\pm$0.05 &    3.38$\pm$0.25 &   19.50$\pm$0.61 &  6.72$\pm$0.62 &	16.17$\pm$0.47 \\
\hline        
NGC 2903 & nuc &       4.68$\pm$0.35 &    1.62$\pm$0.23 &   17.78$\pm$0.03 &  8.17$\pm$0.04 &	15.47$\pm$0.27 \\
         & S1  &       4.31$\pm$0.33 &    3.27$\pm$0.16 &   18.09$\pm$0.23 & 10.29$\pm$0.20 &	17.90$\pm$0.65 \\
         & S2  &       2.48$\pm$0.26 &    9.40$\pm$0.10 &   20.68$\pm$0.10 &  9.95$\pm$0.03 &	23.54$\pm$0.28 \\
         & S3  &   	      ------ &    5.57$\pm$0.10 &   22.50$\pm$0.54 &  9.72$\pm$0.67 &	14.18$\pm$0.82 \\
         & N1  &       1.64$\pm$0.31 &    4.77$\pm$0.02 &   16.36$\pm$0.03 &  8.03$\pm$0.05 &	18.64$\pm$0.37 \\
         & N2  &       4.67$\pm$0.51 &    2.83$\pm$0.43 &   21.48$\pm$0.08 &  8.85$\pm$0.25 &	15.60$\pm$0.59 \\
         & N3  &       2.67$\pm$0.58 &  	 ------ &   17.16$\pm$0.25 & 11.86$\pm$0.14 &	16.58$\pm$0.78 \\
         & N4  &       5.02$\pm$0.84 &    5.42$\pm$0.29 &   20.89$\pm$0.23 & 10.42$\pm$0.17 &	23.03$\pm$0.52 \\
\hline        
NGC 2950 & nuc &       7.33$\pm$0.20 &  	 ------ &   19.47$\pm$0.28 & 10.12$\pm$0.12 &	18.51$\pm$0.35 \\
         & S1  &       5.48$\pm$0.02 &    3.24$\pm$0.08 &   18.06$\pm$0.33 &  9.18$\pm$0.37 &	20.88$\pm$0.44 \\
         & S2  &       5.05$\pm$0.04 &    8.06$\pm$0.50 &   12.15$\pm$0.62 &  8.00$\pm$0.85 &	17.05$\pm$1.61 \\
         & N1  &       6.04$\pm$0.17 &    1.92$\pm$0.16 &   18.30$\pm$0.32 & 10.62$\pm$0.35 &	19.02$\pm$0.65 \\
         & N2  &       3.92$\pm$0.56 &  	 ------ &   11.80$\pm$0.58 &	  ------    &	20.00$\pm$0.66 \\
\hline        
NGC 2964 & nuc &       6.43$\pm$0.13 &  	 ------ &   13.86$\pm$0.76 & 11.41$\pm$0.64 &		------    \\
         & S1  &      16.23$\pm$0.32 &   22.13$\pm$1.13 &    3.42$\pm$2.71 & 29.08$\pm$2.12 &		------    \\
         & N1  &      13.40$\pm$0.75 &   11.44$\pm$0.17 &	    ------ &  5.64$\pm$3.46 &		------    \\
\hline        
NGC 3184 & nuc &       8.06$\pm$0.12 &    2.61$\pm$0.75 &   18.36$\pm$1.23 &  5.53$\pm$1.61 &	23.62$\pm$2.10 \\
         & S1  &       4.94$\pm$0.98 &    2.25$\pm$0.04 &   32.36$\pm$0.16 &  7.94$\pm$0.88 &	24.45$\pm$3.66 \\
         & N1  &   	      ------ &    8.06$\pm$1.43 &	    ------ &	  ------    &		------    \\
\hline        
NGC 4102 & nuc &       6.29$\pm$0.13 &    3.34$\pm$0.09 &   26.84$\pm$0.13 & 11.65$\pm$0.30 &	16.35$\pm$0.46 \\
         & S1  &       7.88$\pm$0.03 &    3.37$\pm$0.33 &   24.88$\pm$0.19 &  9.68$\pm$0.15 &	15.93$\pm$0.57 \\
         & N1  &       8.22$\pm$0.18 &  	 ------ &   21.16$\pm$0.13 &  8.11$\pm$0.30 &	13.22$\pm$0.58 \\
\hline        
NGC 4179 & nuc &       9.38$\pm$0.07 &    6.53$\pm$0.14 &   15.19$\pm$0.32 & 10.39$\pm$0.45 &	21.58$\pm$0.10 \\
         & S1  &      14.22$\pm$0.08 &    6.08$\pm$0.38 &    4.15$\pm$1.46 &  4.79$\pm$1.02 &	10.35$\pm$1.82 \\
         & N1  &       6.20$\pm$0.02 &  	 ------ &    9.85$\pm$0.74 &	  ------    &	21.73$\pm$2.15 \\
\hline        
NGC 4303 & nuc &       7.30$\pm$0.09 &  	 ------ &   17.99$\pm$0.19 & 10.48$\pm$0.22 &	17.93$\pm$0.32 \\
         & S1  &       7.87$\pm$0.04 &    2.39$\pm$0.17 &   22.41$\pm$0.57 & 11.31$\pm$0.10 &	17.37$\pm$0.11 \\
         & S2  &       4.95$\pm$0.08 &    2.05$\pm$0.22 &   17.58$\pm$0.31 & 15.39$\pm$0.14 &	18.18$\pm$1.01 \\
         & S3  &       6.83$\pm$0.33 &  	 ------ &   26.08$\pm$0.75 & 15.40$\pm$0.59 &	13.34$\pm$0.26 \\
         & N1  &       5.23$\pm$0.10 &    2.63$\pm$0.08 &   17.50$\pm$0.29 &  9.18$\pm$0.44 &	15.25$\pm$0.21 \\
         & N2  &       7.92$\pm$0.23 &    4.27$\pm$0.19 &   24.19$\pm$0.32 & 13.61$\pm$0.48 &	15.27$\pm$0.60 \\
         & N3  &       5.04$\pm$0.09 &  	 ------ &   29.01$\pm$0.96 & 20.59$\pm$1.40 &	30.07$\pm$1.46 \\
\hline        
NGC 4461 & nuc &       3.61$\pm$0.28 &    7.72$\pm$0.62 &   13.27$\pm$0.44 & 14.98$\pm$0.55 &	 6.77$\pm$0.44 \\
         & N1  &       6.91$\pm$0.55 &    7.11$\pm$0.31 &   22.20$\pm$1.36 & 33.50$\pm$0.28 &	19.60$\pm$0.45 \\
         & S1  &       2.67$\pm$0.03 &    8.40$\pm$0.70 &   12.45$\pm$0.80 & 20.96$\pm$1.23 &	14.05$\pm$2.67 \\
\hline        
NGC 4845 & nuc &       8.44$\pm$0.08 &    2.53$\pm$0.10 &   23.39$\pm$0.18 & 13.08$\pm$0.11 &	19.43$\pm$0.19 \\
         & S1  &       6.97$\pm$0.08 &    3.38$\pm$0.30 &   19.85$\pm$0.15 & 11.84$\pm$0.52 &	20.72$\pm$0.92 \\
         & S2  &       3.82$\pm$0.21 &    5.09$\pm$0.14 &   25.63$\pm$0.17 & 12.29$\pm$0.61 &	14.26$\pm$1.14 \\
         & S3  &      12.35$\pm$0.41 &  	 ------ &   43.58$\pm$0.41 & 14.14$\pm$1.03 &		------    \\
         & N1  &       6.09$\pm$0.16 &    1.82$\pm$0.38 &   24.52$\pm$0.45 & 16.19$\pm$0.20 &	22.26$\pm$0.26 \\
         & N2  &       9.18$\pm$0.19 &    5.88$\pm$0.36 &   23.74$\pm$0.10 & 13.82$\pm$0.02 &	19.69$\pm$0.01 \\
         & N3  &       5.72$\pm$0.54 &    4.80$\pm$1.55 &   30.38$\pm$0.95 & 27.96$\pm$1.18 &	26.99$\pm$3.27 \\
\hline        
NGC 5457 & nuc &       1.20$\pm$0.26 &    5.66$\pm$0.29 &   12.08$\pm$1.03 & 15.23$\pm$0.81 &	27.62$\pm$1.51 \\
         & S1  &   	      ------ &   12.88$\pm$0.62 &   38.76$\pm$0.45 & 28.88$\pm$0.26 &	 4.64$\pm$2.18 \\
         & N1  &      15.99$\pm$1.87 &   14.23$\pm$1.04 &   24.21$\pm$1.87 & 55.10$\pm$1.06 &	46.12$\pm$2.49 \\
\hline        
NGC 5905 & nuc &       4.69$\pm$0.17 &    1.25$\pm$0.09 &   17.49$\pm$0.18 & 12.20$\pm$0.26 &	22.69$\pm$0.42 \\
         & S1  &       8.31$\pm$0.14 &    4.57$\pm$0.48 &    4.93$\pm$0.92 &  8.98$\pm$0.88 &	33.17$\pm$1.46 \\
         & N1  &       8.34$\pm$0.15 &    2.03$\pm$0.09 &    7.14$\pm$0.48 & 10.24$\pm$0.66 &	20.28$\pm$0.15 \\
\hline         
NGC 6181 & nuc &       4.92$\pm$0.39 &    2.18$\pm$0.31 &   13.56$\pm$0.25 & 13.72$\pm$0.04 &	20.00$\pm$0.01 \\
         & S1  &      14.04$\pm$0.81 &  	 ------ &    3.05$\pm$1.06 & 16.87$\pm$1.24 &	26.62$\pm$0.30 \\
         & N1  &      11.88$\pm$1.18 &    2.51$\pm$1.82 &   17.21$\pm$1.66 & 23.66$\pm$1.48 &	28.33$\pm$1.98 \\
\hline        
NGC 6946 & nuc &       5.61$\pm$0.06 &    4.15$\pm$0.08 &   21.82$\pm$0.12 & 10.48$\pm$0.05 &	17.43$\pm$0.40 \\
\hline        
NGC 7080 & nuc &       6.60$\pm$0.30 &    2.18$\pm$0.34 &   19.02$\pm$0.14 &  5.75$\pm$0.28 &	11.12$\pm$0.45 \\
         & S1  &       3.26$\pm$0.41 &   10.36$\pm$0.96 &   18.50$\pm$0.84 &  9.02$\pm$1.06 &	 8.10$\pm$0.56 \\
         & N1  &      13.71$\pm$0.89 &    3.96$\pm$0.60 &   19.05$\pm$0.71 &	  ------    &	14.95$\pm$1.46 \\
\hline        
NGC 7448 & nuc &       3.12$\pm$0.68 &  	 ------ &    4.80$\pm$1.03 & 13.20$\pm$1.08 &	17.25$\pm$1.86 \\
         & S1  &       2.64$\pm$0.55 &  	 ------ &   13.93$\pm$4.05 & 20.27$\pm$1.77 &	14.44$\pm$2.34 \\
         & N1  &   	      ------ &    9.01$\pm$1.52 &    2.77$\pm$2.69 & 23.01$\pm$1.88 &	14.36$\pm$0.48 \\
\hline
NGC 7798 & nuc &       4.59$\pm$0.22 &  	 ------ &   22.35$\pm$0.48 & 10.44$\pm$0.11 &	23.27$\pm$0.48 \\  
         & S1  &       4.52$\pm$0.71 &  	 ------ &   21.31$\pm$3.17 & 15.31$\pm$0.61 &	58.54$\pm$1.48 \\
         & N1  &       6.75$\pm$1.10 &    3.35$\pm$2.69 &   21.82$\pm$5.36 & 29.24$\pm$0.74 &	39.99$\pm$1.18 \\
\hline        
NGC 7817 & nuc &       7.71$\pm$0.19 &  	 ------ &   10.15$\pm$0.65 &  9.34$\pm$0.82 &	23.20$\pm$0.32 \\ 
         & S1  &   	      ------ &  	 ------ &   19.86$\pm$1.61 &	     ------ &	 5.92$\pm$2.30 \\
         & N1  &       2.00$\pm$0.73 &  	 ------ &	    ------ &	     ------ &	31.76$\pm$1.28 \\
        
\hline

\end{longtable}
\end{center}
\twocolumn

\section{Emission Lines}

The spectra of all the galaxies presented in this work are shaped by their stellar populations,
as can be inferred by the numerous stellar absorption features present along the 0.8-2.4$\mu$m region. 
However, besides the stellar features, these galaxies are classified as H\,{\sc ii}/star-forming
based on their emission lines properties in the optical. The NIR spectra is not much different.
Most sources of our sample display nebular lines of weak to moderate intensity, the most
common being H\,{\sc i} (Brackett and
Pashen series), He\,{\sc i}\,1.083$\micron$, [Fe\,{\sc ii}]\,1.257$\micron$, 
[S\,{\sc iii}]\,0.953$\micron$ and H$_2$\,2.121$\micron$. As expected, the 
spectra of normal galaxies completely lack of such lines. In this section we will set up additional 
constraints on the gas properties of these objects based on the study of these lines. 

In order to properly measure these emission lines, 
it is usually necessary to subtract the stellar continuum. This can be done
by fitting the underlying continuum using stellar population synthesis, as it was done
by \citet{riffel+09} and \citet{martins+10} for Seyfert galaxies.

\citet{maraston05}'s models are the only ones available so far that include the effect of the TP-AGB
phase, crucial to model the stellar population in the NIR, and that have different metallicities.
However, these models are characterised by their very low spectral resolution
at NIR wavelengths (R=100). This contrasts with the moderate
spectral resolution of the data (R$\sim$920). Because of this difference,
in order to perform the stellar population synthesis, it is necessary
to downgrade the spectral resolution of the observations to match that of the models. As a result, the final
subtracted spectra (observations minus the final stellar template) 
will enhance the line fluxes of those features very close to or within the absorption
lines, but the weak lines may disappear in the process.

\begin{figure*} 
\includegraphics [width=160mm]{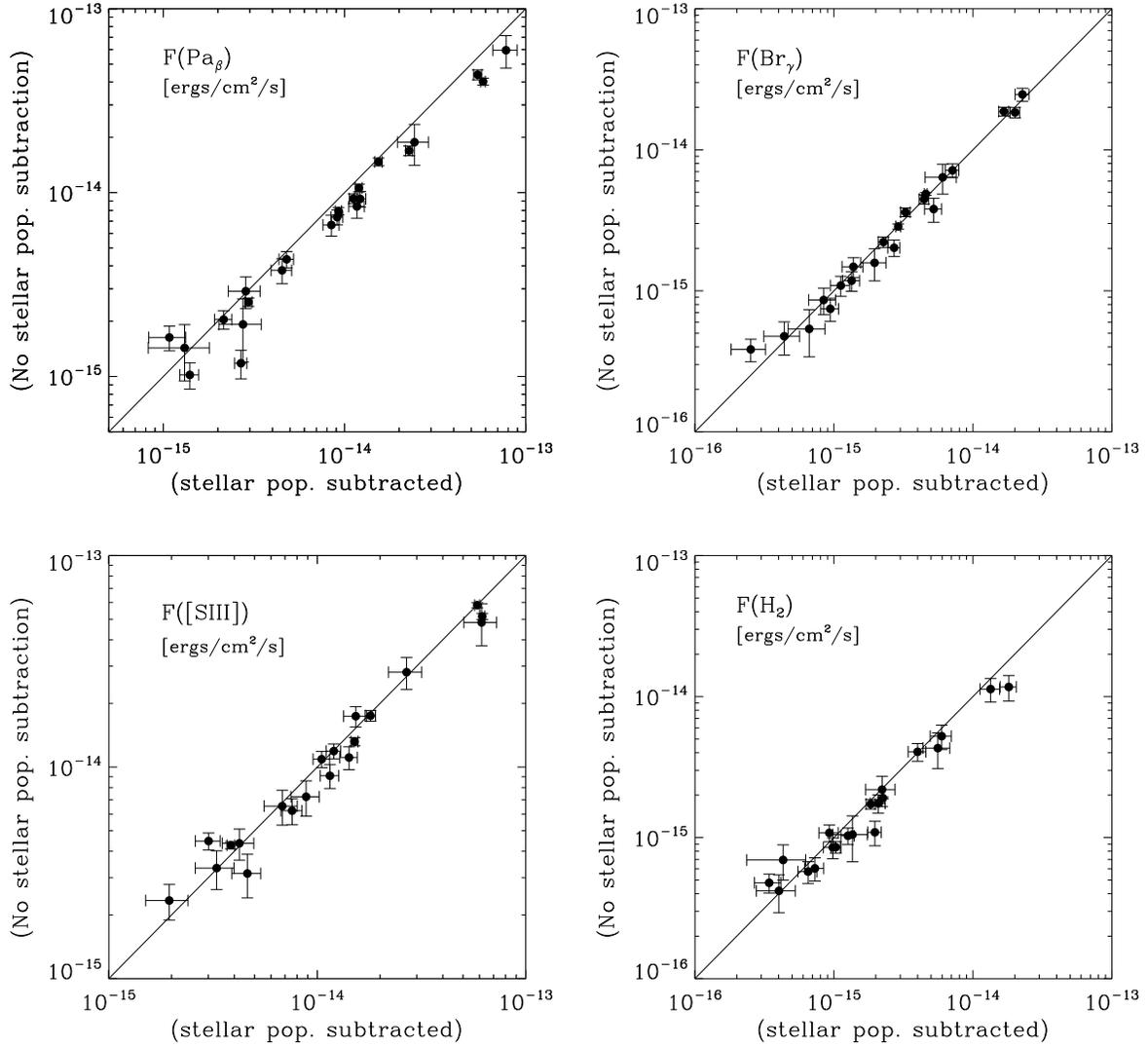}
\caption{Comparison between the line fluxes measured in the low resolution spectra, where the stellar population has been
subtracted, and the high resolution spectra, with no stellar population subtraction. 
The black line shows the one to one correspondence.}
\label{flux_comparison}
\end{figure*}

In order to evaluate how much the stellar population subtraction and loss of spectral
resolution affects the line flux measurements, we performed the stellar population
synthesis (results will be presented in an upcoming paper, Martins et al. 2013) and we measured
the emission line fluxes in both
spectra: the low resolution ones, with the stellar population subtracted, and the
original spectra, without subtraction of the stellar population. Figure~\ref{flux_comparison} shows the
comparison for four of the most common lines - Pa$\beta$, Br$\gamma$, [S\,{\sc iii}]~0.953$\micron$ 
and H$_2$~2.121$\micron$. 
The black line shows the one to one correspondence.
 It can be seen that the subtraction of the stellar population
does significantly affects the measured fluxes relative to those before subtraction.
One possible exception is Pa$\beta$. The underlying stellar
population has a stronger effect on the hydrogen lines. The effect is also larger for higher fluxes. 
The largest differences in fluxes are about 25$\%$, which is around the largest errors measured 
for the lines also. Taking this into account we decided to list the emission lines from the 
spectra with no stellar population subtraction, for which we could measure more lines. We advise 
the reader interested in line ratios to be cautious and remember to take this possible effect into 
account. The emission line flux measurements are presented in Table~\ref{fluxes}.

\onecolumn
\begin{landscape}
\begin{center}
\footnotesize
\setlength{\tabcolsep}{2pt}
\begin{longtable}{cccccccccccccc}
\caption[]{Emission Line Measurements (10$^{-15}$F$_\lambda$ [erg cm$^{-2}$ s$^{-1}$]) } \label{fluxes}\\
\hline
\hline \multicolumn{1}{c}{\textbf{}} & \multicolumn{1}{c}{\textbf{}} 
     & \multicolumn{1}{c}{\textbf{[SIII]}} & \multicolumn{1}{c}{\textbf{[SIII]}} & \multicolumn{1}{c}{\textbf{Pa$\delta$}} 
     & \multicolumn{1}{c}{\textbf{HeI}} & \multicolumn{1}{c}{\textbf{Pa$\gamma$}} 
     & \multicolumn{1}{c}{\textbf{[FeII]}} & \multicolumn{1}{c}{\textbf{Pa$\beta$}} & \multicolumn{1}{c}{\textbf{[FeII]}}
     & \multicolumn{1}{c}{\textbf{Pa$\alpha$}} & \multicolumn{1}{c}{\textbf{H$_2$}} & \multicolumn{1}{c}{\textbf{H$_2$}} & \multicolumn{1}{c}{\textbf{Br$\gamma$}} 
\\

     \multicolumn{1}{c}{\textbf{Galaxy}} & \multicolumn{1}{c}{\textbf{Ap}} 
     & \multicolumn{1}{c}{\textbf{0.907}} & \multicolumn{1}{c}{\textbf{0.953}} & \multicolumn{1}{c}{\textbf{1.005}} 
     & \multicolumn{1}{c}{\textbf{1.083}}& \multicolumn{1}{c}{\textbf{1.093}} 
     & \multicolumn{1}{c}{\textbf{1.257}} & \multicolumn{1}{c}{\textbf{1.282}} & \multicolumn{1}{c}{\textbf{1.646}}
     & \multicolumn{1}{c}{\textbf{1.870}} & \multicolumn{1}{c}{\textbf{1.957}} & \multicolumn{1}{c}{\textbf{2.121}}& \multicolumn{1}{c}{\textbf{2.166}}
\\ \hline \hline
\endfirsthead

\multicolumn{14}{c}%
{{\bfseries \tablename\ \thetable{} -- continued from previous page}} \\
\\
\hline\hline \multicolumn{1}{c}{\textbf{}} & \multicolumn{1}{c}{\textbf{}} 
     & \multicolumn{1}{c}{\textbf{[SIII]}} & \multicolumn{1}{c}{\textbf{[SIII]}} & \multicolumn{1}{c}{\textbf{Pa$\delta$}} 
     & \multicolumn{1}{c}{\textbf{HeI}} & \multicolumn{1}{c}{\textbf{Pa$\gamma$}} 
     & \multicolumn{1}{c}{\textbf{[FeII]}} & \multicolumn{1}{c}{\textbf{Pa$\beta$}} & \multicolumn{1}{c}{\textbf{[FeII]}}
     & \multicolumn{1}{c}{\textbf{Pa$\alpha$}} & \multicolumn{1}{c}{\textbf{H$_2$}} & \multicolumn{1}{c}{\textbf{H$_2$}} & \multicolumn{1}{c}{\textbf{Br$\gamma$}} 
\\

 \multicolumn{1}{c}{\textbf{Galaxy}} & \multicolumn{1}{c}{\textbf{Ap}} 
     & \multicolumn{1}{c}{\textbf{0.907}} & \multicolumn{1}{c}{\textbf{0.953}} & \multicolumn{1}{c}{\textbf{1.005}} 
     & \multicolumn{1}{c}{\textbf{1.083}}& \multicolumn{1}{c}{\textbf{1.093}} 
     & \multicolumn{1}{c}{\textbf{1.257}} & \multicolumn{1}{c}{\textbf{1.282}} & \multicolumn{1}{c}{\textbf{1.646}}
     & \multicolumn{1}{c}{\textbf{1.870}} & \multicolumn{1}{c}{\textbf{1.957}} & \multicolumn{1}{c}{\textbf{2.121}}& \multicolumn{1}{c}{\textbf{2.166}}
\\ \hline \hline
\endhead

\hline\hline
\endfoot

\hline \hline
\endlastfoot
\hline 
 NGC 783  &  S1   &         ------        &         ------        &        ------         &        ------         &        ------         &     0.45 $\pm$ 0.11    &         ------       &          ------        &   2.08 $\pm$ 0.11    &        ------        &     0.22  $\pm$ 0.07  &          ------      \\  
 NGC 783  &  N1   &         ------        &         ------        &        ------         &        ------         &        ------         &        ------           &   0.72 $\pm$ 0.10    &    0.43 $\pm$ 0.09    &   2.98 $\pm$ 0.18    &        ------        &     0.15  $\pm$ 0.05  &          ------      \\  
 NGC 783  &  nuc   &         ------        &         ------        &        ------         &        ------         &        ------         &     0.74 $\pm$ 0.13    &         ------       &          ------        &   2.11 $\pm$ 0.19    &        ------        &          ------       &          ------      \\  
\hline
 NGC 0864 &  S1   &         ------        &     4.79 $\pm$  0.29  &         ------        &         ------        &     2.84 $\pm$  0.68  &         ------        &     2.13 $\pm$  0.17  &     0.48 $\pm$  0.05  &         ------        &         ------        &     0.49 $\pm$  0.05  &     0.35 $\pm$  0.05\\
 NGC 0864 &  N1   &         ------        &     4.68 $\pm$  0.52  &         ------        &         ------        &         ------        &         ------        &         ------        &         ------        &         ------        &         ------        &         ------        &         ------      \\
 NGC 0864 &  nuc  &     7.97 $\pm$  1.31  &    18.70 $\pm$  1.37  &     2.45 $\pm$  0.58  &     5.53 $\pm$  0.65  &     4.67 $\pm$  0.73  &     2.09 $\pm$  0.32  &     6.94 $\pm$  0.60  &     2.38 $\pm$  0.41  &         ------        &         ------        &     0.70 $\pm$  0.13  &     1.50 $\pm$  0.16\\
\hline
 NGC 1174  & nuc  &    10.06 $\pm$  1.35  &    14.48 $\pm$  0.76  &         ------        &     8.52 $\pm$  0.98  &     4.66 $\pm$  1.39  &     4.15 $\pm$  0.20  &     8.92 $\pm$  0.27  &     2.67 $\pm$  0.46  &    29.57 $\pm$  1.21  &         ------        &     1.14 $\pm$  0.11  &     3.01 $\pm$  0.08\\
 \hline
 NGC 1482 &  S1   &         ------        &         ------        &         ------        &         ------        &    77.99 $\pm$  9.64  &    15.43 $\pm$  2.87  &    43.60 $\pm$  8.74  &         ------        &         ------        &         ------        &         ------        &    15.75 $\pm$  1.03\\
 NGC 1482 &  N1   &         ------        &         ------        &         ------        &         ------        &         ------        &         ------        &    15.73 $\pm$  1.73  &         ------        &         ------        &         ------        &         ------        &         ------      \\
 NGC 1482 &  nuc  &         ------        &         ------        &         ------        &    52.14 $\pm$  3.84  &    74.76 $\pm$ 10.39  &    17.42 $\pm$  1.79  &    61.60 $\pm$  8.21  &    19.11 $\pm$  3.35  &         ------        &         ------        &    11.27 $\pm$  2.28  &    25.03 $\pm$  1.75\\
\hline
 NGC 2339  & S1   &         ------        &         ------        &         ------        &         ------        &     2.86 $\pm$  0.37  &     1.91 $\pm$  0.30  &     2.05 $\pm$  0.51  &     1.77 $\pm$  0.12  &     4.61 $\pm$  0.22  &     1.81 $\pm$  0.30  &     0.88 $\pm$  0.05  &     0.89 $\pm$  0.12\\
 NGC 2339  & N1   &         ------        &         ------        &     2.66 $\pm$  0.42  &         ------        &         ------        &         ------        &     3.12 $\pm$  0.41  &     1.39 $\pm$  0.22  &    11.57 $\pm$  0.40  &     1.06 $\pm$  0.10  &     1.07 $\pm$  0.09  &     1.22 $\pm$  0.12\\
 NGC 2339 &  nuc  &     6.35 $\pm$  2.13  &    14.99 $\pm$  1.76  &         ------        &    18.13 $\pm$  2.23  &     5.46 $\pm$  1.84  &    10.30 $\pm$  0.79  &    18.14 $\pm$  0.73  &     9.64 $\pm$  1.30  &    68.89 $\pm$  3.30  &     6.40 $\pm$  0.80  &     5.43 $\pm$  0.71  &     7.36 $\pm$  0.55\\
\hline
 NGC 2342  & nuc  &     7.81 $\pm$  6.65  &    19.42 $\pm$  0.75  &         ------        &     8.73 $\pm$  0.38  &     6.07 $\pm$  0.43  &     5.48 $\pm$  0.24  &    11.22 $\pm$  0.36  &     6.03 $\pm$  0.36  &    57.37 $\pm$  2.13  &     3.19 $\pm$  0.80  &     1.97 $\pm$  0.18  &     3.71 $\pm$  0.17\\
\hline
 NGC 2903  & S1   &         ------        &     5.07 $\pm$  0.57  &         ------        &     1.44 $\pm$  0.32  &     1.83 $\pm$  0.22  &     1.70 $\pm$  0.14  &     2.27 $\pm$  0.28  &     1.25 $\pm$  0.26  &         ------        &         ------        &     0.74 $\pm$  0.13  &     0.86 $\pm$  0.04\\
 NGC 2903  & S2   &     3.81 $\pm$  0.47  &     6.84 $\pm$  0.85  &     0.95 $\pm$  0.14  &     6.25 $\pm$  0.54  &     3.99 $\pm$  0.68  &     3.25 $\pm$  0.35  &     7.56 $\pm$  0.46  &     2.30 $\pm$  0.43  &         ------        &         ------        &     1.06 $\pm$  0.25  &     1.60 $\pm$  0.27\\
 NGC 2903  & S3   &         ------        &     1.05 $\pm$  0.21  &         ------        &     2.07 $\pm$  0.42  &     0.72 $\pm$  0.25  &     1.40 $\pm$  0.16  &     1.53 $\pm$  0.24  &         ------        &         ------        &         ------        &     0.27 $\pm$  0.05  &     0.22 $\pm$  0.09\\
 NGC 2903  & N1   &         ------        &     5.38 $\pm$  1.00  &         ------        &     4.91 $\pm$  0.68  &     7.60 $\pm$  1.60  &     1.80 $\pm$  0.32  &     3.72 $\pm$  0.33  &         ------        &         ------        &     1.81 $\pm$  0.25  &     0.93 $\pm$  0.14  &     0.83 $\pm$  0.62\\
 NGC 2903  & N2   &     7.48 $\pm$  1.01  &     9.52 $\pm$  0.82  &     1.81 $\pm$  0.16  &     7.95 $\pm$  0.64  &     6.02 $\pm$  0.61  &     3.53 $\pm$  0.28  &     9.49 $\pm$  0.62  &     3.82 $\pm$  0.75  &     9.40 $\pm$  0.61  &         ------        &     1.73 $\pm$  0.17  &     2.25 $\pm$  0.13\\
 NGC 2903  & N3   &     4.52 $\pm$  0.58  &    11.35 $\pm$  0.67  &     2.77 $\pm$  0.53  &     8.61 $\pm$  5.94  &     6.87 $\pm$  0.63  &     4.07 $\pm$  0.29  &    15.10 $\pm$  0.48  &     3.99 $\pm$  0.41  &         ------        &         ------        &     1.75 $\pm$  0.09  &     4.92 $\pm$  0.07\\
 NGC 2903  & N4   &         ------        &     3.28 $\pm$  0.51  &     1.01 $\pm$  0.31  &     3.37 $\pm$  0.32  &     2.13 $\pm$  0.44  &     1.26 $\pm$  0.23  &     4.48 $\pm$  0.30  &         ------        &         ------        &         ------        &     0.62 $\pm$  0.08  &     1.09 $\pm$  0.12\\
 NGC 2903  & nuc  &         ------        &     3.36 $\pm$  0.58  &         ------        &         ------        &         ------        &     2.67 $\pm$  0.33  &     2.32 $\pm$  0.43  &         ------        &         ------        &     3.72 $\pm$  0.37  &     1.37 $\pm$  0.38  &     0.76 $\pm$  0.38\\
\hline
 NGC 2964  & S1   &         ------        &    28.89 $\pm$  3.33  &         ------        &         ------        &         ------        &         ------        &         ------        &         ------        &         ------        &         ------        &         ------        &         ------      \\
 NGC 2964  & nuc  &    22.65 $\pm$  0.68  &    49.64 $\pm$  7.46  &         ------        &    55.65 $\pm$  8.40  &    40.72 $\pm$  5.35  &     9.46 $\pm$  1.33  &    19.14 $\pm$  3.19  &         ------        &         ------        &     6.89 $\pm$  0.69  &     4.34 $\pm$  0.82  &     6.43 $\pm$  1.03\\
\hline
 NGC 3184  & S1   &         ------        &     2.40 $\pm$  0.31  &         ------        &         ------        &         ------        &         ------        &     1.45 $\pm$  0.32  &         ------        &         ------        &         ------        &     0.42 $\pm$  0.08  &         ------      \\
 NGC 3184  & N1   &         ------        &         ------        &         ------        &         ------        &         ------        &         ------        &     1.02 $\pm$  0.11  &         ------        &         ------        &         ------        &         ------        &         ------      \\
 NGC 3184  & nuc  &     1.68 $\pm$  0.39  &     6.37 $\pm$  0.59  &         ------        &     4.38 $\pm$  0.68  &     3.13 $\pm$  0.61  &     0.77 $\pm$  0.11  &     3.83 $\pm$  0.40  &     1.52 $\pm$  0.23  &         ------        &         ------        &     0.86 $\pm$  0.09  &     0.75 $\pm$  0.09\\
\hline
 NGC 4102  & S1   &     5.54 $\pm$  0.07  &    13.55 $\pm$  0.41  &         ------        &    11.42 $\pm$  1.20  &     5.95 $\pm$  1.04  &     6.48 $\pm$  0.54  &     9.43 $\pm$  0.38  &     5.35 $\pm$  1.14  &         ------        &     3.09 $\pm$  0.30  &     2.21 $\pm$  0.36  &     4.49 $\pm$  0.23\\
 NGC 4102 &  N1   &     3.86 $\pm$  0.30  &    11.47 $\pm$  0.95  &         ------        &     6.01 $\pm$  0.54  &     4.78 $\pm$  0.44  &     6.02 $\pm$  0.53  &     8.56 $\pm$  0.78  &         ------        &         ------        &     4.29 $\pm$  0.49  &     4.09 $\pm$  0.39  &     3.83 $\pm$  0.49\\
 NGC 4102 &  nuc  &    18.97 $\pm$  0.73  &    60.06 $\pm$  1.10  &     5.71 $\pm$  0.55  &    27.42 $\pm$  2.12  &    20.63 $\pm$  2.15  &    28.47 $\pm$  1.15  &    44.59 $\pm$  1.84  &    20.84 $\pm$  5.06  &         ------        &    21.96 $\pm$  1.60  &    11.36 $\pm$  1.45  &    18.50 $\pm$  1.04\\
\hline
 NGC 4303 &  S1   &         ------        &     2.53 $\pm$  0.72  &         ------        &         ------        &         ------        &         ------        &         ------        &         ------        &         ------        &         ------        &     0.40 $\pm$  0.06  &         ------      \\
 NGC 4303 &  S2   &     0.99 $\pm$  0.22  &     4.14 $\pm$  0.59  &         ------        &     1.19 $\pm$  0.23  &         ------        &     0.65 $\pm$  0.12  &     0.58 $\pm$  0.12  &         ------        &         ------        &         ------        &         ------        &         ------      \\
 NGC 4303 &  S3   &         ------        &         ------        &         ------        &     1.14 $\pm$  0.15  &         ------        &         ------        &         ------        &         ------        &         ------        &     0.53 $\pm$  0.05  &     0.21 $\pm$  0.05  &         ------      \\
 NGC 4303 &  N1   &     4.71 $\pm$  0.60  &     2.27 $\pm$  0.53  &         ------        &         ------        &         ------        &     0.74 $\pm$  0.17  &         ------        &         ------        &         ------        &         ------        &     0.16 $\pm$  0.04  &         ------      \\
 NGC 4303 &  N2   &         ------        &         ------        &         ------        &     1.08 $\pm$  0.24  &     0.81 $\pm$  0.10  &     0.80 $\pm$  0.09  &         ------        &         ------        &         ------        &         ------        &     0.40 $\pm$  0.06  &         ------      \\
 NGC 4303 &  N3   &         ------        &     0.80 $\pm$  0.29  &         ------        &         ------        &         ------        &         ------        &         ------        &         ------        &         ------        &         ------        &     0.29 $\pm$  0.04  &         ------      \\
 NGC 4303  & nuc  &     5.14 $\pm$  1.06  &     7.88 $\pm$  1.28  &         ------        &         ------        &         ------        &         ------        &         ------        &         ------        &         ------        &     5.99 $\pm$  1.09  &     1.23 $\pm$  0.21  &         ------      \\
\hline
 NGC 4845 &  S1   &         ------        &         ------        &         ------        &     2.07 $\pm$  0.49  &         ------        &         ------        &         ------        &         ------        &         ------        &     1.52 $\pm$  0.18  &     0.64 $\pm$  0.12  &         ------      \\
 NGC 4845 &  S2   &         ------        &         ------        &         ------        &     2.04 $\pm$  0.37  &         ------        &         ------        &         ------        &         ------        &         ------        &         ------        &     0.52 $\pm$  0.06  &         ------      \\
 NGC 4845 &  S3   &         ------        &         ------        &         ------        &         ------        &         ------        &         ------        &     1.90 $\pm$  0.25  &         ------        &         ------        &         ------        &     0.77 $\pm$  0.11  &         ------      \\
 NGC 4845 &  N1   &         ------        &         ------        &         ------        &         ------        &         ------        &         ------        &         ------        &         ------        &         ------        &         ------        &     0.70 $\pm$  0.14  &         ------      \\
 NGC 4845 &  N2   &         ------        &         ------        &         ------        &     1.07 $\pm$  0.17  &     1.08 $\pm$  0.12  &         ------        &         ------        &         ------        &         ------        &     0.54 $\pm$  0.11  &         ------        &         ------      \\
 NGC 4845 &  N3   &         ------        &         ------        &         ------        &         ------        &         ------        &         ------        &         ------        &         ------        &         ------        &     0.91 $\pm$  0.18  &         ------        &         ------      \\
 NGC 4845 &  nuc  &         ------        &    16.71 $\pm$  1.30  &         ------        &    10.47 $\pm$  0.94  &         ------        &         ------        &         ------        &         ------        &         ------        &    13.45 $\pm$  1.30  &     3.17 $\pm$  0.63  &         ------      \\
\hline
 NGC 5457 &  N1   &         ------        &     9.06 $\pm$  0.60  &         ------        &         ------        &         ------        &         ------        &     5.24 $\pm$  0.34  &         ------        &         ------        &         ------        &     1.02 $\pm$  0.11  &         ------      \\
 NGC 5457 &  nuc  &         ------        &         ------        &         ------        &         ------        &         ------        &     2.61 $\pm$  0.47  &         ------        &         ------        &         ------        &         ------        &         ------        &         ------      \\
\hline
 NGC 5905 &  S1   &         ------        &     3.40 $\pm$  0.47  &         ------        &         ------        &     1.78 $\pm$  0.21  &         ------        &     2.57 $\pm$  0.09  &         ------        &     8.91 $\pm$  0.56  &         ------        &     0.57 $\pm$  0.07  &     0.48 $\pm$  0.08\\
 NGC 5905 &  N1   &         ------        &     2.37 $\pm$  0.41  &         ------        &         ------        &         ------        &         ------        &     1.65 $\pm$  0.17  &         ------        &     6.10 $\pm$  0.32  &         ------        &         ------        &     0.54 $\pm$  0.13\\
 NGC 5905 &  nuc  &         ------        &     4.36 $\pm$  0.37  &         ------        &         ------        &         ------        &     1.04 $\pm$  0.16  &     1.45 $\pm$  0.24  &         ------        &     7.94 $\pm$  0.96  &         ------        &         ------        &     0.71 $\pm$  0.11\\
\hline
 NGC 6946 &  nuc  &    46.56 $\pm$  8.55  &    84.54 $\pm$  1.78  &         ------        &    48.62 $\pm$  1.09  &    14.91 $\pm$  1.64  &    51.52 $\pm$  0.83  &    53.87 $\pm$  1.54  &    44.20 $\pm$  5.62  &         ------        &         ------        &    13.53 $\pm$  1.83  &    21.21 $\pm$  1.03\\
\hline
 NGC 7080 &  S1   &         ------        &         ------        &         ------        &     1.83 $\pm$  0.44  &         ------        &         ------        &         ------        &         ------        &         ------        &         ------        &         ------        &     0.40 $\pm$  0.05\\
 NGC 7080 &  N1   &         ------        &         ------        &         ------        &         ------        &         ------        &         ------        &         ------        &         ------        &     3.02 $\pm$  0.29  &         ------        &         ------        &     0.23 $\pm$  0.04\\
 NGC 7080 &  nuc  &         ------        &     3.46 $\pm$  0.30  &         ------        &     2.91 $\pm$  0.26  &         ------        &         ------        &     1.33 $\pm$  0.16  &         ------        &     5.06 $\pm$  0.29  &     1.94 $\pm$  0.22  &     0.67 $\pm$  0.15  &     0.60 $\pm$  0.18\\
\hline
 NGC 7798 &  S1   &         ------        &         ------        &         ------        &     1.45 $\pm$  0.13  &         ------        &         ------        &         ------        &         ------        &     1.11 $\pm$  0.19  &     0.34 $\pm$  0.10  &         ------        &         ------      \\
 NGC 7798 &  nuc  &     8.73 $\pm$  0.82  &     7.97 $\pm$  0.99  &         ------        &     4.98 $\pm$  0.44  &     2.67 $\pm$  0.40  &     3.74 $\pm$  0.24  &     4.78 $\pm$  0.60  &     2.46 $\pm$  0.44  &    15.38 $\pm$  0.46  &     1.89 $\pm$  0.14  &     1.12 $\pm$  0.14  &     2.06 $\pm$  0.18\\
\hline
 NGC 7817 &  N1   &         ------        &    18.28 $\pm$  1.87  &         ------        &         ------        &         ------        &         ------        &         ------        &         ------        &         ------        &         ------        &         ------        &         ------      \\
 NGC 7817 &  nuc  &         ------        &    22.86 $\pm$  1.54  &         ------        &         ------        &         ------        &         ------        &         ------        &         ------        &         ------        &         ------        &         ------        &         ------      \\

\end{longtable}
\end{center}
\end{landscape}
\twocolumn

\subsection{Extinction}

The effects of dust extinction are crucial to understand the spectra of 
nearby and distant starburst galaxies. In these sources, where
active star formation is taking place, dust grains are not only produced 
and ejected from stars but are also processed in the interstellar medium. Therefore,
NIR signatures should be preferred than their optical counterparts 
in order to penetrate as deep as possible into the dusty starburst cores. 
When nebular emission is present, the effect of dust opacity is usually
removed from the spectra by attributing to dust obscuration any
difference between the observed and intrinsic ratio of hydrogen emission
lines. A few studies in starburst galaxies
found out that the NIR extinction is generally larger than the optical
one, suggesting that the dust may be clumped or heavily
mixed with the emitting gas \citep{moorwood+88, kawara+89, puxley+94}. 

The trade-off between using optical and NIR H\,{\sc i} lines is that
the latter are considerably weaker and display smaller differential
extinction than the former. Thus, a small uncertainty in the measurement
of the NIR lines translates into a large uncertainty in the final 
extinction value. 

Reddening estimates based on NIR H\,{\sc i} lines have been 
determined for many individual H\,{\sc ii}/star-forming galaxies. However, few
works have studied systematically the effect of extinction using
a sample larger than 10 objects \citep{heisler+99}. Our sample,
free of systematics, may contribute on this subject by determining the
extinction not only in the nucleus but also with the distance to
the galaxy nucleus along the slit. Note, however, that the lack of emission
lines at some locations may, in part, hinder this goal.  

\begin{figure} 
\includegraphics [width=86mm]{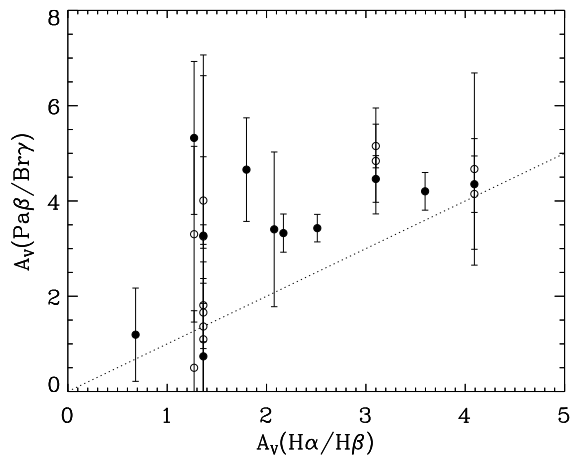}
\caption{Comparison between extinction values measured 
from the Pa$\beta$/Br$\gamma$ ratio in our NIR spectra and the H$\alpha$/H$\beta$ from HO97.}
\label{graf_ext}
\end{figure}

The strongest hydrogen lines observed in the sample are Pa$\beta$ and
Br$\gamma$. We decided not to use the Pa$\alpha$ because,
as mentioned before, it falls inside the gap of very poor atmospheric
transmission, making the measurement of its flux highly unreliable.
Pa$\gamma$ and Pa$\delta$ are weaker than Pa$\beta$ and for this
reason are not employed. Besides that, 
the [Fe\,{\sc ii}] lines at 1.257$\micron$ and 1.644$\micron$ theoretically can be
used as a direct measurement of extinction that is independent 
of electron density and temperature, since they arise from the same upper level. 
However, the A$_V$ values measured from this ratio
will be very uncertain because the intrinsic weakness of the iron lines, making the measurement 
of their emission line flux uncertain. This is specially true for  
[Fe\,{\sc ii}]~1.644 $\micron$, which is particularly affected 
because it is located very close to a CO and Brackett10 absorption 
feature.
For the hydrogen lines we assumed case\,B recombination with an effective temperature
of 10$^4$~K and electron number density of 10$^4$ cm$^{-3}$. The adopted
intrinsic line ratio Pa$\beta$/Br$\gamma$ is 5.89 \citep{osterbrock89}.
To estimate the extinction in visual magnitudes for the NIR we 
used the \citet{calzetti+00} law, appropriated for 
actively star-forming regions  and starburst. 
Table~\ref{tab_ext} shows the
A$_V$ values calculated from the Pa$\beta$/Br$\gamma$ ratio, and the A$_V$ determined
by HO97, based on the H$\alpha$/H$\beta$ ratio. 
Figure~\ref{graf_ext} shows a comparison between these values.

The comparison of the extinction values derived from the Pa$\beta$/Br$\gamma$ ratio with
the ones derived from the H$\alpha$/H$\beta$ ratio (Figure~\ref{graf_ext}) shows that the
NIR values tend to be larger than the optical ones. This result agrees to previous works
in the literature and reflects the fact that the NIR lines sample an extinction 
column larger than the optical lines \citep{heisler+99, moorwood+88}. Other
factors may also play a role here. We have assumed case\,B recombination with
the same temperature and density for all the ratios, 
which could introduce some scatter. Differences in the slit size
should also be important. Since HO95' slit is much larger, the 
extinction they derived is an average value for the galaxy. However, these
values may vary with the distance to the centre, as it can be confirmed
from Table~\ref{tab_ext}. For many sources where A$_V$ could be measured
along the slit, the off-nuclear values are lower than the nuclear ones.
If we imagine that the dust tends to be more concentrated in the nuclear regions, 
smaller apertures would naturally probe higher extinction values, as is our case.

\begin{table}
\centering
\caption{ Extinction measured from NIR line ratios for the nuclear apertures }
\footnotesize
\setlength{\tabcolsep}{8pt}
\begin{tabular}{@{}ccc@{}}
\hline
Galaxy&      A$_V$        &   A$_V$$^a$\\  
      &  (Pa$\beta$/Br$\gamma$)&  (H$\alpha$/H$\beta$)\\
\hline

NGC 0864 &    1.19 &       0.68 \\
NGC 1174 &    3.43 &      2.51 \\
NGC 1482 &    4.36 &      --- \\
NGC 2339 &    4.35 &      4.09 \\
NGC 2342 &    3.33 &       2.17 \\
NGC 2903 &    3.27 &       1.36 \\
NGC 2964 &    3.40 &      2.08 \\
NGC 3184 &    0.74 &    1.36 \\
NGC 4102 &    4.46 &      3.10 \\
NGC 5905 &    5.32 &     1.27 \\
NGC 6946 &    4.20 &       3.60 \\
NGC 7080 &    4.88 &      --- \\
NGC 7798 &    4.66 &      1.80 \\
\hline
\end{tabular}
\raggedright
\\
(a) Obtained from \citet{ho+97}
\label{tab_ext}
\end{table}

\subsection{Diagnostic Diagram in the NIR}

\citet{rodriguez-ardila+04} and \citet{rodriguez-ardila+08} suggested that the
line ratios H$_2$~ 2.121$\micron$/Br$\gamma$ and [Fe\,{\sc ii}]~1.257$\micron$/Pa$\beta$
are suitable to separate emission line objects according to their level of activity. 
Starburst galaxies should be preferentially located in the region with 
[Fe\,{\sc ii}]~1.257$\micron$/Pa$\beta$ and H$_2$~2.121$\micron$/Br$\gamma$ $<$ 0.6, while 
AGNs would have higher values for both ratios.

\begin{figure*} 
\includegraphics [width=170mm]{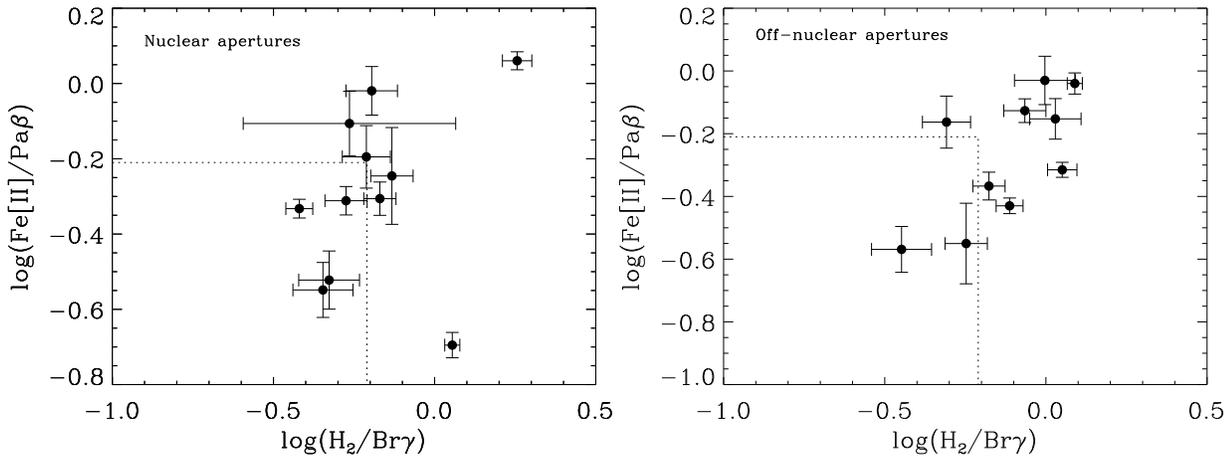}
\caption{The flux ratios H$_2$~2.12$\micron$/Br$\gamma$ vs 
the [Fe\,{\sc ii}]~1.257$\micron$/Pa$\beta$ for the nuclear extractions.
The dotted line marks the literature value for the boundary between AGN and SBs.}
\label{diagnostic}
\end{figure*}

Figure~\ref{diagnostic} shows the flux ratios [Fe\,{\sc ii}]~1.257$\micron$/Pa$\beta$ 
and H$_2$~2.121$\micron$/Br$\gamma$ for the objects of our sample in which these lines could
be measured. Note that because of the proximity in wavelength, both ratios are insensitive
to reddening. The largest source of uncertainty here is the flux associated to the 
H\,{\sc i} lines, as it may be suppressed by underlying absorption.
  
The dotted linen in Figure~\ref{diagnostic} marks the suggested value for the boundary between AGN and SBs.
Although many objects fall inside or very close to this boundary, this figure
clearly shows that the values adopted for this separation have to be reviewed. 
Even if we take into account that the hydrogen lines might be underestimated because of 
the stellar population contribution that was not subtracted, the overall result 
should not be significantly altered after correcting for this effect. 
The most extreme case seems to be the nucleus of
NGC~2903. This might be due to the fact that the emission lines are intrinsically weaker
in the nuclear region, which makes the measurements of the emission lines
more subjected to residual of the telluric lines extraction or residual
noises, and therefore, more uncertain. Note here that this result does not
contradict the results from \citet{rodriguez-ardila+08},
where at least the nuclear region of one SBT, as well as some off-nuclear results 
shows ratios values higher than those limited by 0.6. 
AGNs would 
indeed be located in the region with higher values of [Fe\,{\sc ii}]~1.257$\micron$/Pa$\beta$ 
and H$_2$~2.121$\micron$/Br$\gamma$, as they showed, and galaxies 
falling inside the dotted region in Figure~\ref{diagnostic} are probably
starbursts. However, starbursts can also have higher values for these
ratios and therefore be located outside the dotted boundary in this figure.

\section{Optical properties vs. Near-Infrared properties} \label{optprop}

HO97 classified the galaxies of our sample as H\,{\sc ii}/starburst galaxies 
based on their optical emission line ratios, that should obey 
[O\,{\sc i}]$\lambda$6300/H$\alpha$ $<$ 0.08, [N\,{\sc ii}]$\lambda$6584/H$\alpha$  $<$ 0.06 and 
[S\,{\sc ii}]$\lambda\lambda$6716,6731/H$\alpha$ $<$ 0.04.  
It is then interesting to compare the optical
emission line spectrum with that measured in the NIR. This allows 
us to see, for instance, if evidences of broad permitted H\,{\sc i} 
lines are present in the low-order Brackett or Paschen lines, implying the 
existence of a buried AGN. It also permits to place upper limits to 
the location of the starburst component based on the published measures of the
H$\alpha$ flux. In this discussion, it is important to 
keep in mind the differences in slit sizes: each of our extractions
are typically 0.8$\arcsec \times 2\arcsec$, while HO95's
slit is 2$\arcsec \times 4\arcsec$. That is, 
a factor of 5 in the area covered compared to ours.  

The figures in the Appendix present the optical spectrum of HO95 (upper panel)
and the corresponding NIR apertures extracted for each galaxy (lower panel).  
It is easy to see that all optical spectra but those of the normal
galaxies NGC\,221, NGC\,2950, NGC\,4179 and NGC\,4461 display
prominent H$\alpha$, [N\,{\sc ii}]$\lambda$6584 and 
[S\,{\sc ii}]$\lambda\lambda$6716,6731 emission in the optical. In comparison, 
only a fraction of the sample displays bright emission lines
in the NIR. Moreover, no evidence of the presence of broad components 
in the H\,{\sc i} is observed.  Although we cannot rule out completely
the existence of a hidden AGN based on these pieces of evidence alone,
very likely none of the objects of our sample could be classified
as an AGN. If hidden AGN does exist, the reddening towards the nucleus
would need to be larger than 30 mag.
This implies  that all the observed emission lines are produced 
by stellar processes, being it patronisation by a star or
a star cluster, or even shocks produced by supernova explosions.

Comparing HO95's optical spectra with the corresponding NIR allows us to separate 
the galaxies in 4 main classes, based on the presence and strength of their emission 
lines:

\begin{itemize}

\item Weak emission lines in the optical, no emission lines in the NIR,
either at the nucleus or
in the extended region (class 1): NGC\,514, NGC\,674, NGC\,6181, NGC\,7448.

\item Strong emission lines in the optical, no emission lines in the NIR, 
either at the nucleus or
in the extended region (class 2): NGC\,278, NGC\,7817

\item Strong emission lines in the optical, evidence of weak to moderate-intensity lines 
in the NIR (nucleus or/and extended region - class 3):  
NGC\,783, NGC\,864, NGC\,1174, NGC\,2964, NGC\,3184, NGC\,4303, NGC\,4845, NGC\,5457, NGC\,5905, NGC\,7080,
NGC\,7798


\item Strong emission lines in the optical, moderate to strong emission lines in the NIR 
(class 4):  
NGC\,2339, NGC\,2342, NGC\,2903, NGC\,4102, NGC\,6946

\end{itemize}

It is easy to see that class\,3 concentrates the largest number of objects of our sample,
mid-point between objects with no emission lines in the NIR and those with prominent permitted
and forbidden lines. Moreover, three of the galaxies belonging to this class (NGC\,4303, NGC\,4845 
and NGC\,5457) have no H\,{\sc i} lines at the nucleus, presenting them only in the off-nuclear apertures, 
implying that the region(s) of strong starburst activity is(are) very 
likely located on a ring of star formation around the nuclear region, or in a hot spot 
outside the nucleus. As our slit is considerably narrower than that of HO95, regions
of active star formation outside the 0.8$\arcsec$ strip, with centre at the nucleus, are not 
covered by our spectra but would show up in HO95 work if the angular distance 
of these regions from the centre is $< 2\arcsec$. 

Thus, the lack of NIR emission lines in objects of classes 1 and 2 can be explained by
differences in the slit width between the optical and NIR observations. In addition, differences 
in the position angle between our observations 
and those of HO95 (they oriented the slit along the parallactic angle while ours was always
kept in the 
N-S direction, PA=0 deg) may also lead to the mapping of different portions of 
the circumnuclear region of the host galaxy. A third possibility is insufficient signal above 
the noise to detect them. 

In order to examine the three alternatives above, Table~\ref{tab:emission}
lists the upper limit derived for Pa$\alpha$ (3-$\sigma$, column 2) for the nuclear NIR spectrum
of all objects belonging to classes 1 and 2, objects from class 3 without emission lines in the
nucleus and the flux measured on this line for two of the objects of class 4 with bright Pa$\alpha$ lines. 
The H$\alpha$ flux (column 3) were taken from H095. The expected Pa$\beta$ flux (column 4)
was determined assuming Case-B recombination and an intrinsic Pa$\beta$/H$\alpha$ flux
ratio of 0.05.

\begin{table}
\centering
\caption{Measured and expected Pa$\beta$ fluxes for the galaxy sample}
\footnotesize
\setlength{\tabcolsep}{2pt}
\begin{tabular}{@{}ccccc@{}}

\hline
Galaxy  &  Pa$\beta$  &  H$\alpha^{a}$  &  Pa$\beta^{b}$  & Class \\
        &(upper limit, 3$\sigma$)&         &    (expected)& \\
\hline
NGC\,278  & 5.40x10$^{-16}$  & 3.55x10$^{-16}$ & 1.77x10$^{-15}$ & 2 \\
NGC\,514  & 3.32x10$^{-16}$  & 2.29x10$^{-15}$ & 1.15x10$^{-16}$ & 1 \\
NGC\,674  & 1.11x10$^{-15}$  & 4.90x10$^{-15}$ & 2.45x10$^{-16}$ & 1 \\
NGC\,2903 & 7.90x10$^{-16}$  & 2.88x10$^{-14}$ & 1.44x10$^{-15}$ & 4 \\
NGC\,4303 & 2.43x10$^{-15}$  & 1.45x10$^{-13}$ & 7.23x10$^{-15}$ & 3 \\
NGC\,4845 & 2.54x10$^{-15}$  & 2.45x10$^{-14}$ & 1.23x10$^{-15}$ & 3 \\
NGC\,5457 & 2.03x10$^{-15}$  & 4.68x10$^{-14}$ & 2.34x10$^{-15}$ & 3 \\  
NGC\,6181 & 1.25x10$^{-15}$  & 2.69x10$^{-14}$ & 1.35x10$^{-15}$ & 1 \\
NGC\,7448 & 4.69x10$^{-16}$  & 9.77x10$^{-15}$ & 4.89x10$^{-16}$ & 1 \\
NGC\,7817 & 6.54x10$^{-16}$  & 5.62x10$^{-14}$ & 2.81x10$^{-15}$ & 2 \\
NGC\,2339 & 1.69x10$^{-14}$  & 8.51x10$^{-14}$ & 4.26x10$^{-15}$ & 4 \\
NGC\,2342 & 1.05x10$^{-14}$  & 1.48x10$^{-13}$ & 7.40x10$^{-15}$ & 4 \\

\hline
\end{tabular}
\raggedright
\\
(a) Obtained from \citet{ho+97}\\
(b) Calculated based on the H$\alpha$ value

\label{tab:emission}
\end{table}

When the expected Pa$\beta$
flux is comparable or smaller than the measured 3$\sigma$ upper limit,
the NIR spectrum lacks of sufficient S/N for its adequate detection.
This would be the case of all the four galaxies of class\,1. That
result is expected as the optical spectra of these sources show weaker
H$\alpha$ emission, sometimes still embedded within the stellar absorption depth.
Therefore, the absence of emission lines in these sources is
probably due to the intrinsic weakness of the line.
This contrasts with what is observed in class\,2 sources. The two objects 
belonging to this category have predicted Pa$\beta$ fluxes larger than the
upper limit determined for that line. 
This means that if the region emitting the Pa$\beta$ line were inside the slit, 
it should have been observed.
Therefore, the optical emission lines should come from regions not covered 
by our slit but that were inside HO95's slit.

Considering the galaxies of class~\,3, it should be noted that six of them display 
emission lines only in the nuclear region (NGC~864, NGC~1174, NGC\,2964, NGC\,3184, NGC\,7080 and NGC\,7798), 
while NGC\,4303, NGC~4845, NGC~5457 and NGC\,5905 have emission lines
that are more prominent on the off-nuclear extractions. For 
three of these sources, where Pa$\beta$ was not measured in the nuclear region, Table~\ref{tab:emission} lists the expected Pa$\beta$ flux and the
upper limit flux measured from their nuclear spectra. The results show that  
for two of these galaxies, the predicted Pa$\beta$ is larger than the detection limit, and therefore
the starburst component is very likely located around the
nuclear region. This hypothesis is further supported by the off-nuclear emission 
lines that are clearly detected in our spectra. This would also be the situation 
of\,NGC 5905, which has emission lines on all three extractions, but the emission 
is stronger to the south. Indeed, this galaxy has a star-forming ring \citep{cameron+10} 
that would explain the enhanced emission outside the nucleus.
           
For most of the galaxies with strong nebular lines in both optical and NIR, the lines in the
infrared are either only in the nuclear region, or stronger on the nucleus and with some
emission on the off-nuclear extractions or vice-versa.  

From all said above, the comparison between the optical and NIR
spectra allowed us to conclude that the location of starburst component of the 
objects studied here varies from galaxy to galaxy. Three cases are particularly
relevant: (i) weak starbursts, where the NIR nebular lines are too weak to
show up; (ii) nuclear starbursts, where the star-forming region is located 
within the inner few tens of parsecs of the galaxy nucleus. The NIR spectrum
of these objects shows prominent nuclear nebular lines that decreases in strength
as the distance from the centre increases; (iii) star-forming rings or hot spots, located at
distances larger than 100~pc from the nucleus. The NIR nuclear spectrum of these objects
lacks of nuclear emission lines. The nebular lines become prominent in the
off-nuclear spectra.

\subsection{A NIR template of star-forming galaxies in the NIR}

Spectral templates constructed from moderate to large uniform 
samples have been invaluable 
to understand the class of objects 
in question. Composite spectra enhance common features,
while unusual characteristics are averaged out. Composite spectra can also be 
useful at comparing new subclasses of sources with existing populations.
The lack of a suitable star-forming galaxy template redward of 9000~\AA\ 
motivated us to produce 
such template in the NIR region.

In order to create a mean spectrum it is necessary to shift the objects to
their rest frame using the redshifts from the emission lines and
then rebin everything to a common dispersion. This last step is necessary
as the dispersion changes from the $z-$ to the $K$-bands by more than a 
factor of 2 over the wavelength range of our spectra. Therefore, we chose to rebin
our spectra by picking a dispersion that is proportional to the
wavelength so that R is a constant. We choose $\Delta\lambda$=5~\AA\
as it corresponds to the bin the $K$-band, the largest across the
wavelength region sampled. We linearly 
interpolated the flux and error
arrays onto this wavelength grid. Finally, we normalised the spectra to
unity in the interval 15250$\pm$100~\AA. This wavelength region was 
taken as reference because it approximately coincides with the centre 
of the spectra. 

\begin{figure*} 
\includegraphics [width=190mm]{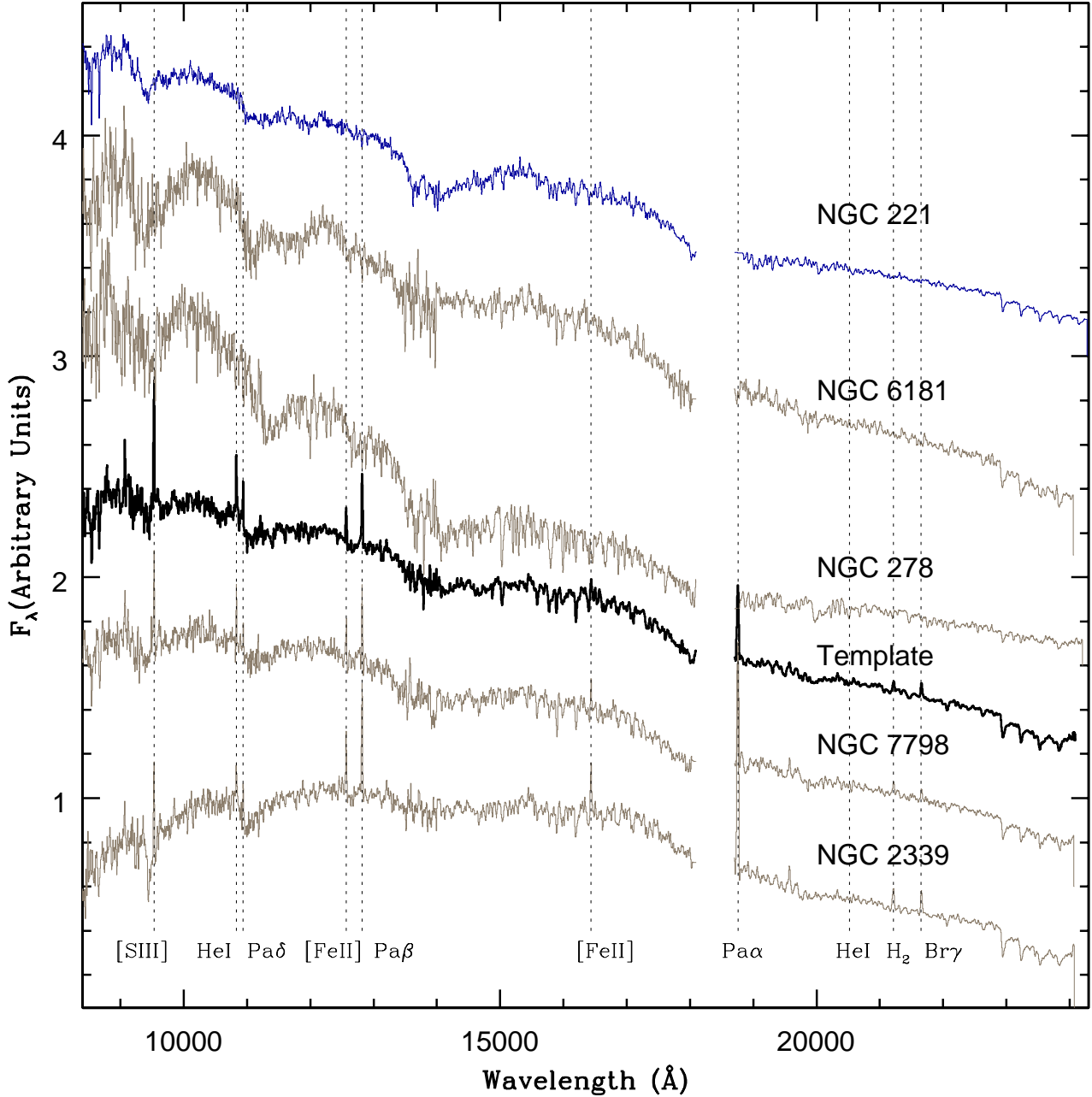}
\caption{NIR star-forming galaxy composite spectrum (black line). Spectra
representative of group 1 (NGC~6181), group 2 (NGC\,278), group 3 (NGC\,7798),
and group 4 (NGC\,2339) are overlaid (gray line). The spectrum in blue is that
of a normal galaxy (NGC\,221). The dotted lines show
the positions of expected prominent emission lines: [S\,{\sc iii}]
$\lambda$9069, He\,{\sc i} $\lambda$10830, Pa$\delta~\lambda$10941, [Fe\,{sc ii}]
$\lambda$12568, Pa$\beta~\lambda$12818, [Fe\,{sc ii}]\,$\lambda$16346, 
Pa$\alpha~\lambda$18756, He\,{\sc i} $\lambda$20520, H$_2~\lambda$21212, and 
Br$\gamma~\lambda$21654.}
\label{fig:template}
\end{figure*}

We built our composite spectrum by averaging the integrated normalised spectra
of the star-forming galaxies. For this purpose, all the signal along the slit
was integrated for each source. Normal galaxies were not included in
this procedure. Figure~\ref{fig:template} displays the final composite star-forming
galaxy spectrum.  Because of the
increase 
in the S/N of this template spectrum, most emission
and absorption features are significantly more prominent than in the
individual sources. 
 
How representative of the star forming galaxies in the NIR is our derived 
template? As we have shown in the previous sections, some of our sources
display little or 
no emission lines, while for others these lines are
Bright and prominent. Figure~\ref{fig:template} compares the template with a 
representative spectrum of each of the four groups defined in the preceding
section. We also compared our template with a normal galaxy spectrum. 
It can be seen that the main differences between star-forming galaxies relative
to the template is found in the slope of the continuum, while stronger
differences 
are observed when the template is compared to normal galaxies.

We expect that our template can be used as a benchmark of stellar population(s) in 
starburst galaxies against which to compare near-IR spectroscopy of different types 
of galaxies, especially those with AGN activity and/or those at high-redshift.

\section{The NIR continuum in H\,{\sc ii}/star-forming galaxies} \label{continuum}

A rapid inspection of the figures in the Appendix shows that despite the fact that the optical 
spectra of most objects of the sample are rather similar, the NIR spectroscopy reveals a wealth of 
absorption features and continuum shapes that contrast to what is observed in the
optical. The analysis carried out in the preceding sections show that the continuum is
intrinsically of stellar origin. Therefore, its shape should reflect the dominant
stellar population in each galaxy.
 
\begin{figure*} 
\includegraphics [width=160mm]{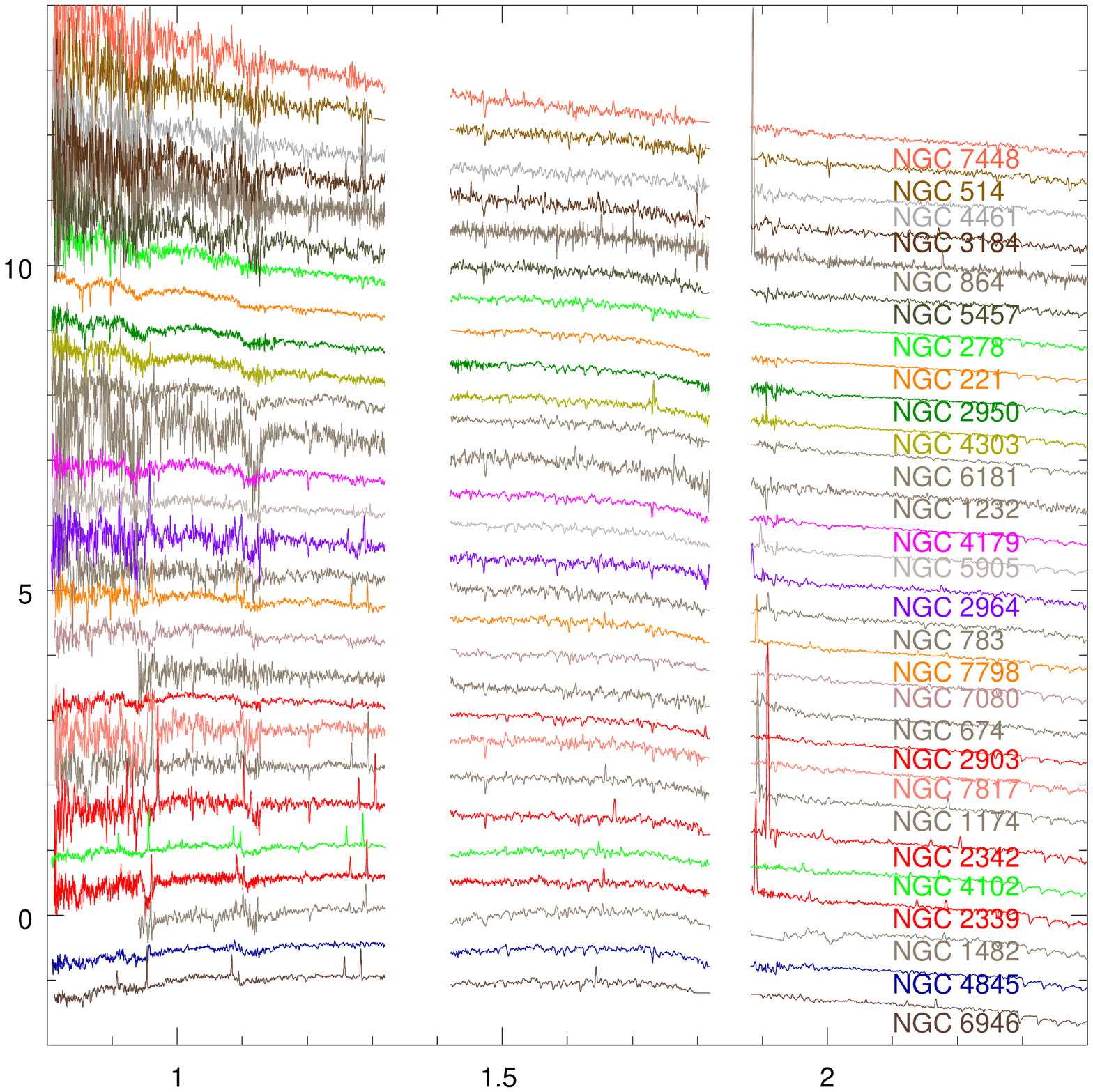}
\caption{The galaxy sample organised by the shape of the observed continuum. At the top are
the sources with the steepest continua, followed by those with a flatter continuum. The bottom
is dominated by objects with the reddest continuum. All galaxies were normalised to unity at
1.5~$\micron$ and displaced by a constant amount for displaying purposes.}
\label{fig:continuum}
\end{figure*}

In order to study more closely this issue, Figure~\ref{fig:continuum} shows the nuclear
spectra of all galaxies (normal and H\,{\sc ii}) sorted out according to the form of
the observed continuum. At the top is NGC\,7748, which shows the steepest continuum, whose
shape resembles that of a power-law usually found in AGN \citep{riffel+06}. However,
several absorption features, particularly the TiO and CN bands discussed in Sect.~\ref{indices},
are easily distinguished. The sequence continues with sources whose continuum emission gets
progressively flatter, particularly in the $z+J-$bands, up to the point where it starts increasing 
its flux with wavelength in the $z-$band, then in the $H-$band reaches a maximum and then 
its flux decreases with wavelength in the $K-$band (from NGC\,4102 up to NGC\,6946, located 
at the bottom of the Figure.  

Although the galaxies in Figure~\ref{fig:continuum} are sorted out by their continuum shape, 
it is easy to see that most sources from the mid- to the top part are populated by
objects with very weak or no emission lines. In contrast, objects located in the
lower part of the plot are dominated by permitted H\,{\sc i} and He\,{\sc i} lines
as well as forbidden [S\,{\sc iii}] and [Fe\,{\sc ii}]. 
Note that the five normal galaxies used as a control sample (NGC\,221, NGC\,1232,
NGC\,2950, NGC\,4179 and NGC\,4461) are all located in the upper portion of the graph.

If the continuum shape follows the absence/presence of emission lines, it can
be concluded, on a first approximation, that those sources with a steep continuum
should be dominated by a population of old cool stars, unable to ionise the
gas surrounding it. Objects located in the lower portion of Figure~\ref{fig:continuum}
should, in contrast, display a significant amount of young to intermediate age
stellar population. Support to this hypothesis comes from the fact that
the strength of the CO bands at 2.3$\mu$m clearly increases from top to bottom. 
It is well-known that the first CO overtone bands in the $K$ band 
($\sim$2.29 - 2.5\,$\micron$) are strongest in supergiants and become progressively weaker 
with decreasing luminosity. 

If our hypothesis is correct, objects located in the lower portion of  
Figure~\ref{fig:continuum} should also display high values of extinction compared to
those of the upper portion. The reason is because a young to intermediate stellar population should
still contain remnant dust from their parent cloud. Column\,3 of Table~\ref{tab_ext}
confirms that 10 out of 12 sources with emission lines have $A_{\rm V}$ values 
between $\sim$4.7 to 6.9. These objects are all located from the mid- to the bottom
part of the Figure. The remaining two objects (NGC\,3184 and NGC\,864) have $A_{\rm V}$
of 1.04 and 1.68, respectively, both located in the top portion of the graph.

\section{Summary and conclusions}

We obtained long-slit NIR spectra of 23 star-forming galaxies with 
the objective of building an homogeneous sample to test  the predictions
of stellar population models
that are and will soon be released in this wavelength region. We also observed
5 normal galaxies for comparison. The spectra
were obtained at IRTF using the SpeX spectrograph. 

We compared our NIR spectra with optical spectra available in the literature. For 
this comparison one have to keep in mind that the aperture sizes are very different. The optical 
aperture is much larger (2" x 4") compared to our long slit aperture (0.8" x 2"),
which means a factor of 5 times in area. This is probably
the main reason for most of the differences observed between the spectra. We found that 26$\%$
of the galaxies classified as star-forming galaxies in the optical show no emission lines
in the NIR. 
For the ones that have weak optical emission lines this is explained by our
lack of sensibility to detect the emission lines in the NIR, where they would be
intrinsically weaker. However, for the galaxies with strong
emission lines in the optical that have no signs of emission in the NIR
we believe that the star-formation region in these galaxies
is not nuclear and was missed by our slit.  

The extinction in the NIR due to the gas was calculated using the hydrogen 
ratio Pa$\beta$/Br$\gamma$, and was found to be generally
larger than the extinction calculated in the optical by
HO97. This result agrees with what was found in the literature,
and might be due to a combination of several effects. Since the slit
size used for the optical observations is much larger than ours, the 
extinction derived by HO97 is an average value for the galaxy.
If the dust tends to be more
concentrated in the nuclear regions, smaller apertures would naturally probe
higher extinction values. Along with that, the NIR intrinsically
should probe larger optical depths than the optical range.

As suggested in the literature, we constructed a diagnostic diagram based
on the [Fe\,{\sc ii}]$\lambda$1.257$\micron$/Pa$\beta$ and H$_2\lambda$2.121$\micron$/Br$\gamma$
ratios, that was suggested in the literature to 
be able to separate star-forming dominated systems from the ones
dominated by AGN activity. Although several of our galaxies 
fall in the region where star-forming systems should be, some of
the points fall in the regions where AGN activity is supposed to dominate
the ionisation process. This result seems to indicate that the values of
[Fe\,{\sc ii}]$\lambda$1.257$\micron$/Pa$\beta$ and H$_2\lambda$2.121$\micron$/Br$\gamma$
ratios limiting the starburst zone should be reviewed.
Although
the region defined by the lower values of these line ratios would be occupied by
true starbursts only, and AGNs would always fall in the region for higher values 
of these ratios, starbursts might also have high values for these ratios and be mixed
with the AGNs.  

We also measured NIR indexes according to the definitions of \citet{riffel+08}. 
We found no
correlation between these indexes and optical indicators of 
age or activity obtained from the literature.
We detected important signatures predicted for a stellar population
dominated by the TP-AGBs, like CN 1.1 $\mu$m and CO 2.3$\mu$m. We detected in 
at least one galaxy (NGC~4102) the CN 1.4 $\mu$m, which is
 the first time this signature has been detected in an extragalactic source. We also
detect TiO and ZrO bands that, again
have never been reported before 
in galaxies.

With our sample we created a NIR template of star-forming galaxies
that can be used as a benchmark of stellar populations in future
studies.

The NIR continuum of the galaxies behaves differently from what is seen 
in the optical. While in the optical the continuum of most of the sources
presented here is very similar, in the NIR they show a diversity of 
absorption signatures and continuum shapes which we believe are related to
the presence  
or absence of dust, as well as of young stellar populations: sources with steep continuum
would be dominated by a population of old cool stars. Sources where there is a 
maximum in the H-band and the flux decreases towards the K-band would have
a significant fraction of young and/or intermediate age population.

\section*{Acknowledgments}
The authors thank Jari Kotilainen, the referee of this paper, for his valuable comments. 
This research has been partially supported by the Brazilian agency FAPESP (2011/00171-4).
ARA thanks to CNPq for financial support through grant 308877/2009-8.
SD thanks INCT-A and CAPES for finnacial support.

\appendix
\section{Spectra}

\begin{figure*}
\includegraphics [width=160mm]{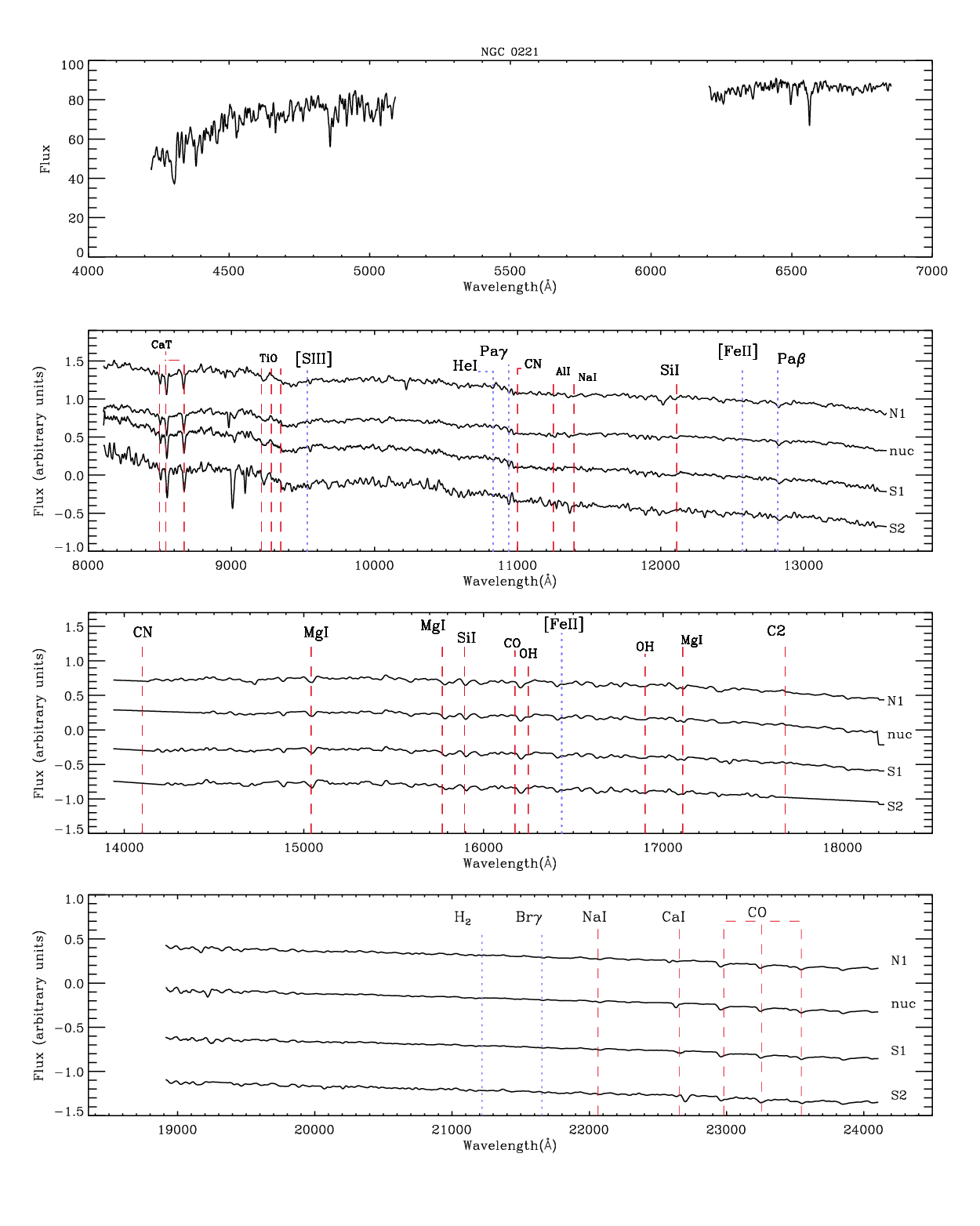}
\caption{ Comparison between the optical spectrum of NGC 221 from Ho et al. (1997)
and our near-infrared spectra.}
\end{figure*}

\begin{figure*} 
\includegraphics [width=160mm]{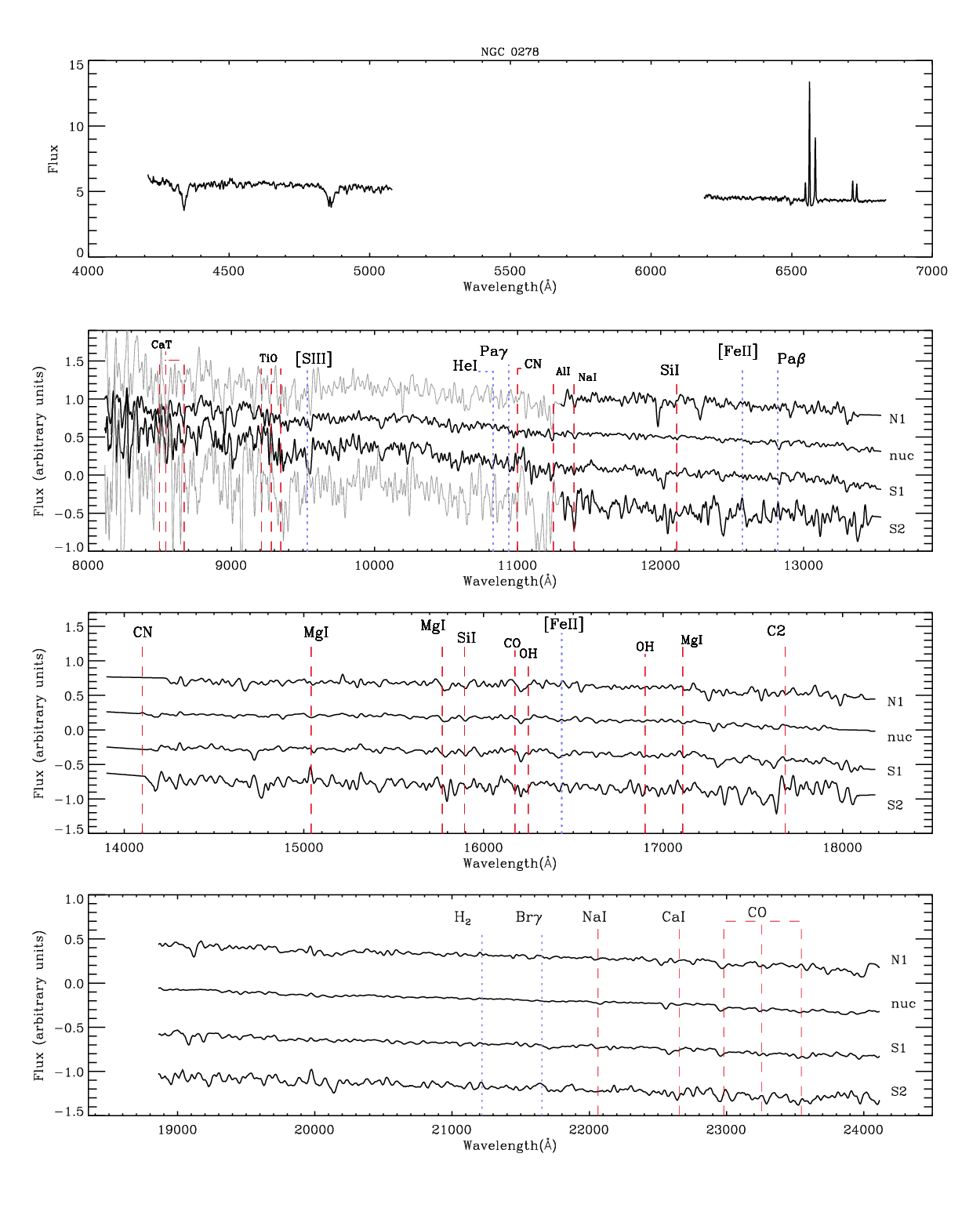}
\caption{Comparison between the optical spectrum of NGC 278 from Ho et al. (1997)
and our near-infrared spectra. }
\end{figure*}

\begin{figure*} 
\includegraphics [width=160mm]{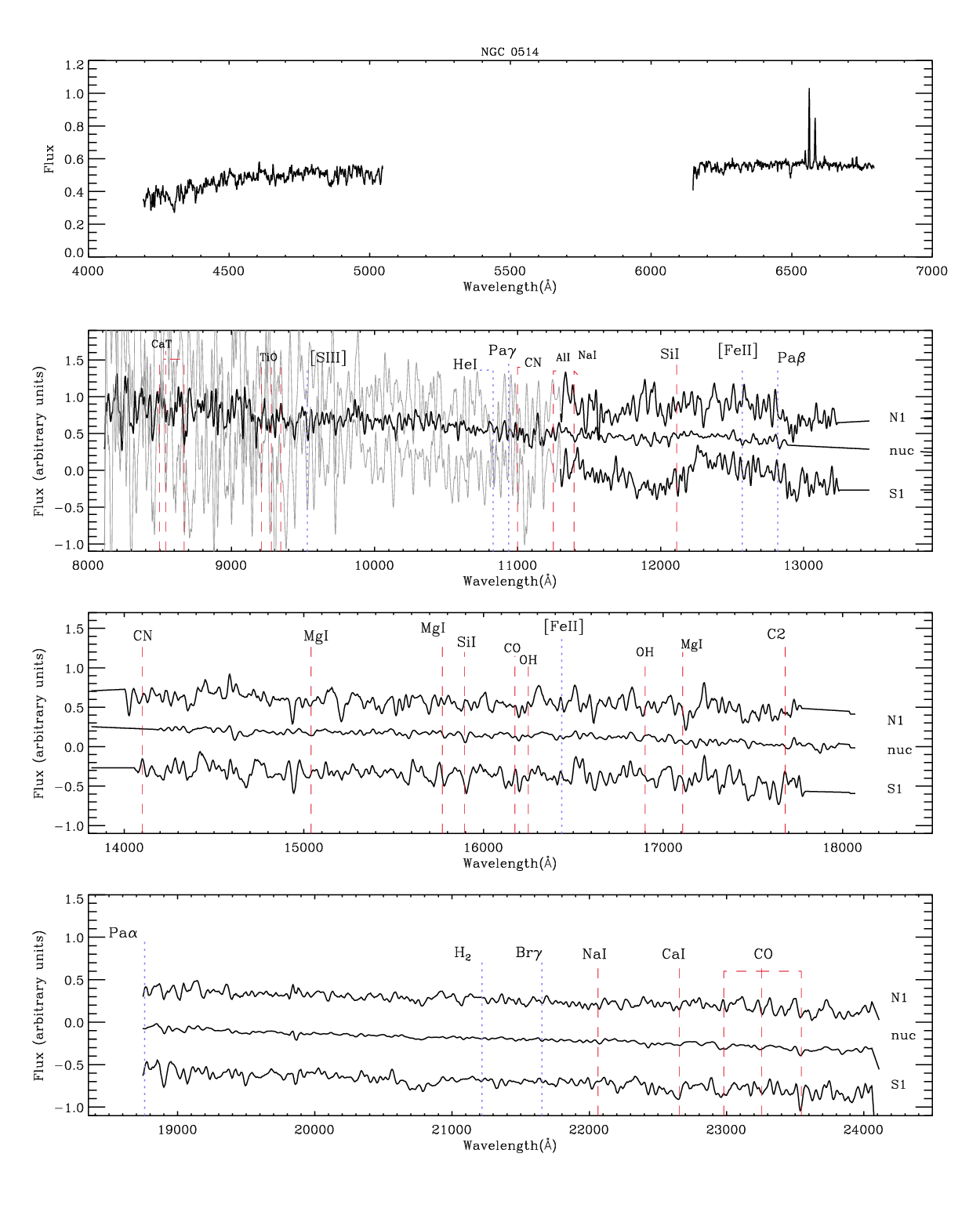}
\caption{Comparison between the optical spectrum of NGC 514 from Ho et al. (1997)
and our near-infrared spectra. }
\end{figure*}

\begin{figure*} 
\includegraphics [width=160mm]{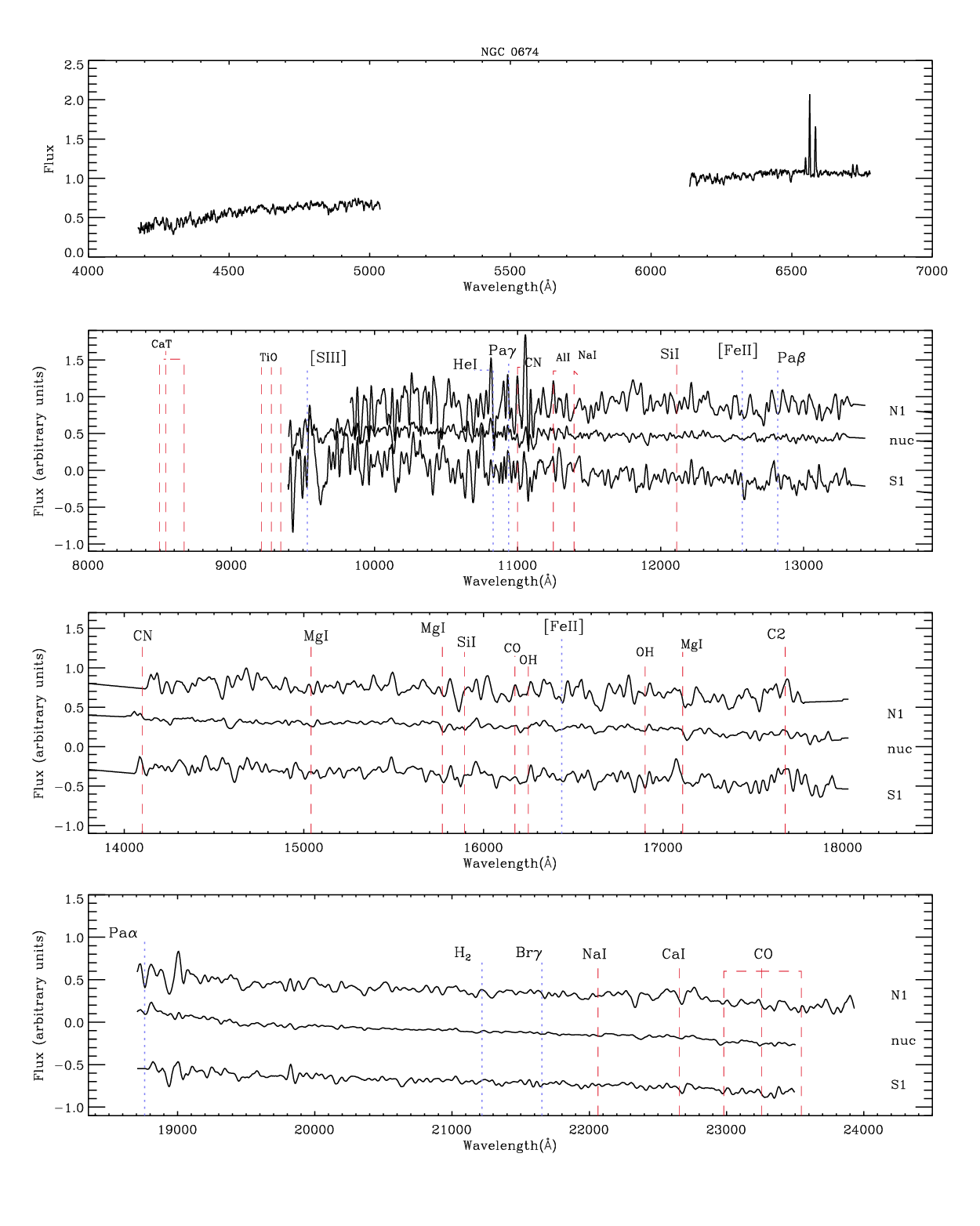}
\caption{Comparison between the optical spectrum of NGC 674 from Ho et al. (1997)
and our near-infrared spectra.  }
\end{figure*}

\begin{figure*} 
\includegraphics [width=160mm]{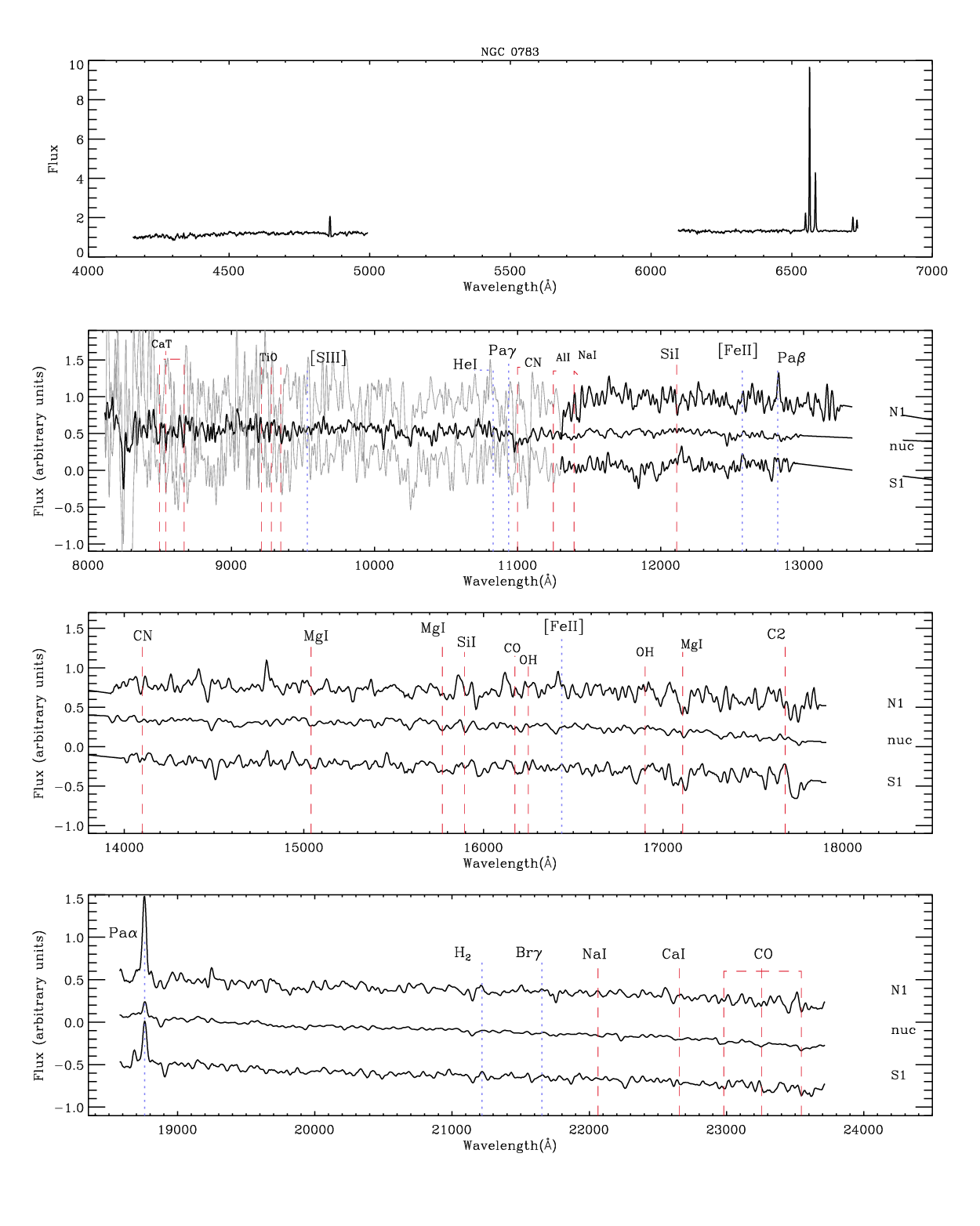}
\caption{Comparison between the optical spectrum of NGC 783 from Ho et al. (1997)
and our near-infrared spectra.  }
\end{figure*}

\begin{figure*} 
\includegraphics [width=160mm]{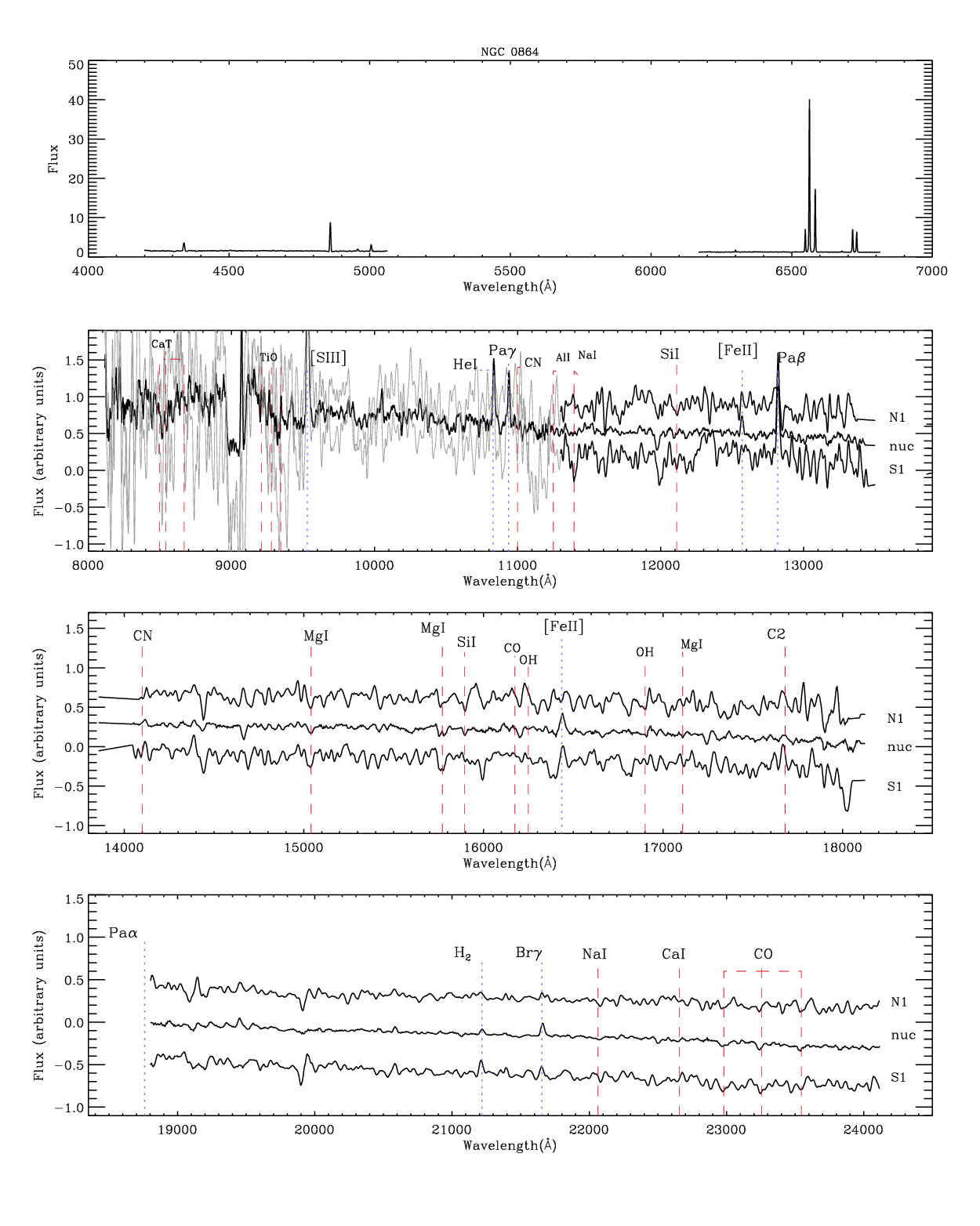}
\caption{Comparison between the optical spectrum of NGC 864 from Ho et al. (1997)
and our near-infrared spectra.  }
\end{figure*}

\begin{figure*} 
\includegraphics [width=160mm]{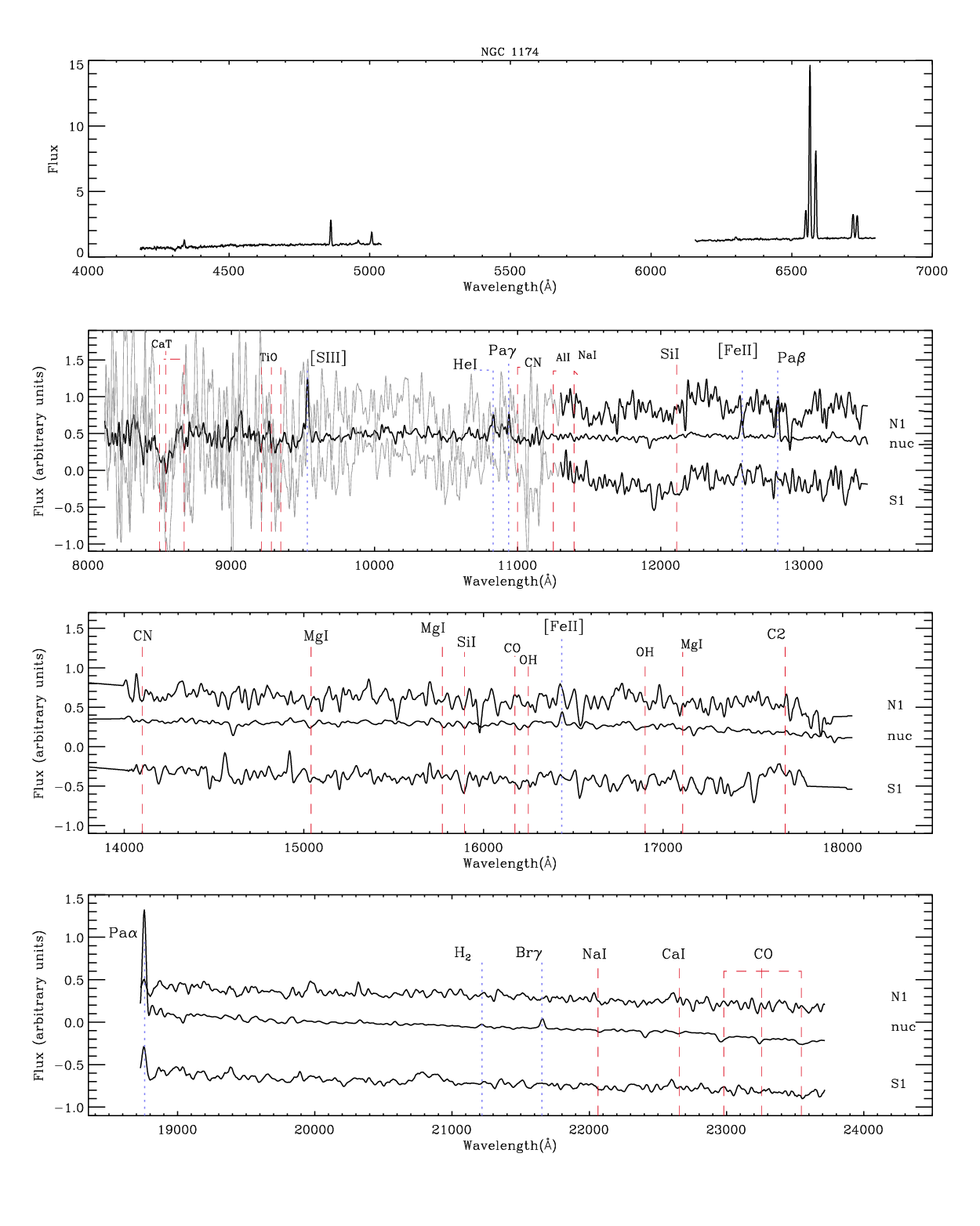}
\caption{Comparison between the optical spectrum of NGC~1174 from Ho et al. (1997)
and our near-infrared spectra.  }
\end{figure*}

\begin{figure*} 
\includegraphics [width=160mm]{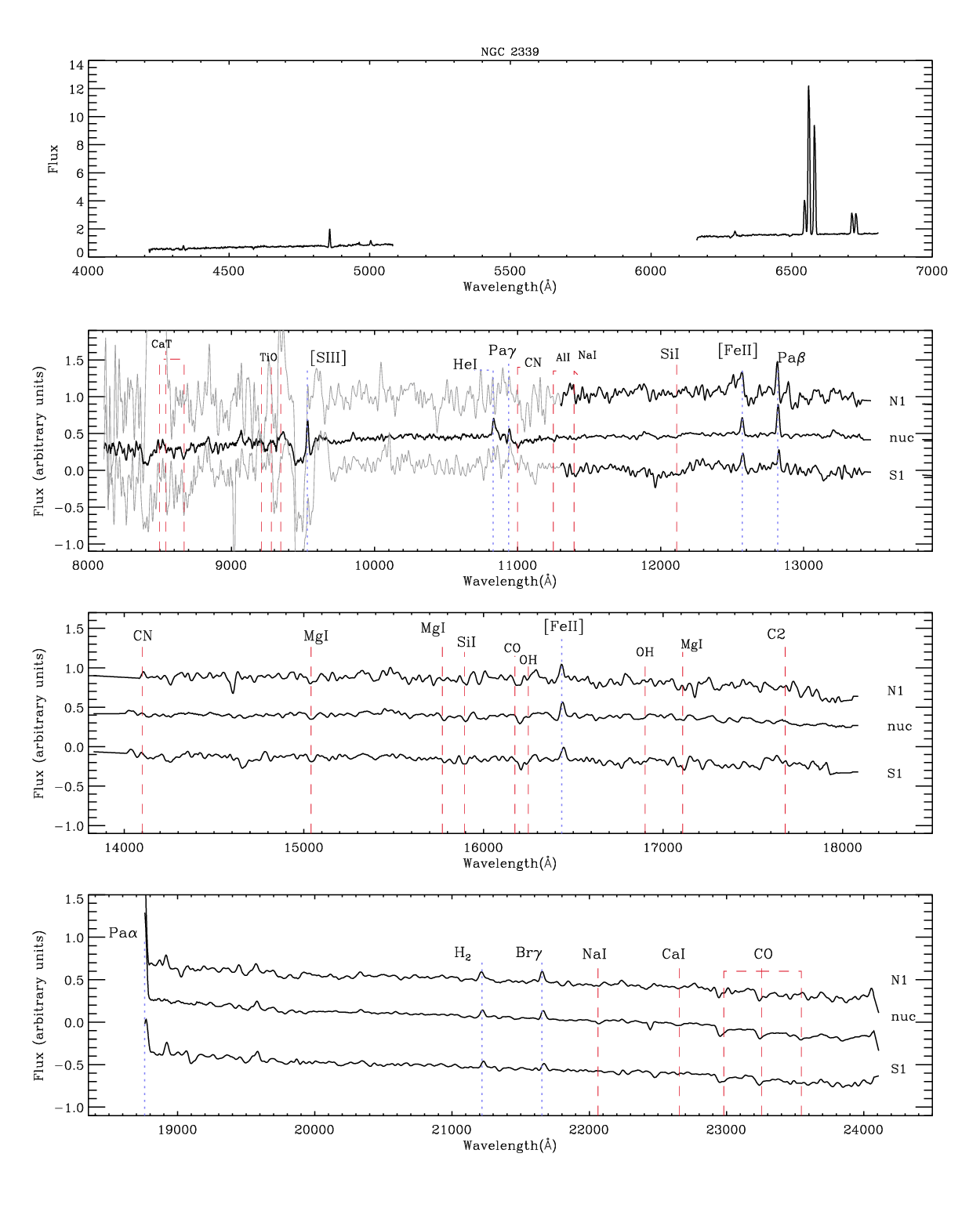}
\caption{Comparison between the optical spectrum of NGC~2339 from Ho et al. (1997)
and our near-infrared spectra.  }
\end{figure*}

\begin{figure*} 
\includegraphics [width=160mm]{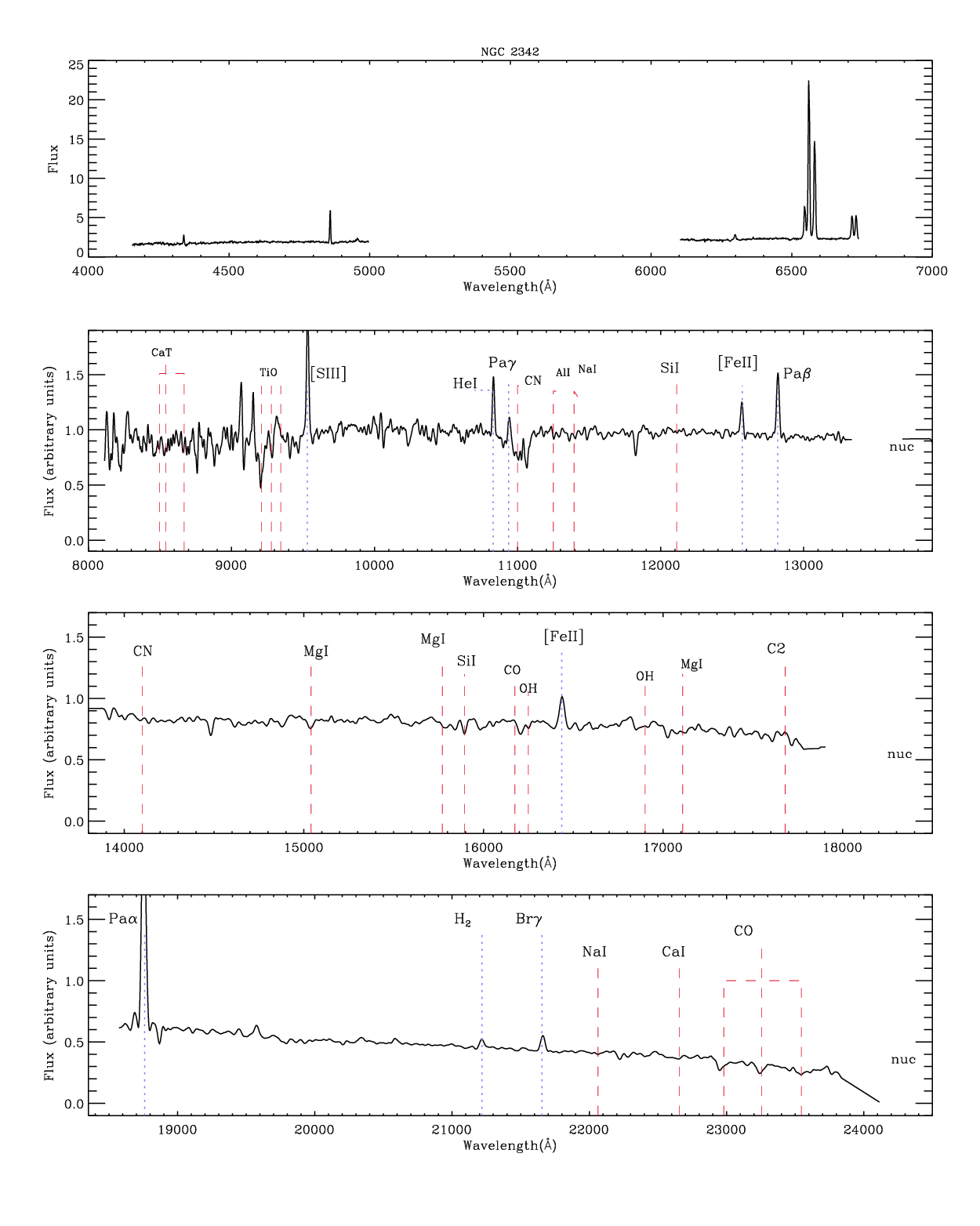}
\caption{Comparison between the optical spectrum of NGC~2342 from Ho et al. (1997)
and our near-infrared spectra.  }
\end{figure*}

\begin{figure*} 
\includegraphics [width=160mm]{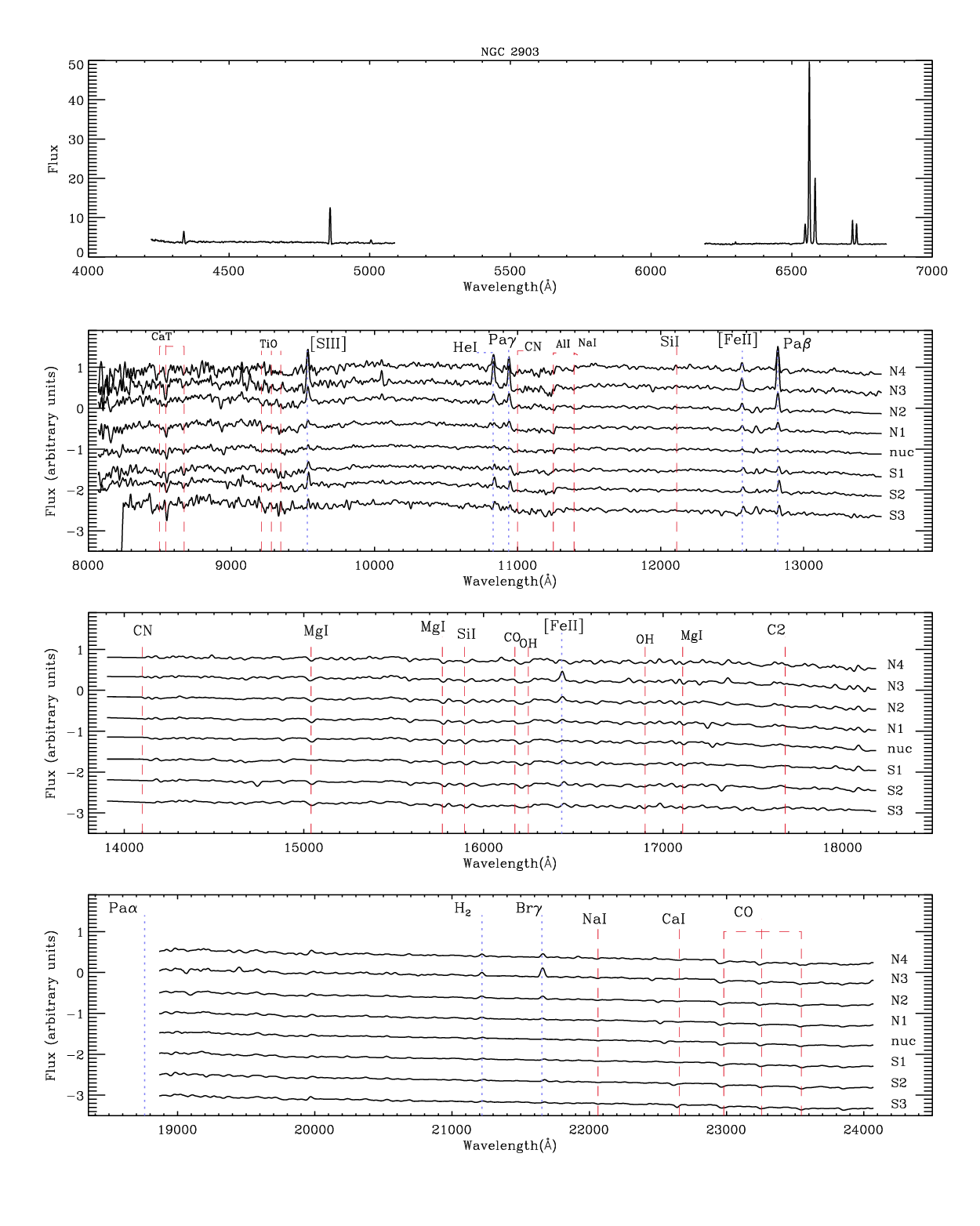}
\caption{ Comparison between the optical spectrum of NGC~2903 from Ho et al. (1997)
and our near-infrared spectra. }
\end{figure*}

\begin{figure*} 
\includegraphics [width=160mm]{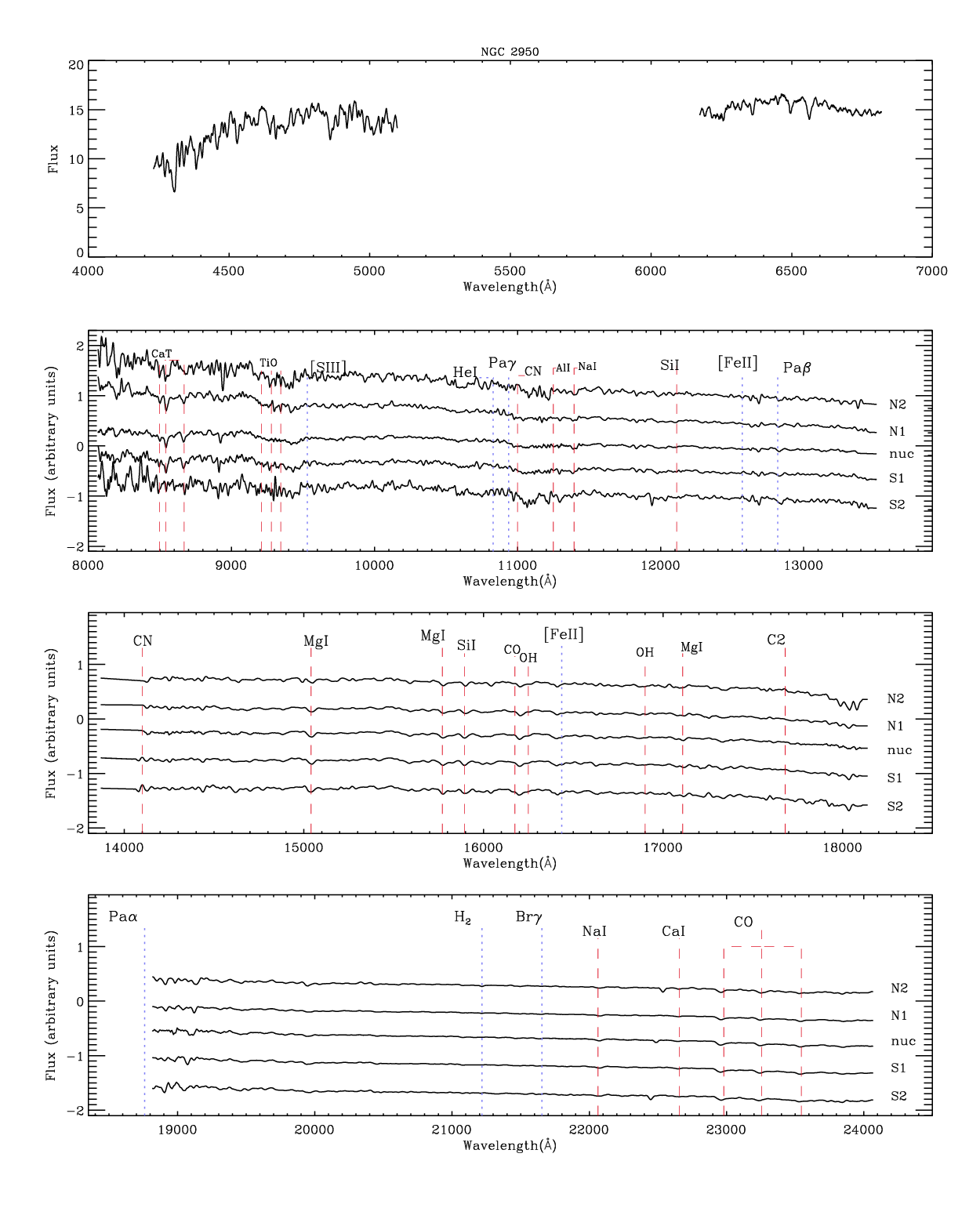}
\caption{Comparison between the optical spectrum of NGC~2950 from Ho et al. (1997)
and our near-infrared spectra.  }
\end{figure*}

\begin{figure*} 
\includegraphics [width=160mm]{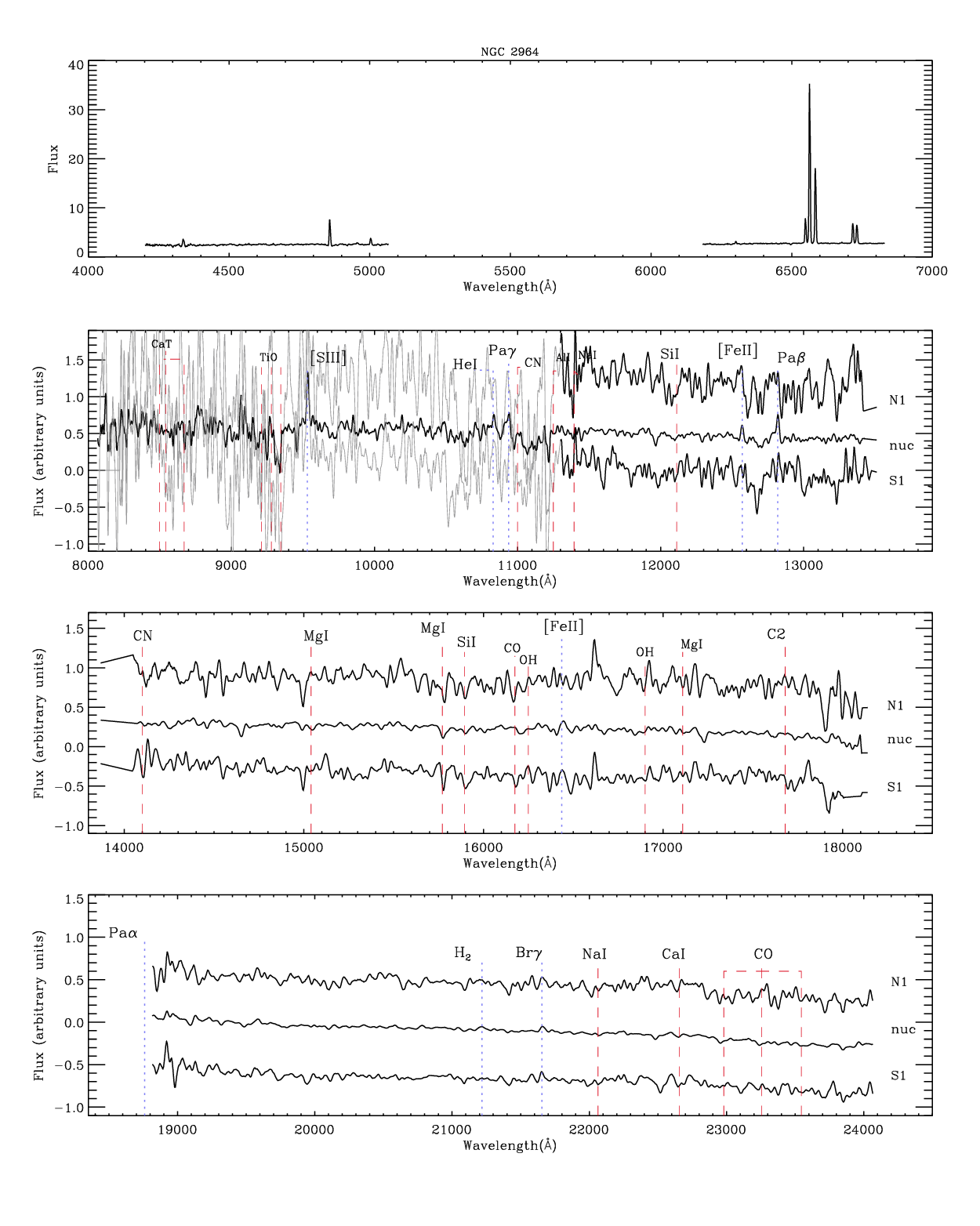}
\caption{Comparison between the optical spectrum of NGC~2964 from Ho et al. (1997)
and our near-infrared spectra.  }
\end{figure*}

\begin{figure*} 
\includegraphics [width=160mm]{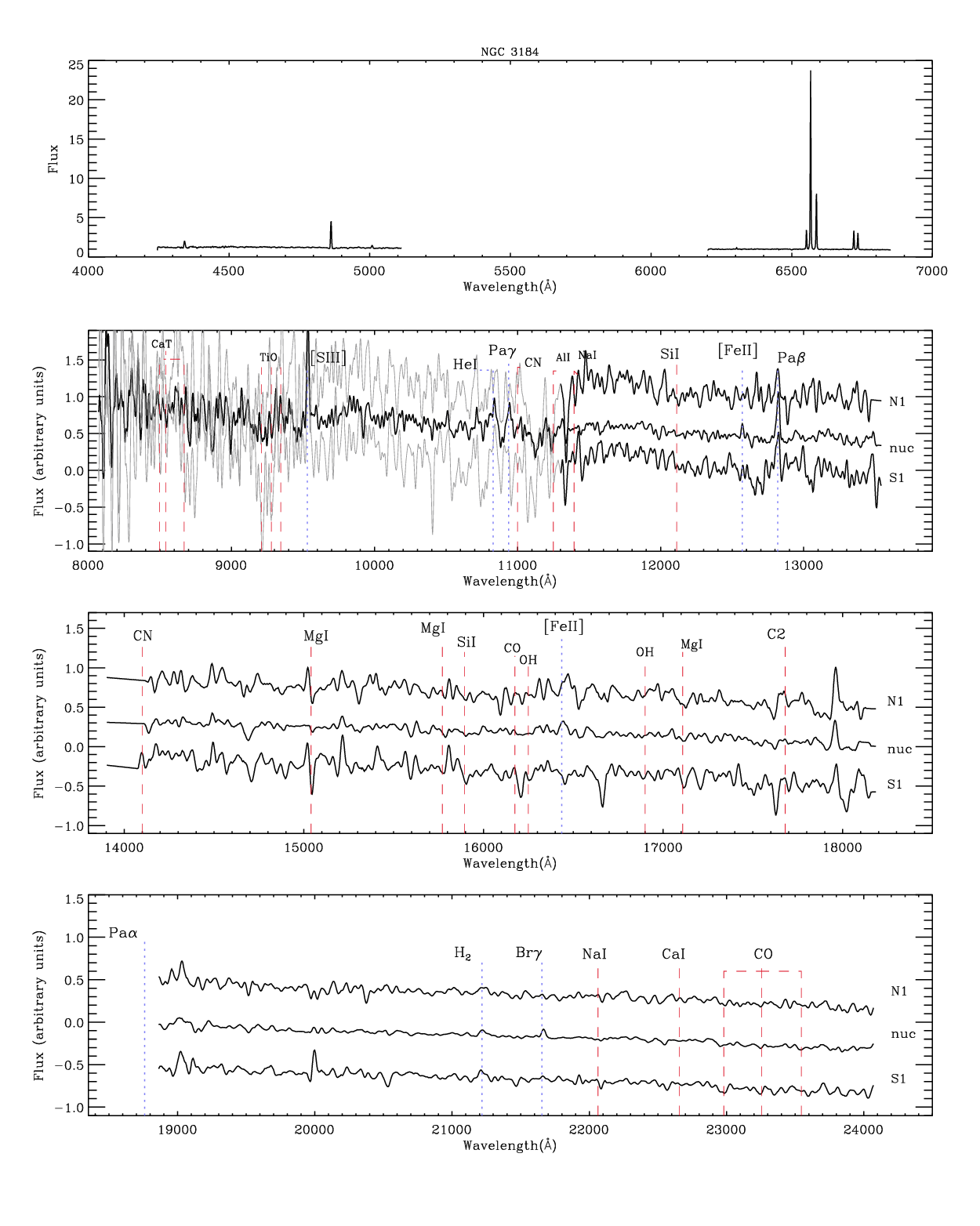}
\caption{ Comparison between the optical spectrum of NGC~3184 from Ho et al. (1997)
and our near-infrared spectra. }
\end{figure*}

\begin{figure*} 
\includegraphics [width=160mm]{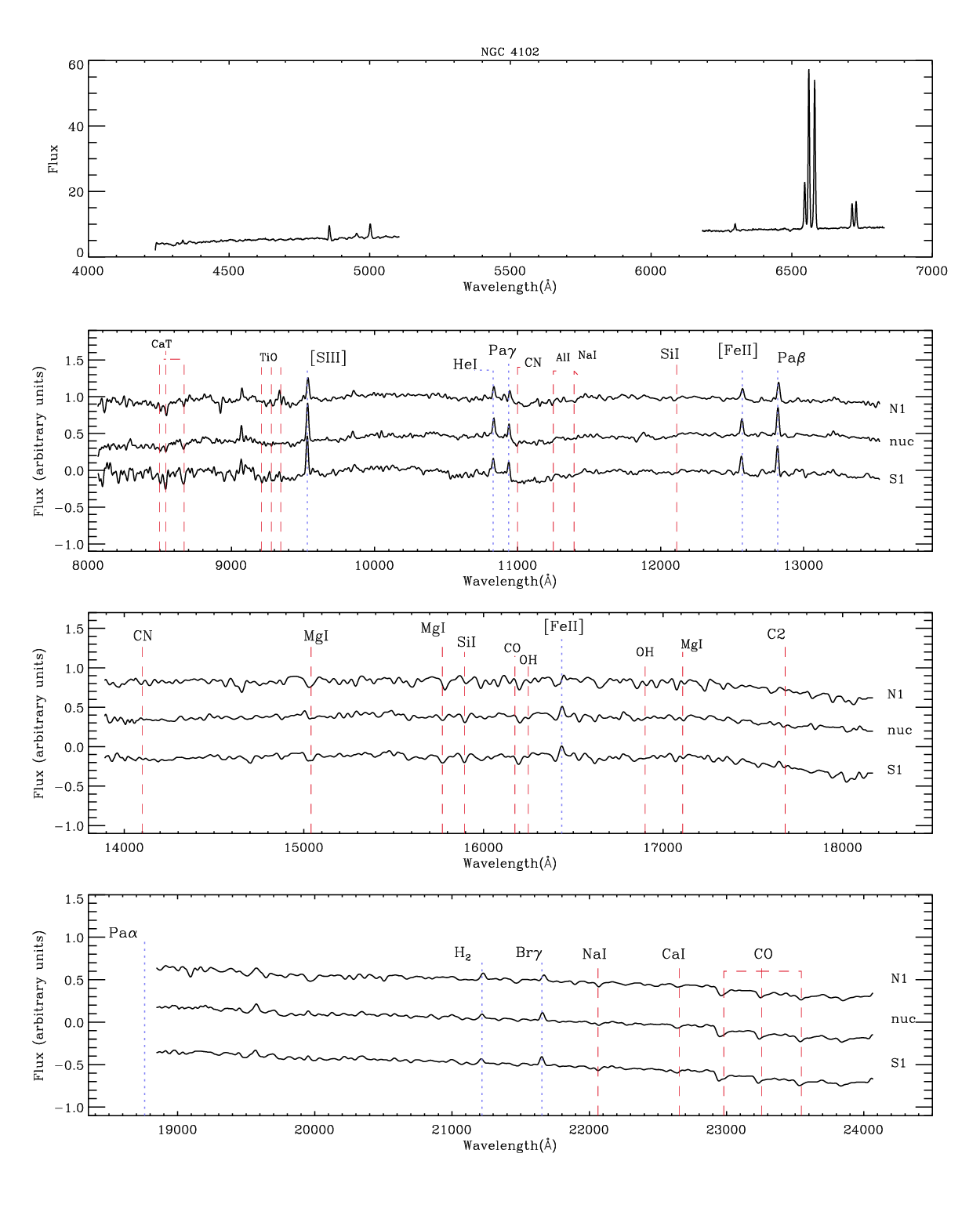}
\caption{Comparison between the optical spectrum of NGC~4102 from Ho et al. (1997)
and our near-infrared spectra. }
\label{cn_proof}
\end{figure*}

\begin{figure*} 
\includegraphics [width=160mm]{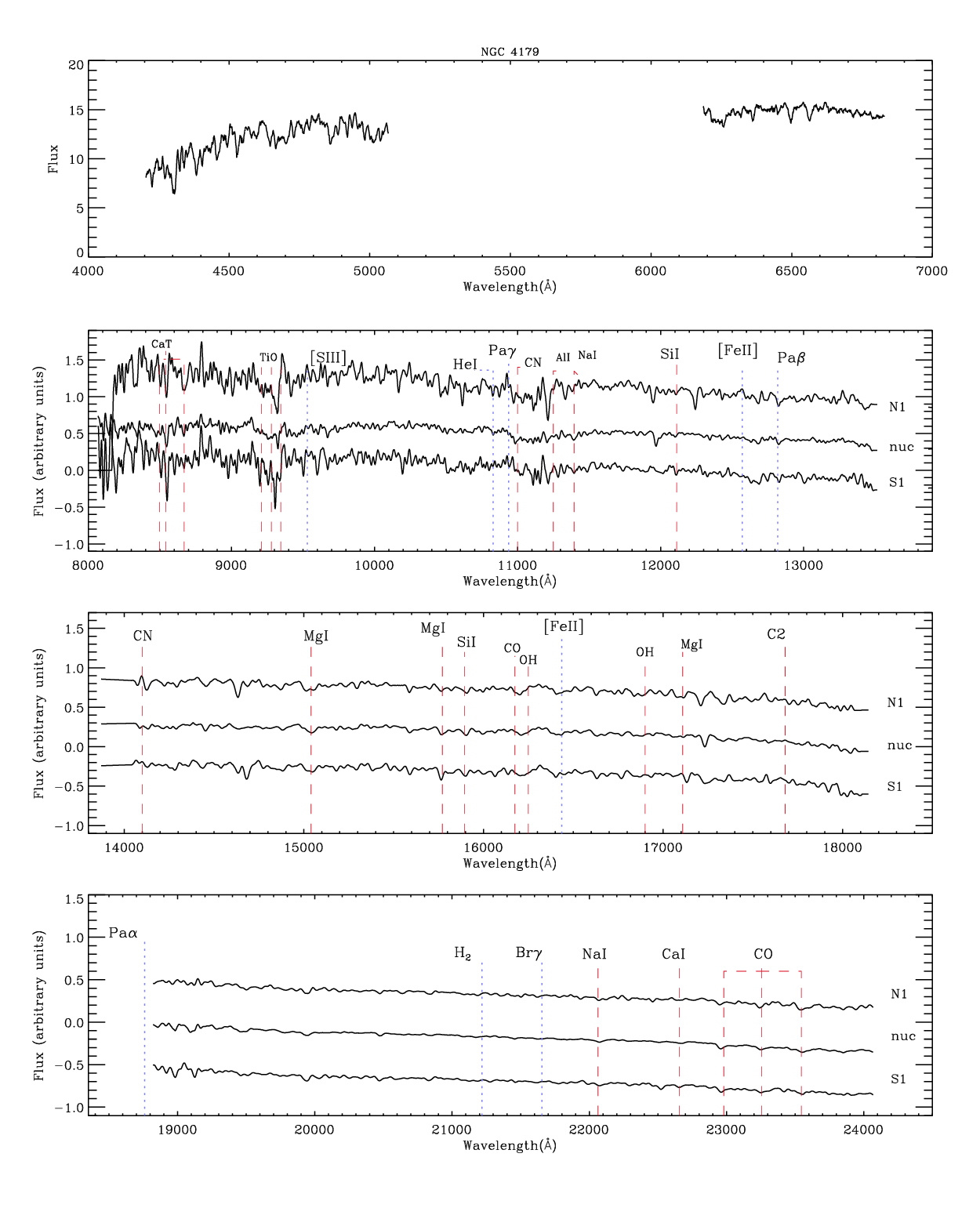}
\caption{ Comparison between the optical spectrum of NGC~4179 from Ho et al. (1997)
and our near-infrared spectra. }
\end{figure*}

\begin{figure*} 
\includegraphics [width=160mm]{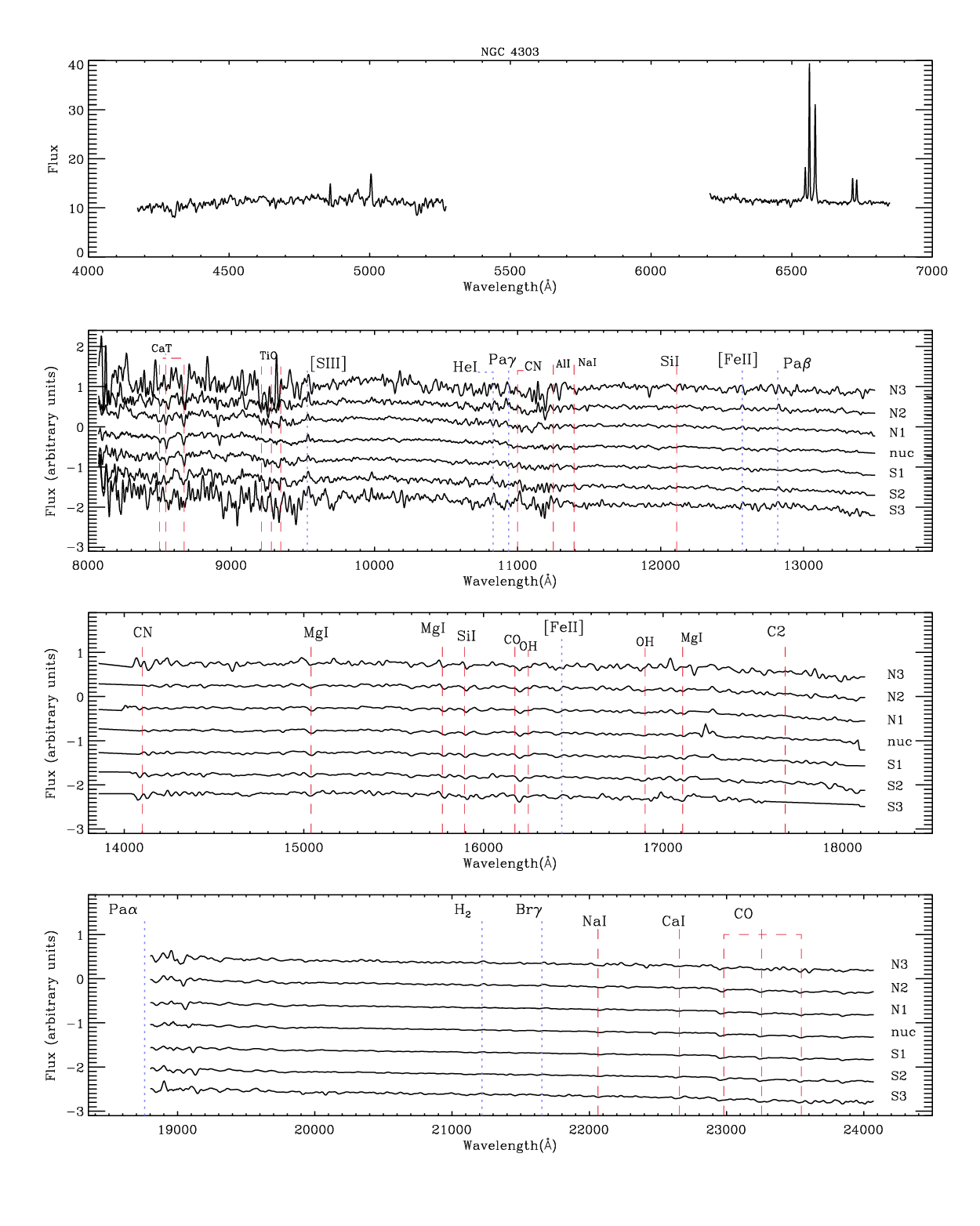}
\caption{ Comparison between the optical spectrum of NGC~4303 from Ho et al. (1997)
and our near-infrared spectra. }
\end{figure*}

\clearpage

\begin{figure*} 
\includegraphics [width=160mm]{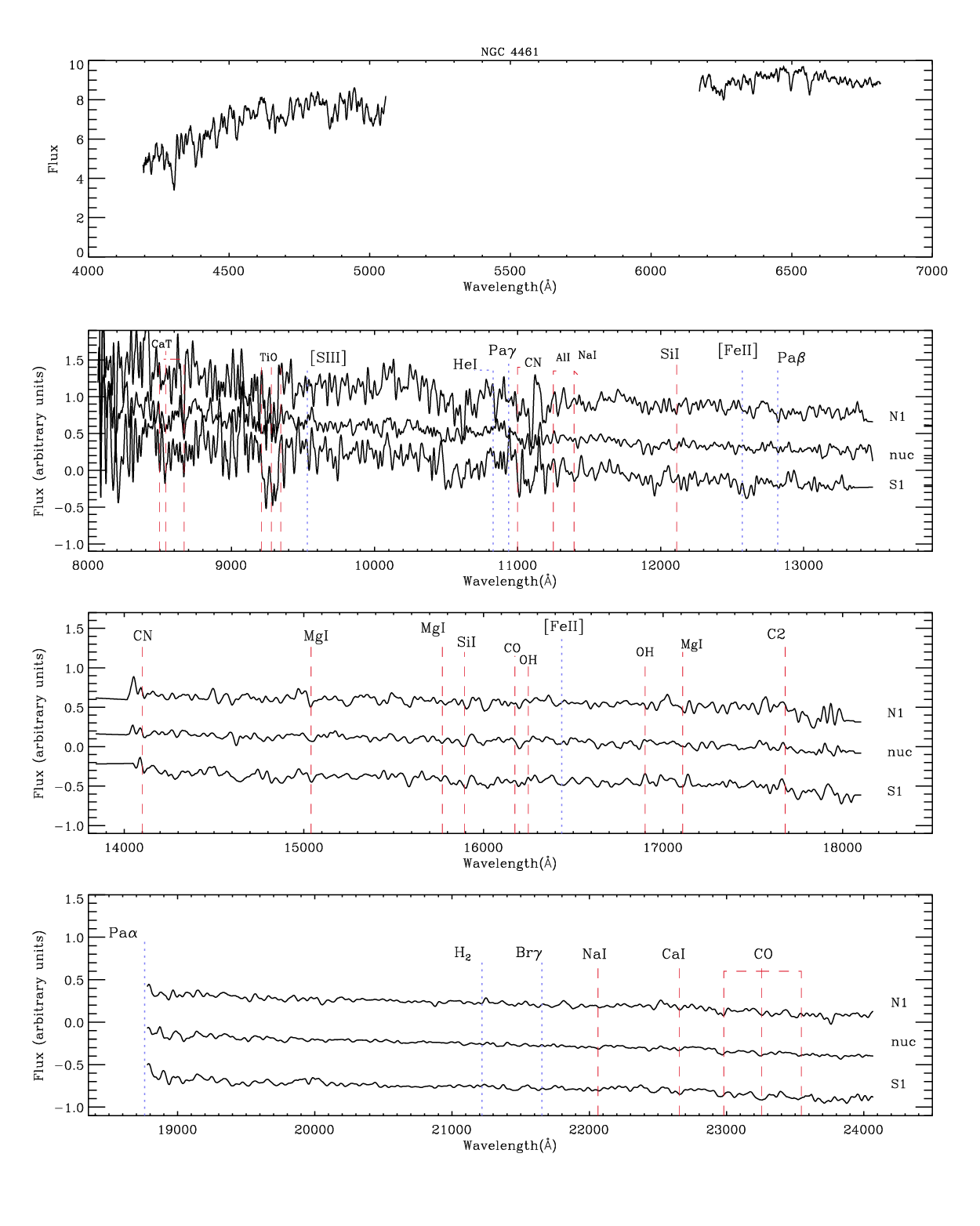}
\caption{ Comparison between the optical spectrum of NGC~4461 from Ho et al. (1997)
and our near-infrared spectra. }
\end{figure*}

\begin{figure*} 
\includegraphics [width=160mm]{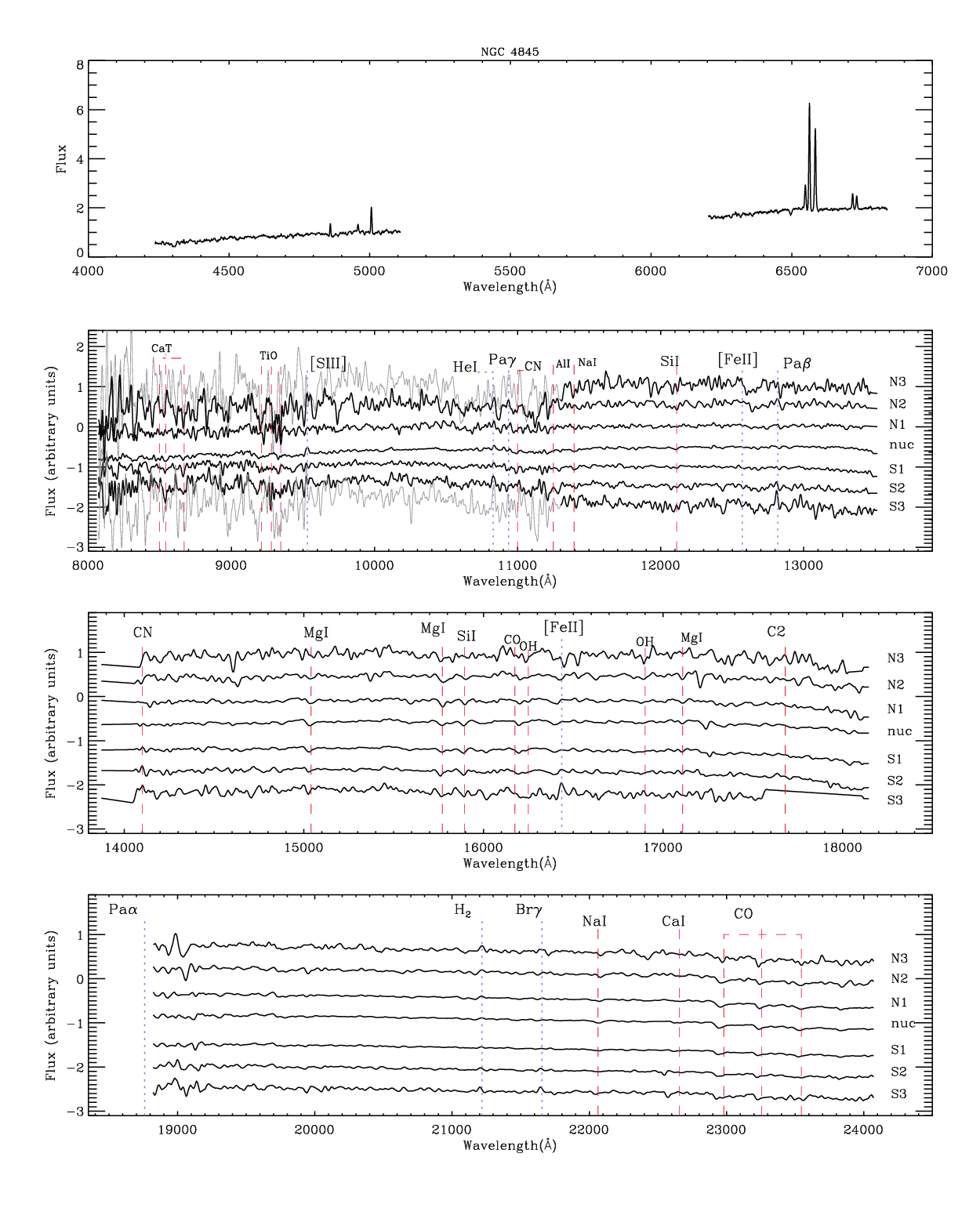}
\caption{ Comparison between the optical spectrum of NGC~4845 from Ho et al. (1997)
and our near-infrared spectra. }
\end{figure*}

\begin{figure*} 
\includegraphics [width=160mm]{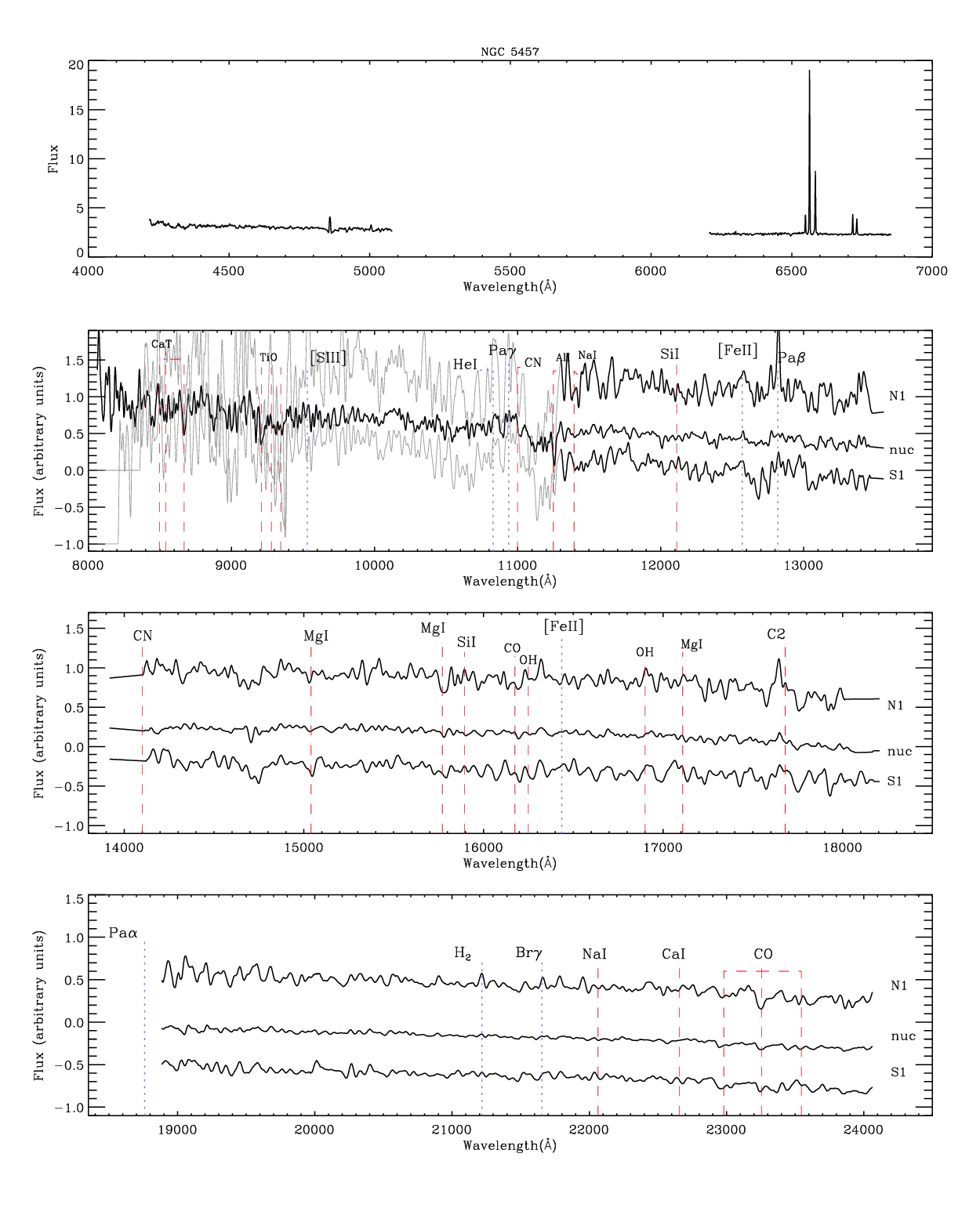}
\caption{Comparison between the optical spectrum of NGC~5457 from Ho et al. (1997)
and our near-infrared spectra.  }
\end{figure*}

\begin{figure*} 
\includegraphics [width=160mm]{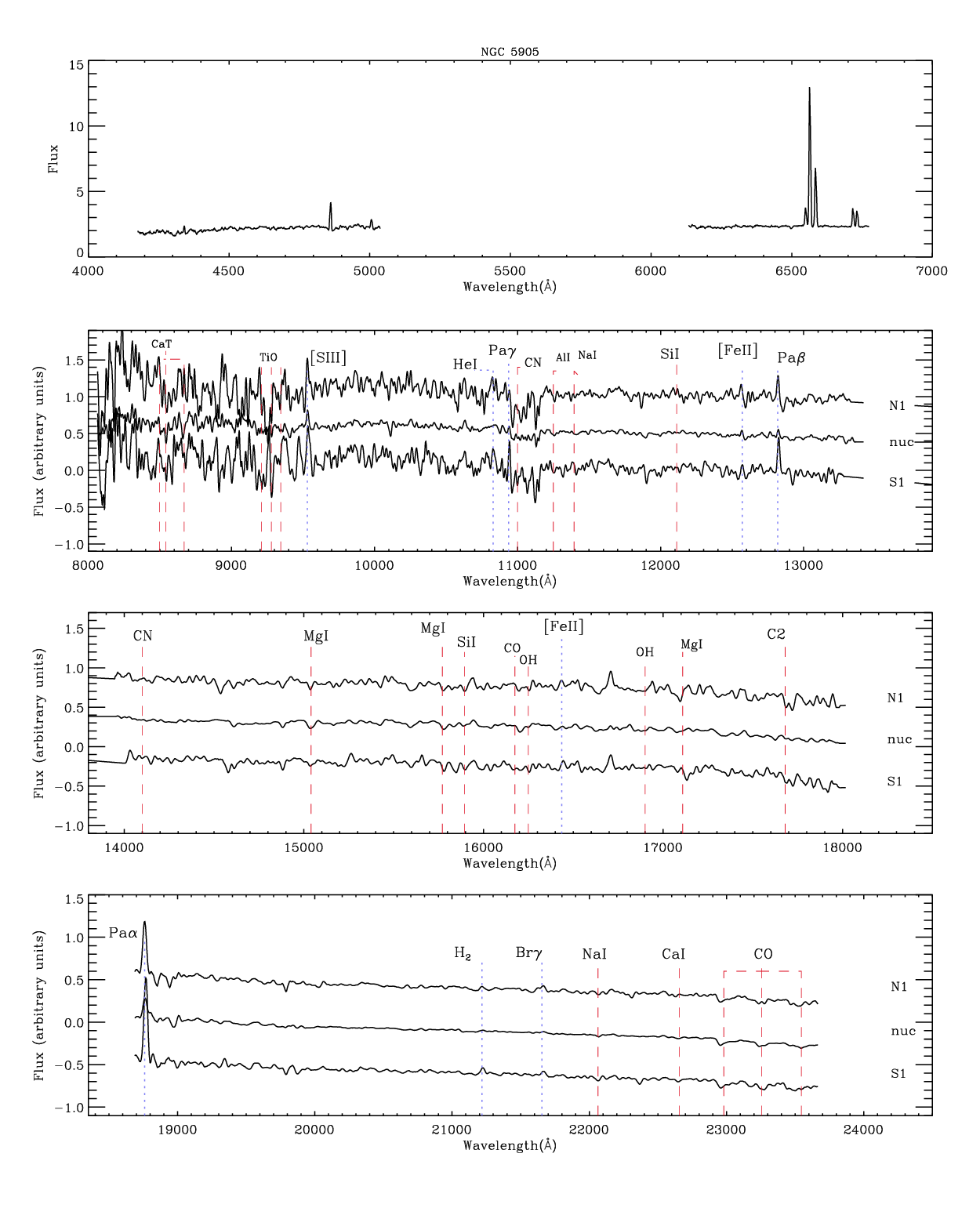}
\caption{ Comparison between the optical spectrum of NGC~5905 from Ho et al. (1997)
and our near-infrared spectra. }
\end{figure*}

\begin{figure*} 
\includegraphics [width=160mm]{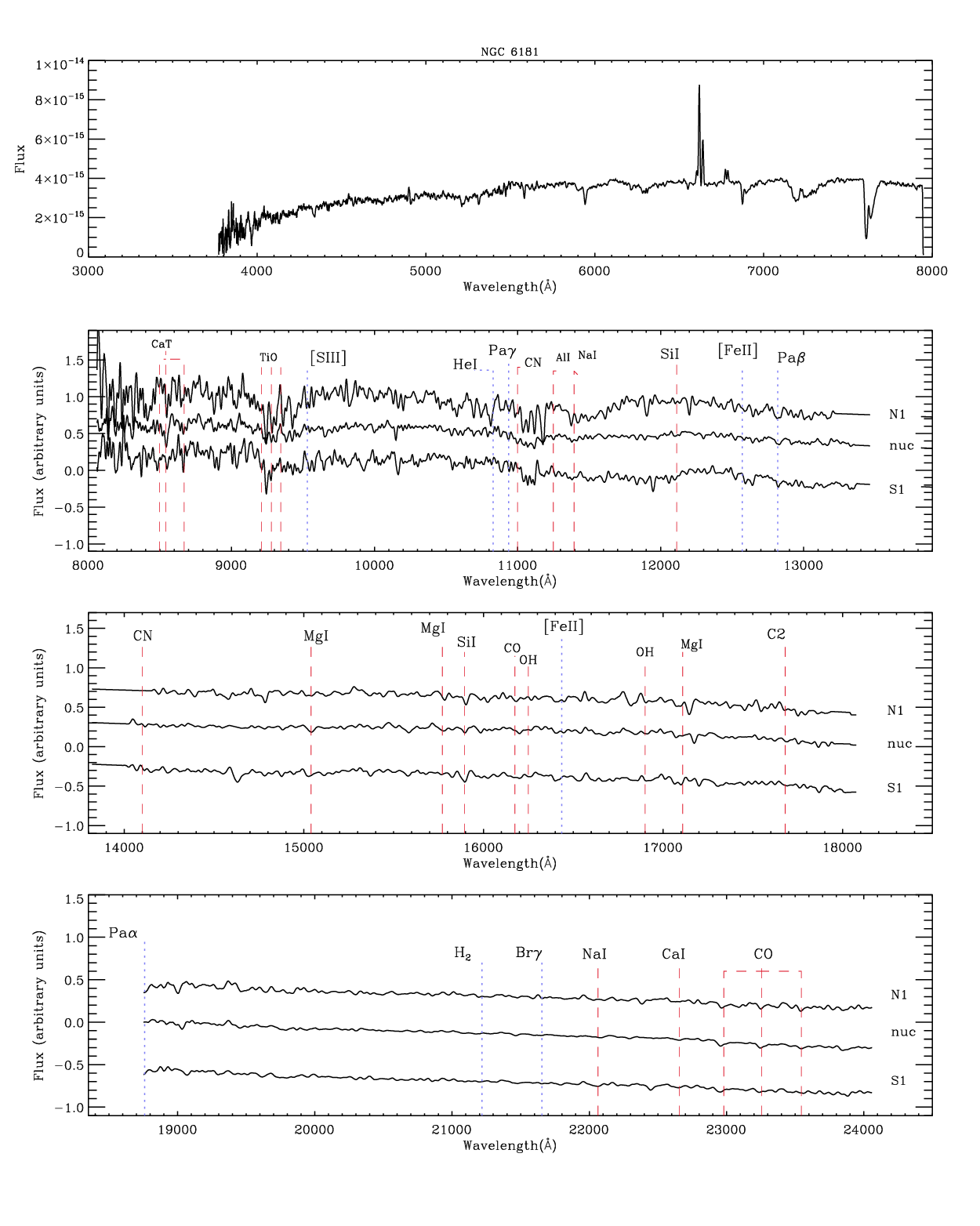}
\caption{ Comparison between the optical spectrum of NGC~6181 from Ho et al. (1997)
and our near-infrared spectra. }
\end{figure*}

\begin{figure*} 
\includegraphics [width=160mm]{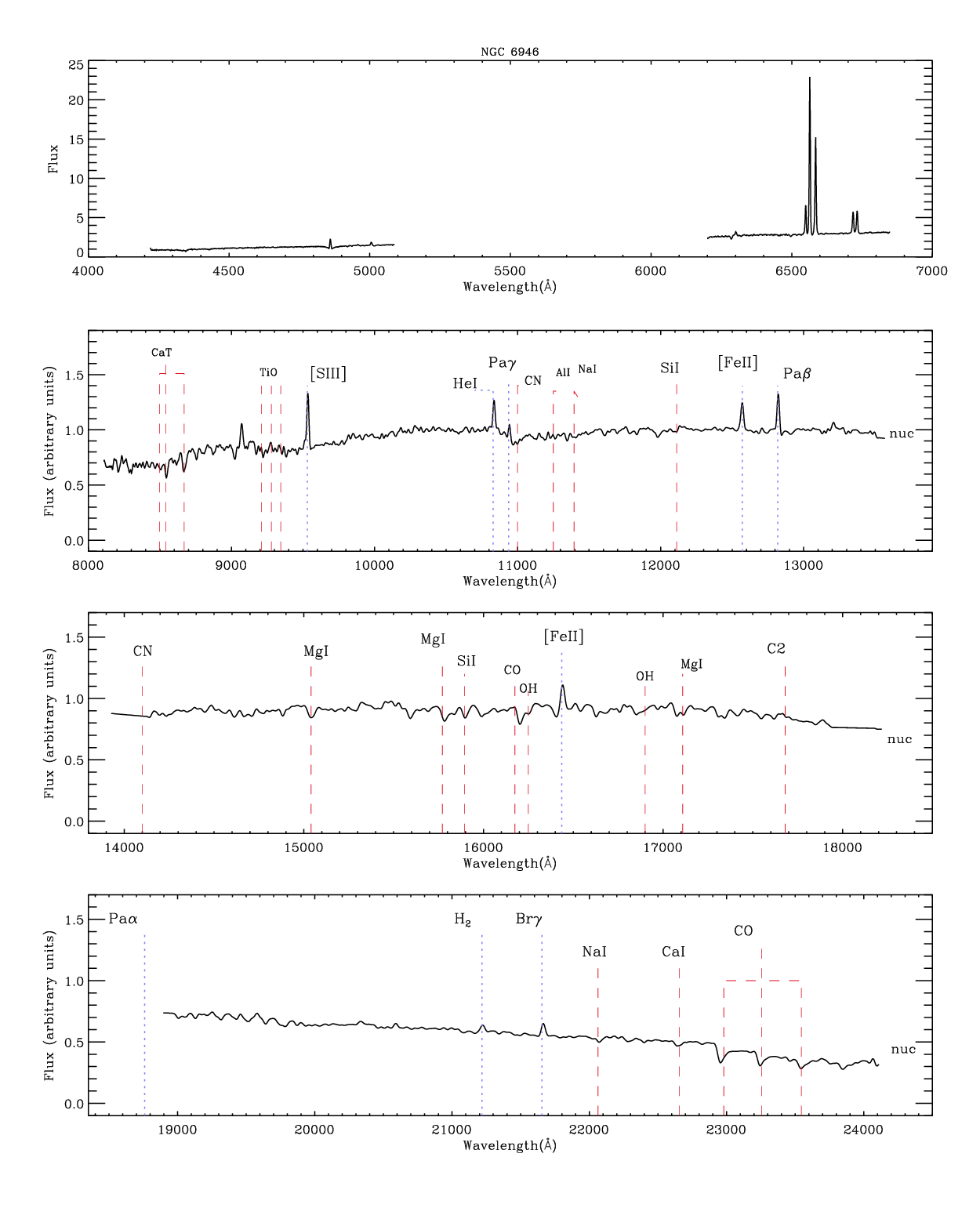}
\caption{Comparison between the optical spectrum of NGC~6946 from Ho et al. (1997)
and our near-infrared spectra. }
\end{figure*}

\begin{figure*} 
\includegraphics [width=160mm]{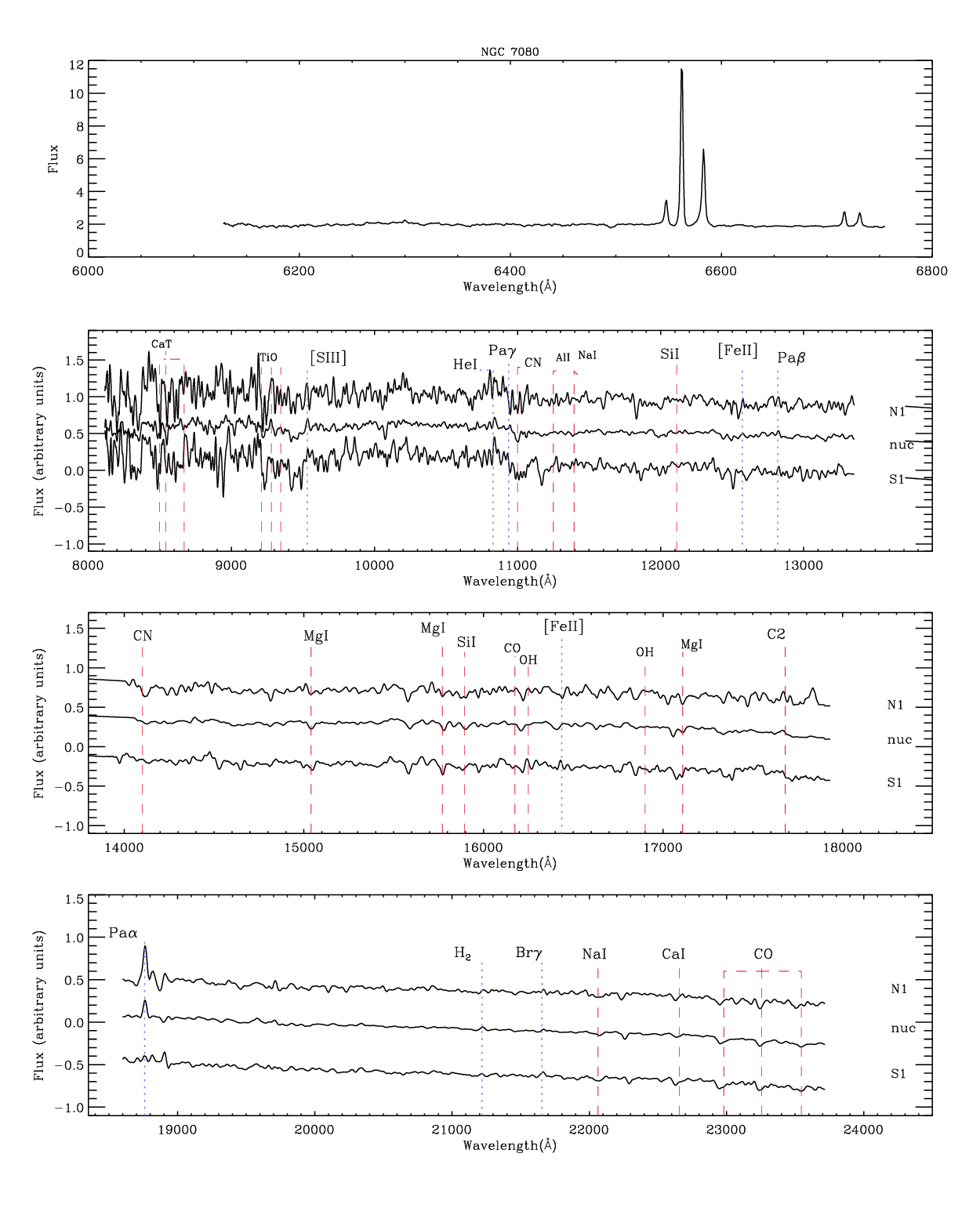}
\caption{ Comparison between the optical spectrum of NGC~7080 from Ho et al. (1997)
and our near-infrared spectra. }
\end{figure*}

\begin{figure*} 
\includegraphics [width=160mm]{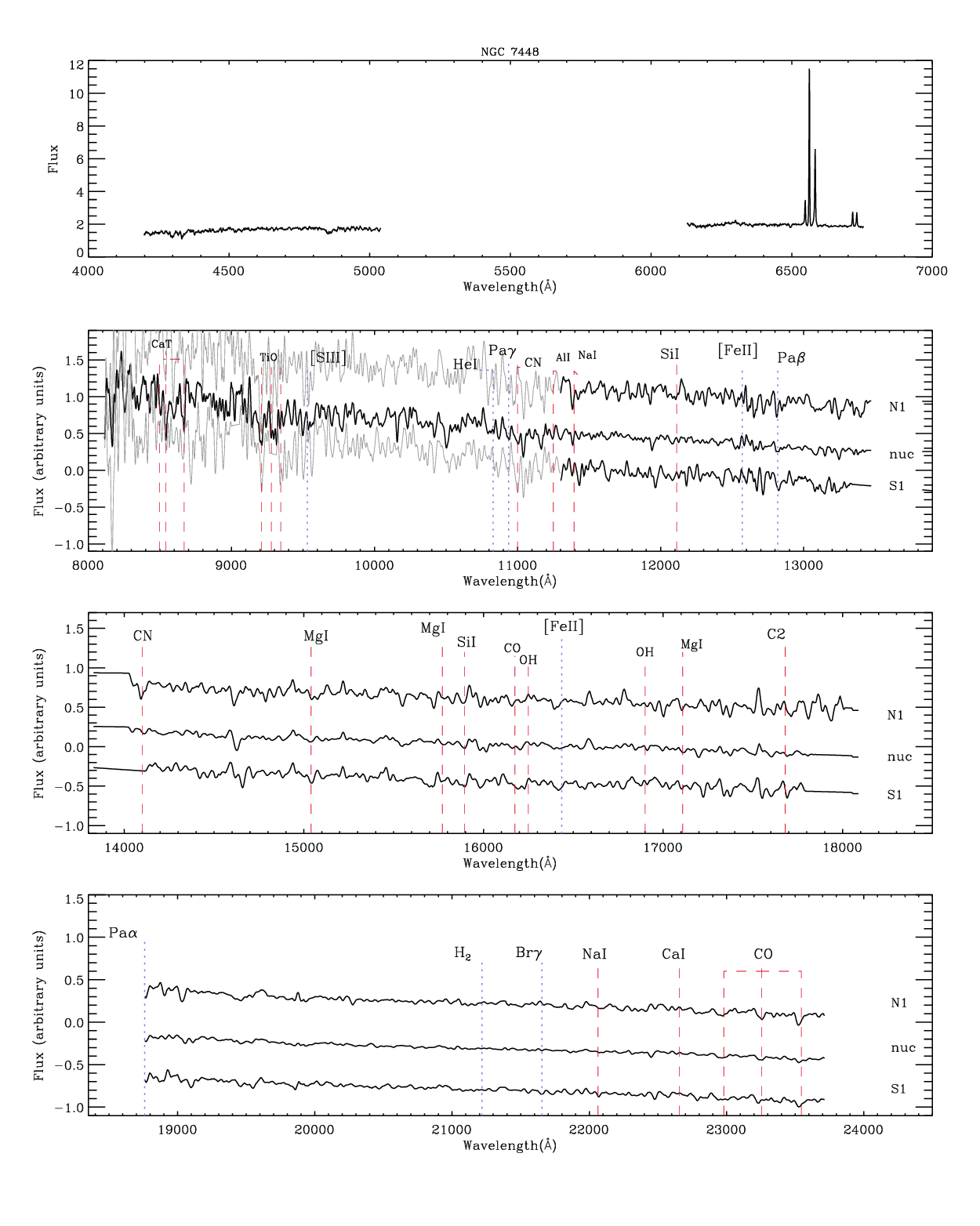}
\caption{Comparison between the optical spectrum of NGC~7448 from Ho et al. (1997)
and our near-infrared spectra.  }
\end{figure*}

\begin{figure*} 
\includegraphics [width=160mm]{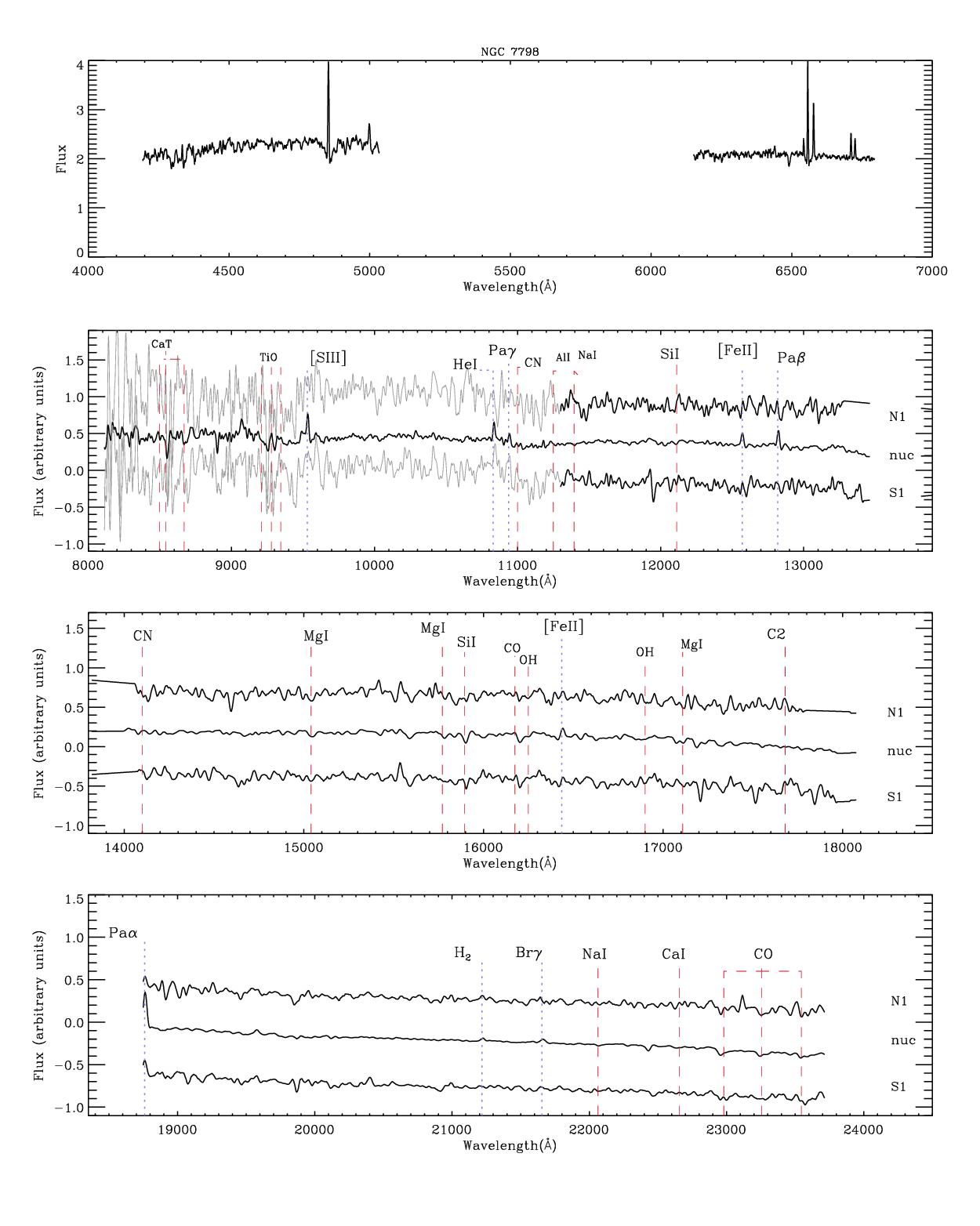}
\caption{ Comparison between the optical spectrum of NGC~7798 from Ho et al. (1997)
and our near-infrared spectra. }
\end{figure*}

\begin{figure*} 
\includegraphics [width=160mm]{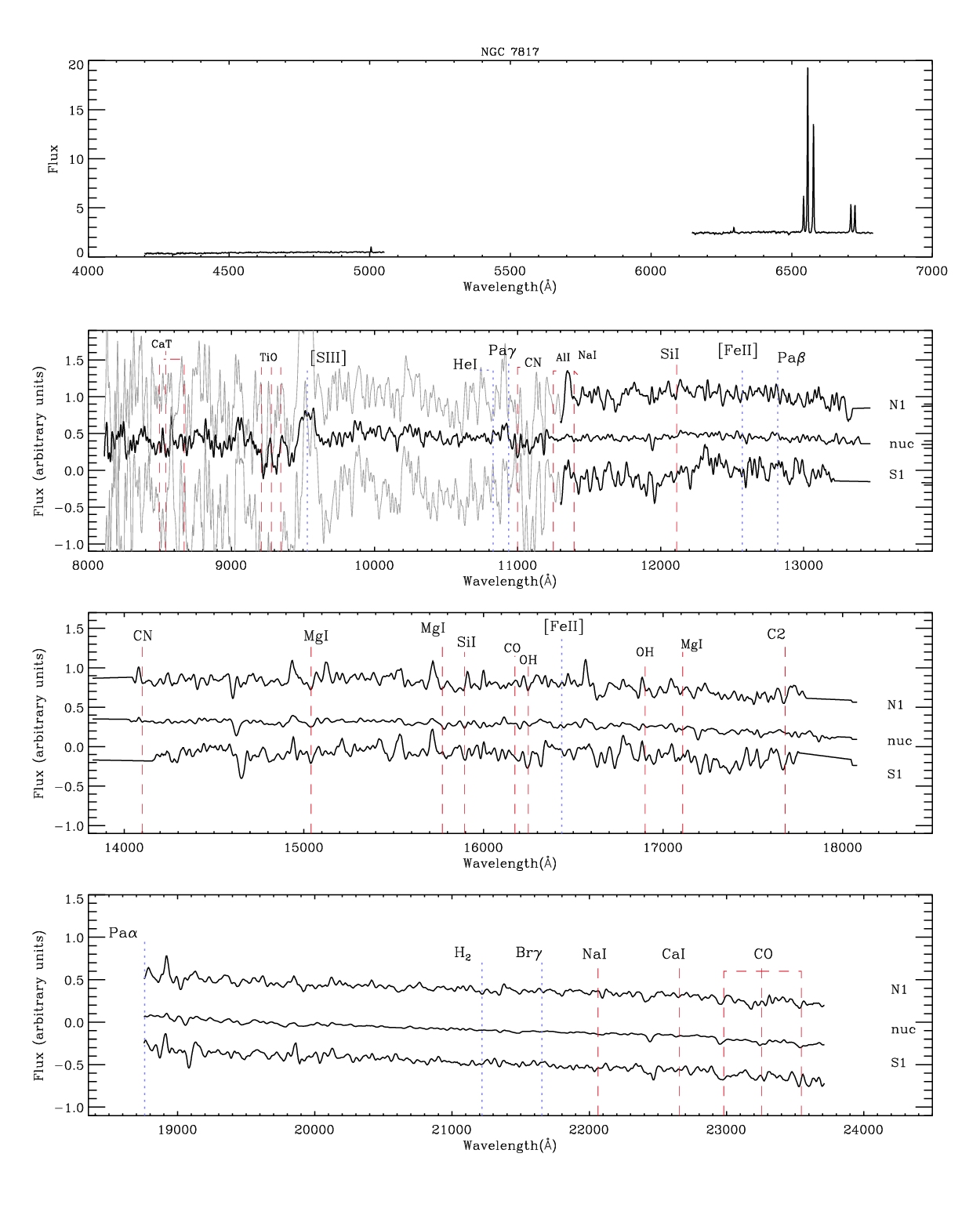}
\caption{ Comparison between the optical spectrum of NGC~7817 from Ho et al. (1997)
and our near-infrared spectra. }
\end{figure*}

\begin{figure*} 
\includegraphics [width=160mm]{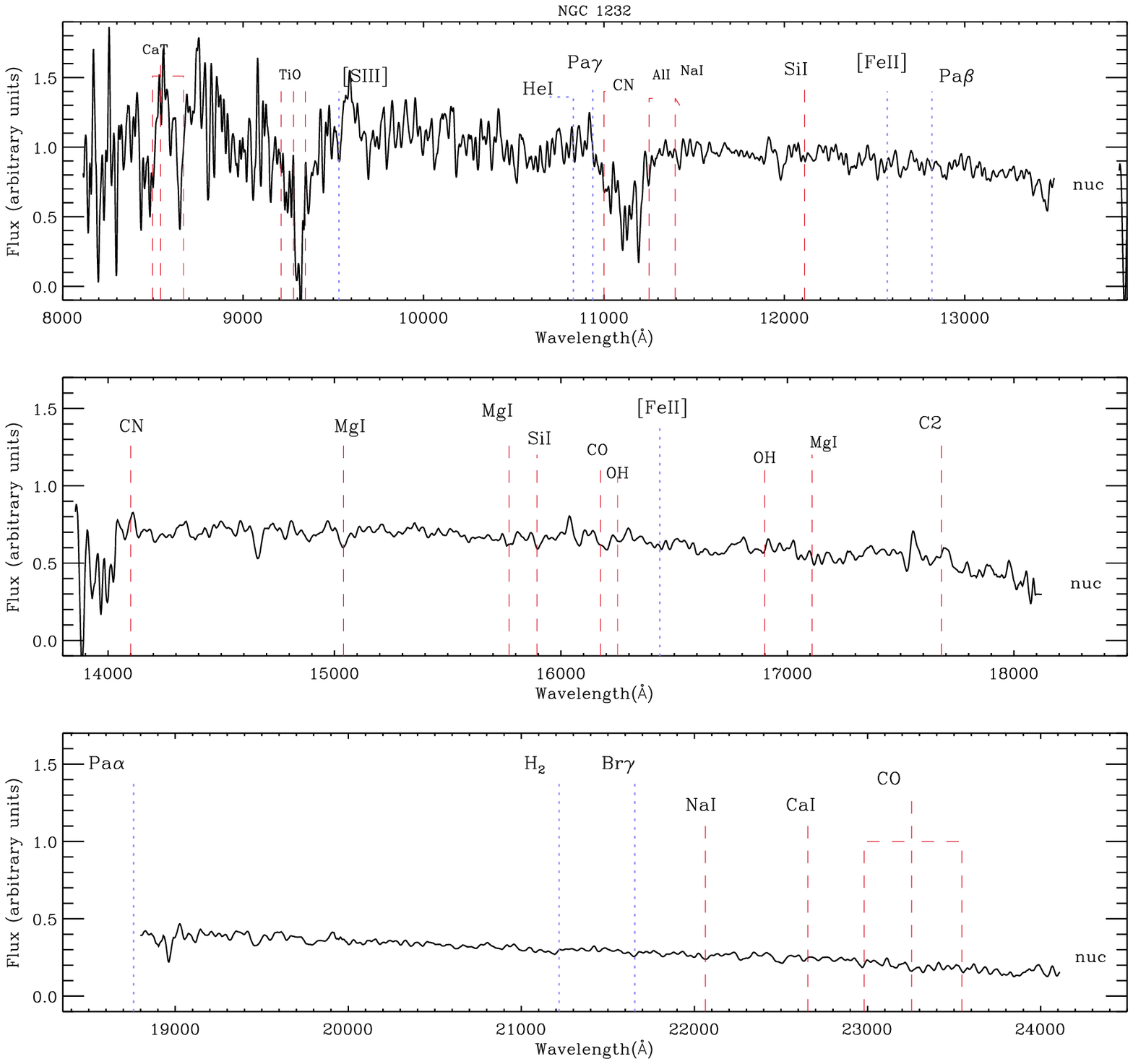}
\caption{The near-infrared spectra of NGC~1232. For this galaxy there is no optical spectrum available in the literature. }
\end{figure*}

\begin{figure*} 
\includegraphics [width=160mm]{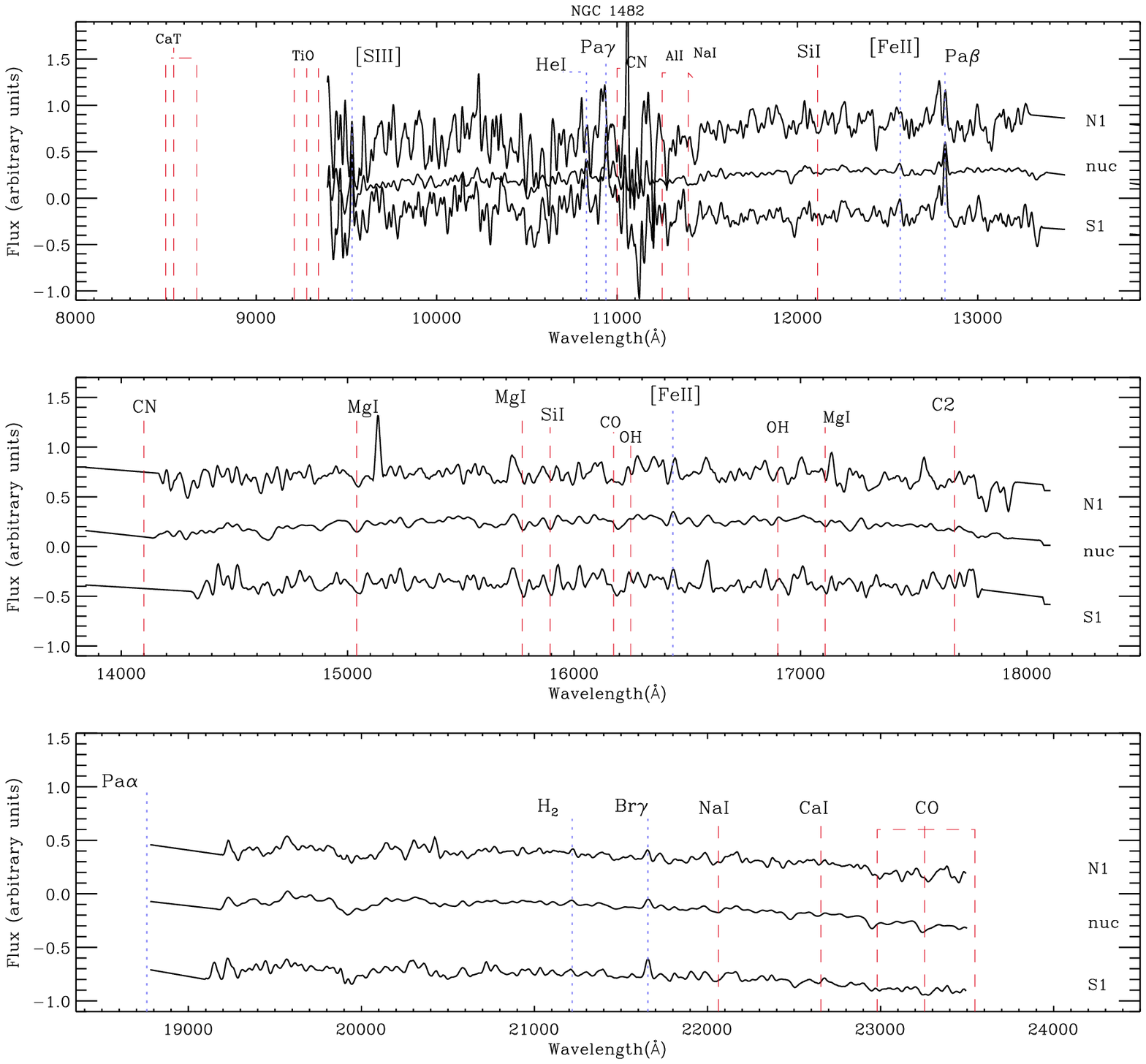}
\caption{The near-infrared spectra of NGC~1482. For this galaxy there is no optical spectrum available in the literature.}
\end{figure*}

\bsp

\label{lastpage}


\begin{thebibliography}{55}
\expandafter\ifx\csname natexlab\endcsname\relax\def\natexlab#1{#1}\fi

\expandafter\ifx\csname natexlab\endcsname\relax\def\natexlab#1{#1}\fi

\bibitem[{{Alonso-Herrero} {et~al}\mbox{.}(2000){Alonso-Herrero}, {Rieke},
  {Rieke}, \& {Shields}}]{alonso-herrero+00}
{Alonso-Herrero} A., {Rieke} M.~J., {Rieke} G.~H., {Shields} J.~C., 2000, \apj,
  530, 688

\bibitem[{{Balogh} {et~al}\mbox{.}(1997){Balogh}, {Morris}, {Yee}, {Carlberg},
  \& {Ellingson}}]{balogh+97}
{Balogh} M.~L., {Morris} S.~L., {Yee} H.~K.~C., {Carlberg} R.~G., {Ellingson}
  E., 1997, \apjl, 488, L75

\bibitem[{{Bendo} \& {Joseph}(2004)}]{bendo+04}
{Bendo} G.~J., {Joseph} R.~D., 2004, \aj, 127, 3338

\bibitem[{{Bernard-Salas} {et~al}\mbox{.}(2009){Bernard-Salas}, {Spoon},
  {Charmandaris}, {Lebouteiller}, {Farrah}, {Devost}, {Brandl}, {Wu}, {Armus},
  {Hao}, {Sloan}, {Weedman}, \& {Houck}}]{bernard-salas+09}
{Bernard-Salas} J. {et~al.}, 2009, \apjs, 184, 230

\bibitem[{{Blain} {et~al}\mbox{.}(2002){Blain}, {Smail}, {Ivison}, {Kneib}, \&
  {Frayer}}]{blain+02}
{Blain} A.~W., {Smail} I., {Ivison} R.~J., {Kneib} J.-P., {Frayer} D.~T., 2002,
  \physrep, 369, 111

\bibitem[{{Calzetti} {et~al}\mbox{.}(2000){Calzetti}, {Armus}, {Bohlin},
  {Kinney}, {Koornneef}, \& {Storchi-Bergmann}}]{calzetti+00}
{Calzetti} D., {Armus} L., {Bohlin} R.~C., {Kinney} A.~L., {Koornneef} J.,
  {Storchi-Bergmann} T., 2000, \apj, 533, 682

\bibitem[{{Cameron} {et~al}\mbox{.}(2010){Cameron}, {Carollo}, {Oesch},
  {Aller}, {Bschorr}, {Cerulo}, {Aussel}, {Capak}, {Le Floc'h}, {Ilbert},
  {Kneib}, {Koekemoer}, {Leauthaud}, {Lilly}, {Massey}, {McCracken}, {Rhodes},
  {Salvato}, {Sanders}, {Scoville}, {Sheth}, {Taniguchi}, \&
  {Thompson}}]{cameron+10}
{Cameron} E. {et~al.}, 2010, \mnras, 409, 346

\bibitem[{{Cardelli} {et~al}\mbox{.}(1989){Cardelli}, {Clayton}, \&
  {Mathis}}]{cardelli+89}
{Cardelli} J.~A., {Clayton} G.~C., {Mathis} J.~S., 1989, \apj, 345, 245

\bibitem[{{Cesetti} {et~al}\mbox{.}(2009){Cesetti}, {Ivanov}, {Morelli},
  {Pizzella}, {Buson}, {Corsini}, {Dalla Bont{\`a}}, {Stiavelli}, \&
  {Bertola}}]{cesetti+09}
{Cesetti} M. {et~al.}, 2009, \aap, 497, 41

\bibitem[{{Coziol} {et~al}\mbox{.}(2001){Coziol}, {Doyon}, \&
  {Demers}}]{coziol+01}
{Coziol} R., {Doyon} R., {Demers} S., 2001, \mnras, 325, 1081

\bibitem[{{Coziol} {et~al}\mbox{.}(1998){Coziol}, {Torres}, {Quast}, {Contini},
  \& {Davoust}}]{coziol+98}
{Coziol} R., {Torres} C.~A.~O., {Quast} G.~R., {Contini} T., {Davoust} E.,
  1998, \apjs, 119, 239

\bibitem[{{Cushing} {et~al}\mbox{.}(2004){Cushing}, {Vacca}, \&
  {Rayner}}]{cushing+04}
{Cushing} M.~C., {Vacca} W.~D., {Rayner} J.~T., 2004, \pasp, 116, 362

\bibitem[{{Elbaz} \& {Cesarsky}(2003)}]{elbaz+03}
{Elbaz} D., {Cesarsky} C.~J., 2003, Science, 300, 270

\bibitem[{{Engelbracht} {et~al}\mbox{.}(1998){Engelbracht}, {Rieke}, {Rieke},
  {Kelly}, \& {Achtermann}}]{engelbracht+98}
{Engelbracht} C.~W., {Rieke} M.~J., {Rieke} G.~H., {Kelly} D.~M., {Achtermann}
  J.~M., 1998, \apj, 505, 639

\bibitem[{{Frogel} {et~al}\mbox{.}(1990){Frogel}, {Mould}, \&
  {Blanco}}]{frogel+90}
{Frogel} J.~A., {Mould} J., {Blanco} V.~M., 1990, \apj, 352, 96

\bibitem[{{Garcia-Rissmann} {et~al}\mbox{.}(2005){Garcia-Rissmann}, {Vega},
  {Asari}, {Cid Fernandes}, {Schmitt}, {Gonz{\'a}lez Delgado}, \&
  {Storchi-Bergmann}}]{garcia-rissman+05}
{Garcia-Rissmann} A., {Vega} L.~R., {Asari} N.~V., {Cid Fernandes} R.,
  {Schmitt} H., {Gonz{\'a}lez Delgado} R.~M., {Storchi-Bergmann} T., 2005,
  \mnras, 359, 765

\bibitem[{{Goldader} {et~al}\mbox{.}(1997){Goldader}, {Goldader}, {Joseph},
  {Doyon}, \& {Sanders}}]{goldader+97}
{Goldader} J.~D., {Goldader} D.~L., {Joseph} R.~D., {Doyon} R., {Sanders}
  D.~B., 1997, \aj, 113, 1569

\bibitem[{{Gu} {et~al}\mbox{.}(2006){Gu}, {Melnick}, {Cid Fernandes}, {Kunth},
  {Terlevich}, \& {Terlevich}}]{gu+06}
{Gu} Q., {Melnick} J., {Cid Fernandes} R., {Kunth} D., {Terlevich} E.,
  {Terlevich} R., 2006, \mnras, 366, 480

\bibitem[{{Heisler} \& {De Robertis}(1999)}]{heisler+99}
{Heisler} C.~A., {De Robertis} M.~M., 1999, \aj, 118, 2038

\bibitem[{{Ho} {et~al}\mbox{.}(1995){Ho}, {Filippenko}, \& {Sargent}}]{ho+95}
{Ho} L.~C., {Filippenko} A.~V., {Sargent} W.~L., 1995, \apjs, 98, 477

\bibitem[{{Ho} {et~al}\mbox{.}(1997){Ho}, {Filippenko}, \& {Sargent}}]{ho+97}
{Ho} L.~C., {Filippenko} A.~V., {Sargent} W.~L.~W., 1997, \apjs, 112, 315

\bibitem[{{Hunt} {et~al}\mbox{.}(2003){Hunt}, {Thuan}, \& {Izotov}}]{hunt+03}
{Hunt} L.~K., {Thuan} T.~X., {Izotov} Y.~I., 2003, \apj, 588, 281

\bibitem[{{Kawara} {et~al}\mbox{.}(1989){Kawara}, {Nishida}, \&
  {Phillips}}]{kawara+89}
{Kawara} K., {Nishida} M., {Phillips} M.~M., 1989, \apj, 337, 230

\bibitem[{{Kennicutt}(1988)}]{kennicutt+88}
{Kennicutt}, Jr. R.~C., 1988, \apj, 334, 144

\bibitem[{{Kennicutt}(1992)}]{kennicutt92}
{Kennicutt}, Jr. R.~C., 1992, \apj, 388, 310

\bibitem[{{Kotilainen} {et~al}\mbox{.}(2012){Kotilainen}, {Hyv{\"o}nen},
  {Reunanen}, \& {Ivanov}}]{kotilainen+12}
{Kotilainen} J.~K., {Hyv{\"o}nen} T., {Reunanen} J., {Ivanov} V.~D., 2012,
  \mnras, 425, 1057

\bibitem[{{Lan{\c c}on} {et~al}\mbox{.}(2001){Lan{\c c}on}, {Goldader},
  {Leitherer}, \& {Gonz{\'a}lez Delgado}}]{lancon+01}
{Lan{\c c}on} A., {Goldader} J.~D., {Leitherer} C., {Gonz{\'a}lez Delgado}
  R.~M., 2001, \apj, 552, 150

\bibitem[{{Lyubenova} {et~al}\mbox{.}(2012){Lyubenova}, {Kuntschner},
  {Rejkuba}, {Silva}, {Kissler-Patig}, \& {Tacconi-Garman}}]{lyubenova+12}
{Lyubenova} M., {Kuntschner} H., {Rejkuba} M., {Silva} D.~R., {Kissler-Patig}
  M., {Tacconi-Garman} L.~E., 2012, \aap, 543, A75

\bibitem[{{Maraston}(2005)}]{maraston05}
{Maraston} C., 2005, \mnras, 362, 799

\bibitem[{{Martins} {et~al}\mbox{.}(2010){Martins}, {Riffel},
  {Rodr{\'{\i}}guez-Ardila}, {Gruenwald}, \& {de Souza}}]{martins+10}
{Martins} L.~P., {Riffel} R., {Rodr{\'{\i}}guez-Ardila} A., {Gruenwald} R., {de
  Souza} R., 2010, \mnras, 406, 2185

\bibitem[{{Melbourne} {et~al}\mbox{.}(2010){Melbourne}, {Williams},
  {Dalcanton}, {Ammons}, {Max}, {Koo}, {Girardi}, \& {Dolphin}}]{melbourne+10}
{Melbourne} J., {Williams} B., {Dalcanton} J., {Ammons} S.~M., {Max} C., {Koo}
  D.~C., {Girardi} L., {Dolphin} A., 2010, \apj, 712, 469

\bibitem[{{Melbourne} {et~al}\mbox{.}(2012){Melbourne}, {Williams},
  {Dalcanton}, {Rosenfield}, {Girardi}, {Marigo}, {Weisz}, {Dolphin}, {Boyer},
  {Olsen}, {Skillman}, \& {Seth}}]{melbourne+12}
{Melbourne} J. {et~al.}, 2012, \apj, 748, 47

\bibitem[{{Micheva} {et~al}\mbox{.}(2012){Micheva}, {{\"O}stlin}, {Zackrisson},
  {Bergvall}, {Marquart}, {Masegosa}, {Marquez}, {Cumming}, \&
  {Durret}}]{micheva+12}
{Micheva} G. {et~al.}, 2012, ArXiv e-prints

\bibitem[{{Miner} {et~al}\mbox{.}(2011){Miner}, {Rose}, \& {Cecil}}]{miner+11}
{Miner} J., {Rose} J.~A., {Cecil} G., 2011, in Bulletin of the American
  Astronomical Society, Vol.~43, American Astronomical Society Meeting
  Abstracts \#217, p. 211.02

\bibitem[{{Moorwood} \& {Glass}(1982)}]{moorwood+82}
{Moorwood} A.~F.~M., {Glass} I.~S., 1982, \aap, 115, 84

\bibitem[{{Moorwood} \& {Oliva}(1988)}]{moorwood+88}
{Moorwood} A.~F.~M., {Oliva} E., 1988, \aap, 203, 278

\bibitem[{{Oliva} {et~al}\mbox{.}(1995){Oliva}, {Origlia}, {Kotilainen}, \&
  {Moorwood}}]{oliva+95}
{Oliva} E., {Origlia} L., {Kotilainen} J.~K., {Moorwood} A.~F.~M., 1995, \aap,
  301, 55

\bibitem[{{Origlia} {et~al}\mbox{.}(1993){Origlia}, {Moorwood}, \&
  {Oliva}}]{origlia+93}
{Origlia} L., {Moorwood} A.~F.~M., {Oliva} E., 1993, \aap, 280, 536

\bibitem[{{Osterbrock}(1989)}]{osterbrock89}
{Osterbrock} D.~E., 1989, {Astrophysics of gaseous nebulae and active galactic
  nuclei}

\bibitem[{{Persson} {et~al}\mbox{.}(1983){Persson}, {Aaronson}, {Cohen},
  {Frogel}, \& {Matthews}}]{persson+83}
{Persson} S.~E., {Aaronson} M., {Cohen} J.~G., {Frogel} J.~A., {Matthews} K.,
  1983, \apj, 266, 105

\bibitem[{{Puxley} \& {Brand}(1994)}]{puxley+94}
{Puxley} P.~J., {Brand} P.~W.~J.~L., 1994, \mnras, 266, 431

\bibitem[{{Rayner} {et~al}\mbox{.}(2009){Rayner}, {Cushing}, \&
  {Vacca}}]{rayner+09}
{Rayner} J.~T., {Cushing} M.~C., {Vacca} W.~D., 2009, \apjs, 185, 289

\bibitem[{{Rayner} {et~al}\mbox{.}(2003){Rayner}, {Toomey}, {Onaka}, {Denault},
  {Stahlberger}, {Vacca}, {Cushing}, \& {Wang}}]{rayner+03}
{Rayner} J.~T., {Toomey} D.~W., {Onaka} P.~M., {Denault} A.~J., {Stahlberger}
  W.~E., {Vacca} W.~D., {Cushing} M.~C., {Wang} S., 2003, \pasp, 115, 362

\bibitem[{{Reunanen} {et~al}\mbox{.}(2002){Reunanen}, {Kotilainen}, \&
  {Prieto}}]{reunanen+02}
{Reunanen} J., {Kotilainen} J.~K., {Prieto} M.~A., 2002, \mnras, 331, 154

\bibitem[{{Reunanen} {et~al}\mbox{.}(2003){Reunanen}, {Kotilainen}, \&
  {Prieto}}]{reunanen+03}
{Reunanen} J., {Kotilainen} J.~K., {Prieto} M.~A., 2003, \mnras, 343, 192

\bibitem[{{Rieke} {et~al}\mbox{.}(1980){Rieke}, {Lebofsky}, {Thompson}, {Low},
  \& {Tokunaga}}]{rieke+80}
{Rieke} G.~H., {Lebofsky} M.~J., {Thompson} R.~I., {Low} F.~J., {Tokunaga}
  A.~T., 1980, \apj, 238, 24

\bibitem[{{Riffel} {et~al}\mbox{.}(2009){Riffel}, {Pastoriza},
  {Rodr{\'{\i}}guez-Ardila}, \& {Bonatto}}]{riffel+09}
{Riffel} R., {Pastoriza} M.~G., {Rodr{\'{\i}}guez-Ardila} A., {Bonatto} C.,
  2009, \mnras, 400, 273

\bibitem[{{Riffel} {et~al}\mbox{.}(2008){Riffel}, {Pastoriza},
  {Rodr{\'{\i}}guez-Ardila}, \& {Maraston}}]{riffel+08}
{Riffel} R., {Pastoriza} M.~G., {Rodr{\'{\i}}guez-Ardila} A., {Maraston} C.,
  2008, \mnras, 388, 803

\bibitem[{{Riffel} {et~al}\mbox{.}(2006){Riffel}, {Rodr{\'{\i}}guez-Ardila}, \&
  {Pastoriza}}]{riffel+06}
{Riffel} R., {Rodr{\'{\i}}guez-Ardila} A., {Pastoriza} M.~G., 2006, \aap, 457,
  61

\bibitem[{{Rodr{\'{\i}}guez-Ardila}
  {et~al}\mbox{.}(2004){Rodr{\'{\i}}guez-Ardila}, {Pastoriza}, {Viegas},
  {Sigut}, \& {Pradhan}}]{rodriguez-ardila+04}
{Rodr{\'{\i}}guez-Ardila} A., {Pastoriza} M.~G., {Viegas} S., {Sigut} T.~A.~A.,
  {Pradhan} A.~K., 2004, \aap, 425, 457

\bibitem[{{Rodr{\'{\i}}guez-Ardila}
  {et~al}\mbox{.}(2008){Rodr{\'{\i}}guez-Ardila}, {Riffel}, \&
  {Carvalho}}]{rodriguez-ardila+08}
{Rodr{\'{\i}}guez-Ardila} A., {Riffel} R., {Carvalho} E.~A., 2008, in Revista
  Mexicana de Astronomia y Astrofisica Conference Series, Vol.~32, Revista
  Mexicana de Astronomia y Astrofisica Conference Series, pp. 77--79

\bibitem[{{Schlafly} \& {Finkbeiner}(2011)}]{schlafly+11}
{Schlafly} E.~F., {Finkbeiner} D.~P., 2011, \apj, 737, 103

\bibitem[{{Vega} {et~al}\mbox{.}(2009){Vega}, {Asari}, {Cid Fernandes},
  {Garcia-Rissmann}, {Storchi-Bergmann}, {Gonz{\'a}lez Delgado}, \&
  {Schmitt}}]{vega+09}
{Vega} L.~R., {Asari} N.~V., {Cid Fernandes} R., {Garcia-Rissmann} A.,
  {Storchi-Bergmann} T., {Gonz{\'a}lez Delgado} R.~M., {Schmitt} H., 2009,
  \mnras, 393, 846

\bibitem[{{Worthey} \& {Ottaviani}(1997)}]{worthey+97}
{Worthey} G., {Ottaviani} D.~L., 1997, \apjs, 111, 377

\bibitem[{{Zibetti} {et~al}\mbox{.}(2012){Zibetti}, {Gallazzi}, {Charlot},
  {Pierini}, \& {Pasquali}}]{zibetti+12}
{Zibetti} S., {Gallazzi} A., {Charlot} S., {Pierini} D., {Pasquali} A., 2012,
  ArXiv e-prints

\end{thebibliography}
\end{document}